\documentclass[twocolumn,amsmath,amssymb,nofootinbib,floatfix]{revtex4}

\usepackage[dvipdfmx]{graphicx} 
\usepackage{dcolumn}
\usepackage{bm}
\usepackage{amsmath,amssymb}
\usepackage{color}
\usepackage{longtable}
\usepackage{ulem}

\newcommand\Ocal{\mathcal{O}}

\newcommand{\ovli}[1]{{\overline{#1}}}

\newcommand\QQbar{Q\overline{Q}}
\newcommand\Qbar{\overline{Q}}
\newcommand\mQ{m_{Q}}

\newcommand\SdotS{{\bm S}_Q\cdot{\bm S}_{\overline{Q}}}

\newcommand\SQ{{\bm S}_Q}
\newcommand\SQbar{{\bm S}_{\overline{Q}}}
\newcommand\tsrc{t_{\rm s}}

\usepackage[colorlinks=true, linkcolor=blue, filecolor=blue, urlcolor=blue, citecolor=blue, pdftex=true, plainpages=false]{hyperref}

\begin{document}
\title{
  Potential description of charmonium and charmed-strange mesons from lattice QCD
}


\author{Taichi Kawanai${}^{1}$} \email{t.kawanai@fz-juelich.de}
\author{Shoichi Sasaki${}^{2,3}$} \email{ssasaki@nucl.phys.tohoku.ac.jp}
\affiliation{${}^{1}$J\"ulich Supercomputing Center, J\"ulich D-52425, Germany} 
\affiliation{${}^{2}$Department of Physics, Tohoku University, Sendai 980-8578, Japan}
\affiliation{${}^{3}$Theoretical Research Division, Nishina Center, RIKEN, Wako 351-0198, Japan}


\date{\today}

\begin{abstract}

We present spin-independent and spin-spin interquark potentials for 
the charmonium and charmed-strange mesons, which are
calculated in 2+1 flavor lattice QCD simulations using the PACS-CS 
gauge configurations generated at the lightest pion mass ($M_\pi \approx 156(7)$~MeV)
with a lattice cutoff of $a^{-1}\approx 2.2$~GeV and a spatial volume of $(3~{\rm fm})^3$.
For the charm quark, we use a relativistic heavy quark (RHQ) action with fine tuned RHQ
parameters, which closely reproduce both the experimental spin-averaged mass
and hyper-fine splitting of the $1S$ charmonium.  
The interquark potential and the quark kinetic mass, both of which are
key ingredients within the potential description
of heavy-heavy and heavy-light mesons, are determined from 
the equal-time Bethe-Salpeter (BS) amplitude. 
The charmonium potentials are obtained from the BS wave function of
$1S$ charmonia ($\eta_c$ and $J/\psi$ mesons),
while the charmed-strange
potential are calculated from the $D_s$ and $D_s^{\ast}$ heavy-light
mesons. We then use resulting potentials and quark masses as purely theoretical inputs so as to 
solve the nonrelativistic Schr\"odinger equation for calculating accessible energy levels 
of charmonium and charmed-strange mesons without unknown parameters. 
The resultant spectra below the $D\bar{D}$ and $DK$ thresholds excellently agree 
with well-established experimental data. 
\end{abstract}

\maketitle

%
%
\section{\label{intro} Introduction}
The heavy-quark~($Q$)-antiquark~($\Qbar$) potential is 
an important quantity to understand properties of the heavy quarkonium states.
Because the dynamics of heavy quarks with masses much larger 
than the QCD scale~($\Lambda_{\rm QCD}$) is well described
within the framework of nonrelativistic quantum mechanics~\cite{Close:1979bt}.
Indeed the constituent quark potential models with a QCD-motivated $\QQbar$ potential have 
successfully predicted the heavy quarkonium spectra and its decay rates
below open charm thresholds~\cite{Eichten:1974af, Godfrey:1985xj, Barnes:2005pb}.

In such nonrelativistic potential~(NRp) models, the conventional heavy quarkonium states 
such as charmonium and bottomonium 
are well understood to be quark-antiquark pair bound by the 
Coulombic potential induced by a perturbative one-gluon exchange that dominates in short range,
plus linearly rising potential that describes the phenomenology of 
confining quark interactions at large distances~\cite{Eichten:1974af}.
This potential is called the Cornell potential and its functional form is given by\begin{equation}
 V(r) = - \frac{4}{3} \frac{\alpha_s}{r} + \sigma r + V_0
\end{equation}
where $\alpha_s$ is the strong coupling constant, $\sigma$ denotes the string tension 
and $V_0$ is the constant term associated with a self-energy contribution of the color sources.
In addition to the spin-independent potential, 
the NRp models include spin-dependent interactions, which
resolve the degeneracy among spin-multiplets.
The spin-dependent potentials appear as relativistic corrections 
in powers of the relative velocity of quarks, and their functional forms are also 
determined by perturbative one-gluon exchange as the 
Fermi-Breit type potential~\cite{Eichten:1980mw}. 
A more direct connection to
QCD is established by the modern approach of effective field theory 
called potential nonrelativistic QCD (pNRQCD)~\cite{Brambilla:2004jw}.

We would like to stress here that 
the functional forms of the $\QQbar$ potentials except at long distances 
are basically deduced by the perturbative approach.
Furthermore all of parameters needed in the NRp models, including a constituent quark mass $m_Q$,
are phenomenologically fixed to reproduce the experimental heavy quarkonium masses~\cite{Barnes:2005pb,Godfrey:1985xj}.
The phenomenological spin-dependent potentials based on the perturbative method 
would have validity only at short distances and also in the vicinity of the heavy quark mass limit.
This fact could cause large uncertainties in predictions for the higher-lying states of the heavy quarkonium
in the NRp models.

In addition, many of the charmonium-like mesons have been announced by 
$B$-factories at KEK and SLAC, which are primarily devoted to the physics of CP violation,
also by Charm factories at BEPC and CESR, and Tevatron at Fermilab.
These newly discovered state above the open charm threshold could not be simply explained 
as conventional charmonium states in the constituent quark description~\cite{Voloshin:2007dx}.
Indeed, the existence of the charged $Z$ states including two charged bottomonium-like states,
$Z_b(10610)$ and $Z_b(10650)$~\cite{Belle:2011aa} indicates that the charmonium-like $XYZ$ mesons are good 
candidates for non-standard quarkonium mesons such as hadronic molecular states, diquark-antidiquark 
bound states (tetraquark states), or hybrid mesons~\cite{Godfrey:2008nc}. 

To discriminate between standard and nonstandard mesons 
in a zoo of the charmonium-like $XYZ$ mesons, it is essential to
investigate the validity of the potential description of the heavy-heavy
and heavy-light mesons directly from first principles of QCD. 
In this paper, we thus aim to provide the central $\QQbar$ 
potentials~(the spin-independent potential and the spin-spin potential), which 
are determined through the Bethe-Salpeter (BS) amplitudes of pseudoscalar and vector mesons 
in dynamical lattice QCD simulations with almost physical quark masses.

In lattice QCD, understanding the properties of $\QQbar$ interactions is one of the great historic 
milestones. The Wilson loop has been originally introduced as a non-local order parameter in $Z_2$ gauge theory
by Wegner~\cite{Wegner:1984qt}.
Subsequently, Wilson generalized it with continuous gauge groups and related it to the static potential 
between infinitely heavy-quark and antiquark in QCD so as to prove the quark confinement 
in the strong coupling limit~\cite{Wilson:1974sk}. 
The static $\QQbar$ potential determined from Wilson loops have been precisely determined 
by lattice QCD in the past decades. The lattice QCD calculations within the Wilson loop formalism 
support a shape of the Cornell potential~\cite{Bali:2000gf}.

On the other hand, the spin-dependent $\QQbar$ potentials 
regarded as the relativistic corrections to the static potential can be determined 
within the framework of pNRQCD.
Although earlier quenched studies~\cite{Bali:1996cj, Bali:1997am} 
and full QCD studies~\cite{Koike:1989jf, Born:1993cp}
did not enable us to determine the functional forms 
of the spin-dependent terms due to large statistical errors, a full set of the spin-dependent terms
({\it i.e.} spin-spin, spin-orbit and tensor terms) have been successfully 
calculated in {\it quenched} QCD with high accuracy 
by using the multilevel algorithm~\cite{Koma:2006fw,Koma:2010zz}.

It is worth mentioning that the multilevel algorithm 
employed in Refs.~\cite{Koma:2006fw,Koma:2010zz} is not easy to be 
implemented in dynamical lattice QCD simulations. 
Furthermore, the leading spin-spin potential determined at $\Ocal(1/m_Q^2)$ in
quenched QCD gives an attractive interaction for the higher spin states 
in the hyperfine multiplet~\cite{Koma:2006fw,Koma:2010zz}. This
contradicts with the spin-spin term of the Fermi-Breit type potential, which
is described by a {\it repulsive} contact interaction.
Although one might think that the inverse of the charm quark mass would be 
far outside the validity region of the $1/\mQ$ expansion, 
this issue still remains even at the bottom quark mass. 

We develop the new method proposed 
in our previous works~\cite{Kawanai:2011xb,Kawanai:2011jt, Kawanai:2013aca}
in order to obtain {\it proper interquark potentials at finite quark masses}, which are 
indispensable for the potential description of the charmonium and charmed-strange mesons.
The interquark potential and the quark kinetic mass, both of which are key ingredients
within the potential description, can be defined by the equal-time and Coulomb gauge BS amplitude 
through an effective Schr\"odinger equation~\cite{Kawanai:2011xb}.
This new method enables us to determine the interquark potentials including spin-dependent terms
at finite quark masses from first principles of QCD, and then fix all parameters needed in the NRp models.
In our previous works with quenched lattice simulations~\cite{Kawanai:2011xb,Kawanai:2013aca}, 
we demonstrated that both spin-independent central potential and spin-spin potential calculated 
in the BS amplitude method reproduce known results calculated within the Wilson loop formalism in the 
$\mQ \rightarrow \infty$ limit. We read off from our $\QQbar$ potentials, which
may encode all orders of the $1/\mQ$ expansion, that the $1/\mQ$ expansion 
scheme may have the convergence behavior up to the bottom sector, while the charm sector 
is far outside the validity region for this expansion~\cite{Kawanai:2013aca}. Furthermore, we found that 
the higher order corrections beyond the next-to-leading order are inevitably required for 
the repulsive feature of the total spin-spin potential even at the bottom sector~\cite{Kawanai:2013aca}.  
In addition, there is no restriction to extend the new method to dynamical calculation~\cite{Kawanai:2011jt}. 
Hereafter we call the new method as {\it BS amplitude method}.

Once one gets the reliable $\QQbar$ potentials, which contain both the spin-dependent 
contributions as well as the spin-independent central one, we can easily verify how well the potential
description is satisfied in the heavy-heavy and heavy-light meson systems through
solving the nonrelativistic Schr\"odinger equation with purely theoretical inputs.
If the potential description is valid, many physical observables such as mass spectrum of 
heavy-heavy and heavy-light mesons and their decay rates are easily accessible 
as is in the NRp models.

In this paper, we extend our previous work~\cite{Kawanai:2011jt} done 
in 2+1 flavor lattice QCD simulations using the PACS-CS gauge 
configurations~\cite{Aoki:2008sm} in order to investigate the validity of 
the potential description of the heavy-heavy and heavy-light mesons.
The simulated pion mass ($M_\pi \approx 156(7)$~MeV)
is close to the physical point, while the simulated $K$ meson mass 
as $M_K \approx 554(2)$~MeV  is about 10\% heavier than the physical value.
Although the strange quark is slightly 
off the physical point, the parameters of clover fermions for the strange quark
are chosen to be equal to those of the strange sea quarks used in gauge field generation.
For the charm quark, we employ the relativistic heavy quark (RHQ) action that 
can control large discretization errors induced by large quark mass~\cite{Aoki:2001ra}.
The RHQ parameters in the action were calibrated to reproduce the experimental
spin-averaged mass and hyper-fine splitting of the $1S$ charmonium.

We first concentrate on the heavy-heavy systems so as to calculate 
the charm quark mass and the charmonium potential from 
the BS amplitudes of $1S$ charmonia ($\eta_c$ and $J/\psi$ mesons).
We reuse the data, which were previously published in Ref.~\cite{Kawanai:2011jt}, 
and then perform a more elaborate analysis proposed in Ref.~\cite{Kawanai:2013aca}.
New analysis significantly reduces systematic uncertainties on the shape of 
the charmonium potential at short distances due to the usage of the highly improved
Laplacian operator~\footnote{Note that the binding energy of the low-lying charmonium states,
which we may consider to be nearly Coulombic bound states, are very sensitive to details of the short-range interaction.}.
Once the charmonium potential and the charm quark mass are precisely
determined, we can numerically solve the nonrelativistic Schr\"odinger equation
with such theoretical inputs and without additional parameters. 

We then extend our research to the $D_s$ heavy-light meson systems
to extract the strange quark mass and the charmed-strange potential
from the BS amplitudes of two lightest charmed-strange mesons
({\it i.e.} the $D_s$ and $D_s^{\ast}$ heavy-light mesons).
We will then discuss the validity of the potential description on 
both charmonium and charmed-strange mesons.

This paper is organized as follows.
In Sec.~\ref{formalism}, we briefly describe the methodology 
to calculate the spin-independent and spin-dependent interquark potentials
from the BS amplitude of heavy-heavy and heavy-light mesons in lattice QCD simulations.
In Sec.~\ref{setup} we give the details of parameters used in our Monte Carlo simulations,
and then discuss the charmonium mass obtained from the standard lattice spectroscopy with 
two point correlation functions of mesons.
In Sec.~\ref{main}, we show numerical results of the BS wave function, 
the quark kinetic mass $\mQ$, the spin-independent central and spin-spin potentials,
calculated from dynamical lattice QCD simulations.
In Sec.~\ref{main2}, we show the charmonium mass spectrum obtained by solving 
the nonrelativistic Schr\"odinger equation with the theoretical inputs determined 
from dynamical lattice QCD simulations at almost physical point, and finally discuss 
possible systematic uncertainties on the resulting energy spectrum of the 
charmonium states.
In Sec.~\ref{Ds}, we present the results from an application to the $D_s$ heavy-light meson systems.
In Sec.~\ref{summary},
we summarize and discuss all results and future perspectives.

%
%
\section{\label{formalism} Formalism}
In this section, we will briefly review the BS amplitude method to
calculate the interquark potential with the finite quark mass.
This is an application based on 
the approach originally used for the hadron-hadron potential, 
which is defined through the equal-time BS amplitude~\cite{Ishii:2006ec,Aoki:2009ji,
Nemura:2008sp,Ikeda:2010zz, Kawanai:2010ev,Doi:2011gq,Aoki:2012tk,Murano:2011nz,Aoki:2011gt,HALQCD:2012aa}.
More details of determination of the interquark potential are given in Ref.~\cite{Kawanai:2013aca}.

For simplicity, we here consider the case of the heavy quarkonium $\QQbar$. 
An extension to the heavy-light meson made of two non-degenerate quarks is easy.
In lattice simulations,
we measure the following equal-time $\QQbar$ 
BS amplitude in the Coulomb gauge for the 
quarkonium states~\cite{{Velikson:1984qw},{Gupta:1993vp}}:
\begin{equation}
  \phi_\Gamma({\bm r})= \sum_{{\bm x}}\langle 0| \overline{Q}
  ({\bm x})\Gamma Q({\bm x}+{\bm r})|
  \QQbar;J^{PC}\rangle \label{eq_phi},
\end{equation}
where ${\bm r}$ is the relative coordinate between quark and antiquark at a certain 
time slice $t$. The operator $\Gamma$ appeared in Eq.~(\ref{eq_phi}) represents 
the Dirac $\gamma$ metrics, which specifies the spin and the parity of meson operators.
For instance, with $\gamma_5$ and $\gamma_i$, one can form   
the pseudoscalar (PS)  and the vector (V) operators
with $J^{P}=0^{-}$ and $J^{P}=1^{-}$, respectively.
A summation over spatial coordinates ${\bm x}$ projects out corresponding states 
with zero total momentum. The ${\bm r}$-dependent amplitude, $\phi_\Gamma({\bm r})$, 
is called {\it BS wave function}.
The BS wave function can be extracted from the four-point correlation function
\begin{eqnarray}
G_{\Gamma}({\bm r}, t, t_{\rm s})&=&
\sum_{{\bm x}, {\bm x}^\prime, {\bm y}^\prime}
\langle 0|\overline{Q}({\bm x}, t)\Gamma Q({\bm x}+{\bm r}, t)  \cr
&&\times \left( 
\overline{Q}({\bm x^\prime}, t_{\rm s})\Gamma Q({\bm y}^\prime, t_{\rm s}) 
\right)^{\dagger}
| 0 \rangle
\label{eq_4ptF}
\end{eqnarray}
at large time separation between the source ($t_{\rm S}$)
and sink ($t$) locations ($|t-t_{\rm S}|/a\gg 1$)~\cite{Kawanai:2013aca}.
Here, the gauge field configurations are necessarily fixed to the Coulomb gauge at both
time slices $t$ and $t_{\rm S}$.
In the limit of ${\bm r}\rightarrow 0$, the four-point correlation functions are reduced to the 
two-point correlation functions of mesons with a wall source.
In this paper, we focus only on the $S$-wave BS wave function ($\eta_c$ and $J/\psi$
for the charmonium and $D_s$ and $D_s^{\ast}$ for the charmed-strange meson),
obtained by an appropriate projection to the $A^{+}_{1}$ representation in cubic group~\cite{Luscher:1990ux}.

Below the inelastic threshold~\footnote{For the charmonium system, the inelastic threshold implies 
the $D\bar{D}$ threshold, while the $DK$ threshold is a counterpart in the charmed-strange meson system.}, 
the BS wave function satisfies an effective Schr\"odinger
equation with a nonlocal and energy-independent 
interquark potential $U$~\cite{Ishii:2006ec,Caswell:1978mt,Ikeda:2011bs} 
\begin{equation}
  -\frac{{\bm \nabla}^2}{2\mu}\phi_\Gamma({\bm r})+
  \int dr'U({\bm r},{\bm r}')\phi_\Gamma({\bm r}')
  =E_\Gamma\phi_\Gamma({\bm r}),
  \label{Eq_schr}
\end{equation}
where $\mu$ is the reduced mass of the $\QQbar$ system.
The energy eigenvalue $E_\Gamma$ of the stationary 
Schr\"odinger equation is supposed to be $M_\Gamma-2\mQ$.
If the relative quark velocity $v=|{{\bm \nabla}}/\mQ|$ is small as $v \ll 1$, 
the nonlocal potential $U$ can generally expand in terms of the velocity $v$ as 
$U({\bm r}',{\bm r})=  \{V(r)+V_{\text{S}}(r)\SdotS +V_{\text{T}}(r)S_{12}+
 V_{\text{LS}}(r){\bm L}\cdot{\bm S} + \mathcal{O}(v^2)\}\delta({\bm r}'-{\bm r})$
where $S_{12}=(\SQ\cdot\hat{\bm r})(\SQbar\cdot\hat{\bm r})-\SdotS/3$
with $\hat{\bm r}={\bm r}/r$, ${\bm S}=\SQ+\SQbar$
and ${\bm L} = {\bm r}\times (-i{\bm \nabla})$~\cite{Ishii:2006ec}.
Here, $V$, $V_{\text{S}}$, $V_{\text{T}}$ and $V_{\text{LS}}$ represent
the spin-independent central, spin-spin, tensor and spin-orbit potentials, 
respectively.
 
The Schr\"odinger equation for $S$-wave states is simplified as 
\begin{equation}
  \left\{
  - \frac{{\bm \nabla}^2}{\mQ}
  +V(r)+\SdotS V_{\text{S}}(r)
  \right\}\phi_{\Gamma}(r)=E_\Gamma \phi_{\Gamma}(r)
  \label{Eq_pot}
\end{equation}
at the leading order of the $v$-expansion.
Here, we essentially follow the NRp models,
where the $J/\psi$ state is purely composed of the $1S$ wave function. 
However, within this method, this assumption can be verified 
by evaluating the size of a mixing between $1S$ and $1D$ wave
functions in principle.  

The spin operator $\SdotS$ can be easily replaced by its expectation values:
$-3/4$ and $1/4$ for the PS and V channels, respectively. 
Then, the spin-independent and spin-spin $\QQbar$ potentials can be  evaluated 
through the following linear combinations of Eq.(\ref{Eq_pot}):
 \begin{eqnarray}
   V(r)
   &=& E_{\text{ave}}+\frac{1}{\mQ}\left\{
   \frac{3}{4}\frac{{\bm \nabla}^2\phi_\text{V}(r)}{\phi_\text{V}(r)}+
    \frac{1}{4}\frac{{\bm \nabla}^2\phi_\text{PS}(r)}{\phi_\text{PS}(r)}
   \right\} \label{Eq_potC}\\
   V_{\text{S}}(r) 
   &=& E_{\text{hyp}} + \frac{1}{\mQ}\left\{
  \frac{{\bm \nabla}^2\phi_\text{V}(r)}{\phi_\text{V}(r)} 
  - \frac{{\bm \nabla}^2\phi_\text{PS}(r)}{\phi_\text{PS}(r)} \right\},\label{Eq_potS}
 \end{eqnarray}
 where $E_{\text{ave}}=M_{\text{ave}}-2\mQ$ 
 and $E_{\text{hyp}}=M_\text{V}-M_\text{PS}$.
 The mass $M_{\text{ave}}$ denotes the spin-averaged mass as  
 $\frac{1}{4}M_\text{PS}+\frac{3}{4}M_\text{V}$.
 The derivative ${\bm \nabla}^2$ is defined by the discrete Laplacian on the lattice.
 
 In the BS amplitude method, there is a room for optimizing the differential
 operator since the discrete Laplacian is itself build in the definition of 
 the interquark potential. In Ref.~\cite{Kawanai:2013aca}, we showed that
 the discrete Laplacian operator defined in the discrete polar coordinates
 called {\it r-Laplacian} is more
 suitable than the naive one defined in the Cartesian coordinates from the viewpoint
 of the reduction of the discretization artifacts on the short-range behavior
 of the interquark potential. The latter was adopted 
 in our earlier works~\cite{Kawanai:2011xb,Kawanai:2011jt}, while 
 we use the {\it r-Laplacian} throughout this paper. 
 For details of the discrete Laplacian operators, we will explain in Sec.~\ref{main}.
 
The quark kinetic mass is also an important quantity in the determination 
of the interquark potentials
since Eq.~(\ref{Eq_potC}) and Eq.~(\ref{Eq_potS})
requires information of the quark kinetic mass $m_Q$.
In Ref.~\cite{Kawanai:2011xb}, 
we propose to calculate the quark kinetic mass through the large-distance behavior 
of the difference of ``quantum kinetic energies'' (the second derivative of the BS wave
function normalized by the BS wave function) between the spin-singlet and -triplet states 
in the hyperfine multiplet. The most simple choice is of course a pair of $^1S_0$ and $^3S_1$ states.
Contributions of the long-range confining force are canceled out
in the difference of ``quantum kinetic energies''. 
Under a simple, but reasonable assumption as $\lim_{r\to\infty} V_S(r)=0$  
which implies there is no long-range correlation and no irrelevant constant term 
in the spin-spin potential, one may expect that the difference of ``quantum kinetic energies''
at long distances stems only from the hyperfine splitting energy $E_\text{hyp}$. 
Therefore, the quark kinetic mass can be read off in the following way: 
\begin{equation}
 m_Q = \lim_{r\to \infty} \frac{-1}{E_\text{hyp}} \left\{ \frac{{\bm \nabla}^2\phi_\text{V}(r)}{\phi_\text{V}(r)} 
  - \frac{{\bm \nabla}^2\phi_\text{PS}(r)}{\phi_\text{PS}(r)} \right\}.
\label{eq_quark_mass}
\end{equation}
The idea has been numerically tested, and the assumption of $\lim_{r\to\infty} V_S(r)=0$
is indeed appropriate in QCD~\cite{Kawanai:2011xb}.
We thus estimate the quark kinetic mass from asymptotic behavior of 
the right-hand side of Eq.~(\ref{eq_quark_mass}) in long-distance region.

\section{\label{setup}Lattice setup and heavy quarkonium  mass}

\subsection{$2+1$ flavor PACS-CS dynamical gauge ensemble}
%
%
\begin{table*}[t]
\caption{Parameters of $2+1$-flavor dynamical QCD gauge field configurations
generated by the PACS-CS collaboration~\cite{Aoki:2008sm}.
The columns list number of flavors, lattice volume, the $\beta$ value,
hopping parameters for light and strange quarks, approximate lattice spacing~(lattice cut-off),
spatial physical volume, pion mass, and number of configurations to be analyzed. }
\label{tab:ensembles_full}
\begin{ruledtabular}                                                                                                                        
 \begin{tabular}{ccccccccc}
  $N_f$ & $N_s^3\times N_t$   &$\beta$ & $\kappa_{ud}$ &$\kappa_{s}$ &$a$~[fm]~($a^{-1}$ [GeV])    
  & $N_sa$~[fm]  & $M_\pi$~[MeV] & \# configs. \\[2pt] \hline
  $2+1$ &$32^3\times 64$  &1.9     & 0.13781 & 0.13640 & 0.0907(13)~($\approx$ 2.18)  & 2.90(4) & $\approx$156 & 198\\ 
 \end{tabular}
\end{ruledtabular} 
\end{table*}
The computation of the interquark potential for the charmonium ($c\bar{c}$)
and charmed-strange ($c\bar{s}$) system is carried out on
a lattice $N_s^3\times N_t=32^3\times 64$ using the $2+1$ flavor PACS-CS gauge 
configurations~\cite{Aoki:2008sm}.
The gauge fields are generated by non-perturbatively $\mathcal{O}(a)$-improved Wilson quark action 
with $c_{SW} = 1.715$~\cite{Aoki:2005et} and Iwasaki gauge action at $\beta=1.90$~\cite{Iwasaki:2011np}, 
which corresponds to a lattice cutoff of $a^{-1} =  2.176(31)$ GeV ($a = 0.0907(13)$fm)~\cite{Aoki:2008sm}.
The spatial lattice size is of about $N_sa \sim 3\;{\rm fm}$.
The hopping parameters for the light sea quarks \{$\kappa_{ud}$,$\kappa_{s}$\}=\{0.13781, 0.13640\}
give a pion mass of $M_\pi=156(7)$ MeV and a kaon mass of $M_K= 554(2)$ MeV~\cite{Aoki:2008sm}.
Simulation parameters of dynamical QCD simulations used in this work is
summarized in Table~\ref{tab:ensembles_full}. 
Although the light sea quark masses are slightly off the physical point,
the systematic uncertainty due to this fact could be extremely small
in this project.
Our results are analyzed on all 198 gauge configurations,
which are available through International Lattice Data Grid and
Japan Lattice Data Grid~\footnote{International Lattice Data Grid/Japan Lattice Data Grid,
http://www.jldg.org.}. Gauge configurations is fixed to the Coulomb gauge.

\subsection{Relativistic charm quark}
\begin{table}
  \caption{
    The hopping parameter $\kappa_Q$ and RHQ parameters ($\nu$, $r_s$, $c_B$ and  $c_E$)
    used for the charm quark. 
      \label{tab:RHQ_para_charmonium}
      }
      \begin{ruledtabular}                                                                              
      \begin{tabular}{ccccc} 
	$\kappa_c$ & $\nu$ & $r_s$  & $c_B$ & $c_E$  \\ \hline
	0.10819 & 1.2153 & 1.2131  & 2.0268 & 1.7911  
      \end{tabular}
     \end{ruledtabular}                                                                                
\end{table}
In order to control discretization errors induced by large charm quark mass,
we employ the relativistic heavy quark~(RHQ) action~\cite{Aoki:2001ra} 
that removes main errors of $\Ocal(|{\bm{p}}|a)$, $\Ocal((m_0 a)^n )$ and 
$\Ocal(|{\bm{p}}|a (m_0 a)^n)$
from the on-shell Green's function.
The RHQ action is the anisotropic version of the $\Ocal (a)$ improved Wilson action 
with five parameters $\kappa_c$, $\nu$, $r_s$, $c_B$ and $c_E$,
called {\it RHQ parameters} (for more details see Ref.~\cite{Aoki:2001ra,Kayaba:2006cg}): 
\begin{multline}
 S_{\text{RHQ}}=  \sum_{x}\Qbar(x)\Big( m_0 a
  +  \gamma_0D_0+\nu\bm{\gamma}\cdot\bm{D} \\
  -\frac{r_t}{2}a(D^0)^2 
  -\frac{r_s}{2}a(\bm{D})^2 \\
  +\sum_{i,j}\frac{i}{4}c_Ba\sigma_{ij}F_{ij}
  +\sum_{i}\frac{i}{2}c_Ea\sigma_{0i}F_{0i}
                                            \Big)Q(x)
\end{multline}              
where the Wilson parameter for the time derivative is set to be $r_t = 1$
and the bare quark mass is related to the hopping parameter $\kappa_c$ as 
$am_0 = \frac{1}{2\kappa_c} - r_t -3r_s$.
The RHQ action utilized here is a variant of the Fermilab
approach~\cite{ElKhadra:1996mp}~(See also Ref.~\cite{Christ:2006us}).

The parameters $r_s$, $c_B$ and $c_E$ in RHQ action are determined
by tadpole improved one-loop perturbation theory~\cite{Kayaba:2006cg}.
For $\nu$, we use a nonperturbatively determined value,
which is tuned by reproducing the effective speed of light $c_{\text{eff}}$
to be unity in the dispersion relation
$E^2({\bm p}^2)=M^2+c^2_{\text{eff}}{\bm p}^2$
for the spin-averaged $1S$-charmonium state, since the parameter $\nu$
is sensitive to the size of hyperfine splitting energy~\cite{Namekawa:2011wt}.
We choose the value of $\kappa_c$ to reproduce the experimental spin-averaged mass
of $1S$-charmonium states $M_\text{ave}^{\text{exp}}(1S)=3.0678(3)$~GeV.
To calibrate RHQ parameters, we employ a gauge invariant
Gauss smearing source for the standard two-point correlation function
with four finite momenta.
As a result, the relevant speed of light in a energy-momentum 
dispersion relation~$E^2 = M^2 + c^2_{\rm eff}{\bm p}^2$ is consistent 
with unity within statistical uncertainties:
$c^2_{\text{eff}}=1.04(5)$ for the spin-averaged state~\cite{Kawanai:2011jt}.
Our chosen RHQ parameters are summarized in Table~\ref{tab:RHQ_para_charmonium}.

Using tuned RHQ parameters, 
we compute the two valence quark propagators with wall sources located
at different time slices $\tsrc/a=6$ and $57$ to increase statistics.
Two sets of two- and four-point correlation functions
are constructed from the corresponding $\Gamma$ operator with the charm quark propagator, and 
folded together to create the single correlation function.
Dirichlet boundary conditions are imposed for the time direction at $t/a=0$ and 63
to eliminate unwanted contributions across time boundaries.

\subsection{Charmonium spectroscopy from two-point functions}
\begin{table}
  \caption{
    Masses of low-lying charmonium states calculated from two-point functions, 
    the spin-averaged mass and hyperfine splitting energy of $1S$ charmonium states.
    Five charmonium states are classified with quantum numbers $J^{PC}$ and corresponding 
    operators $\Gamma$. 
    The fitting ranges and values of $\chi^2/\text{d.o.f.}$ are also included.
    Results are given in units of GeV.
    \label{tab:charmonium_mass}
      }
      \begin{ruledtabular}                                                                              
      \begin{tabular}{cccclc} 
       state & ($J^{PC}$) & $\Gamma$ & fit range & mass~[GeV] & $\chi^2/\text{d.o.f.}$\\ \hline
      $\eta_c$ & ($0^{-+}$)  & $\gamma_5$   & [33:47] & 2.9851(5)  & 0.70\\
      $J/\psi$ & ($1^{-+}$)    & $\gamma_{i}$    & [33:47] & 3.0985(11) & 0.62\\ 
      $M_{\text{ave}}(1S)$ & --- & ---  & --- & 3.0701(9)  & --- \\ 
      $E_{\text{hyp}}(1S)$  & --- & --- & --- & 0.1138(8)  & --- \\
      \hline
      $\chi_{c0}$ & ($0^{++}$)  & $1$ & [14:26] & 3.3928(59) & 0.66\\  
      $\chi_{c1}$ & ($1^{++}$)  & $\gamma_5\gamma_{i}$ & [14:26] &  3.4845(62) & 1.03\\ 
      $h_{c}$ & ($1^{+-}$)      & $\gamma_i\gamma_j$ &  [14:26] & 3.5059(62) & 0.63\\[2pt] 
    \end{tabular}
      \end{ruledtabular}                                                                                
\end{table}
 \begin{figure}
   \centering
  \includegraphics[width=.49\textwidth]{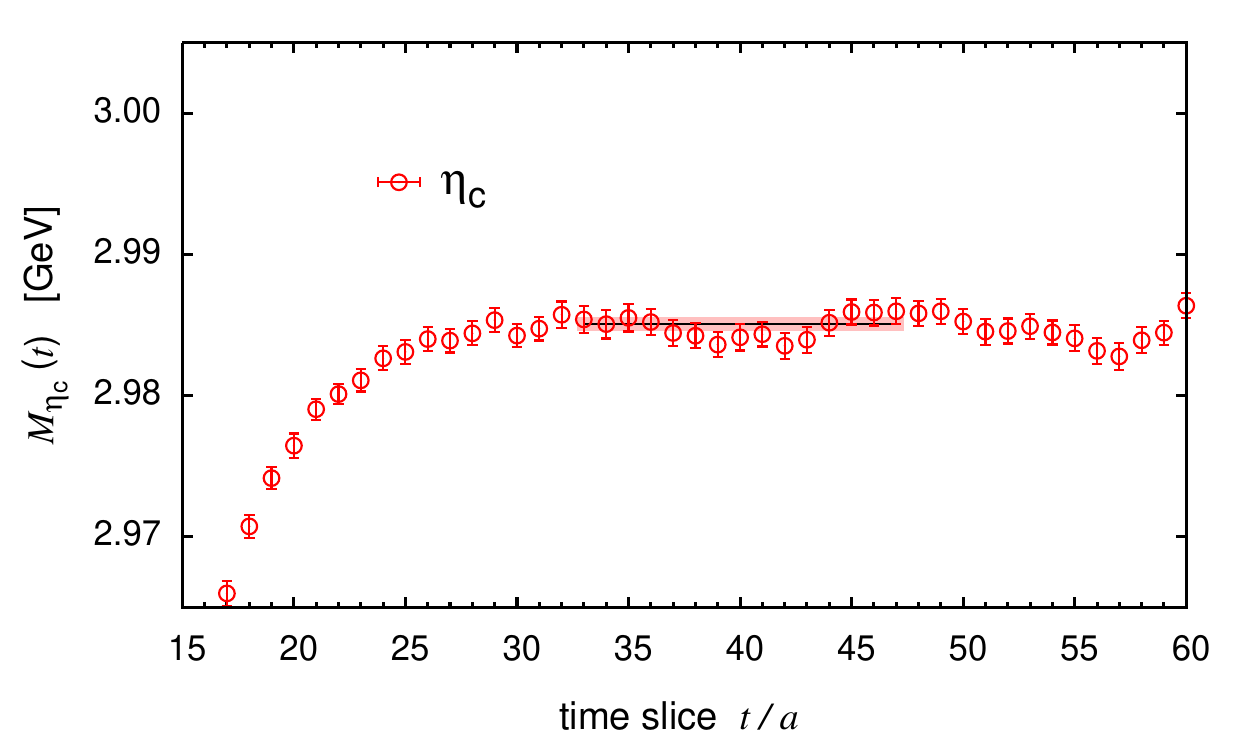}
   \includegraphics[width=.49\textwidth]{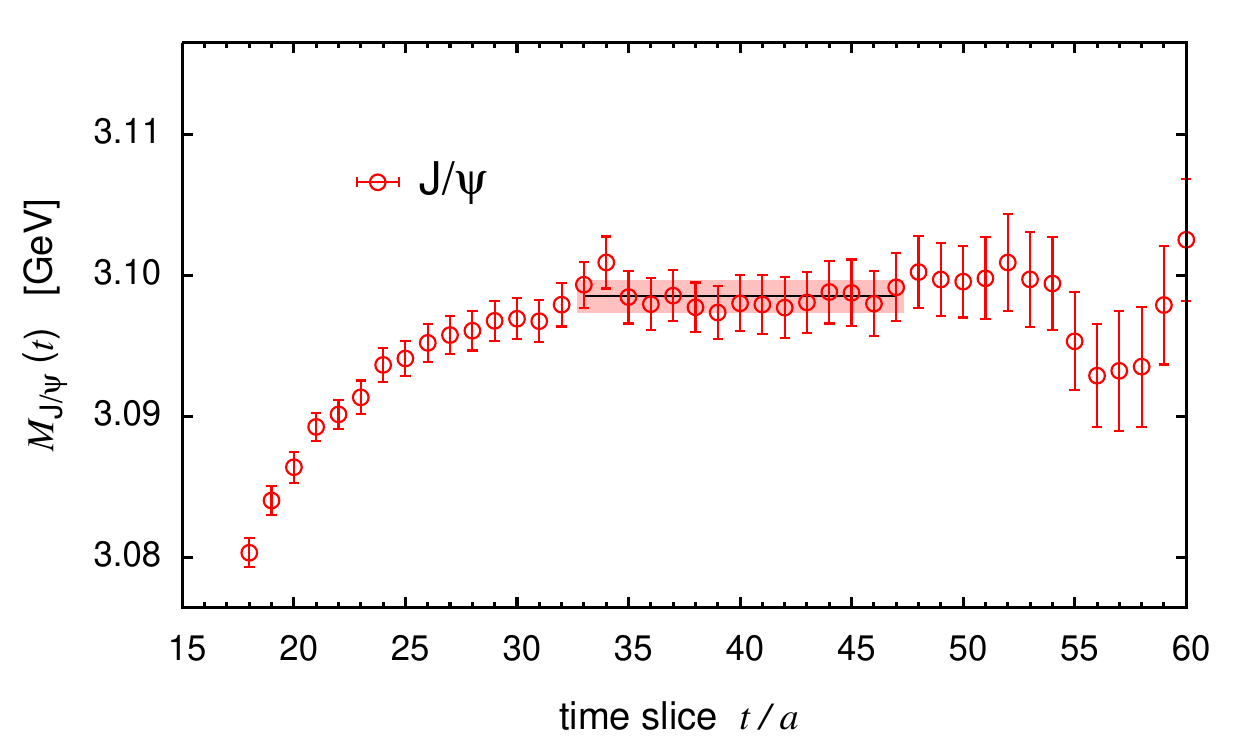}
   \includegraphics[width=.49\textwidth]{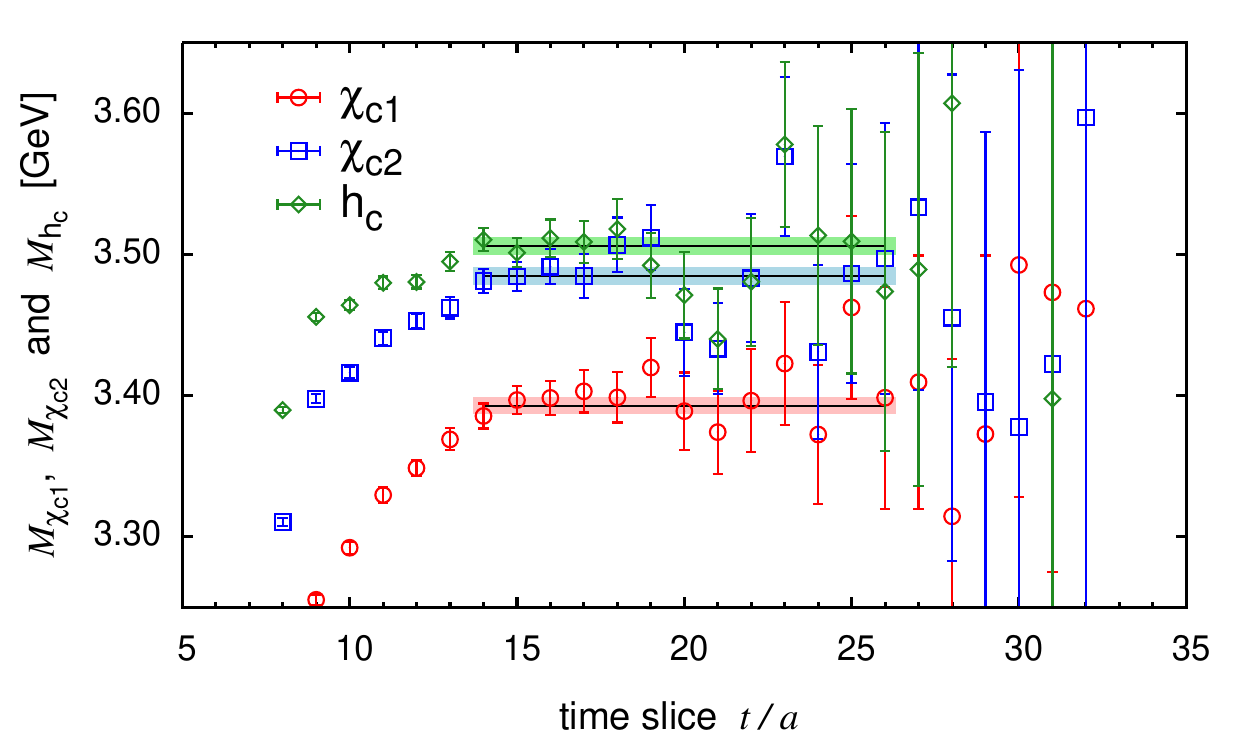}
   \caption{Effective mass plots for $\eta_c$~(upper panel), $J/\psi$~(center pannel)
    and $1P$ charmonium states~($\chi_{c0}$, $\chi_{c1}$ and $h_{c}$)~(lower pannel).
    Charmonium states are specified in legend.
    Solid lines indicate fit results and shaded bands display
    the fitting ranges and one standard deviations estimated by the jackknife method.
  }
  \label{fig:effmass}
 \end{figure}

Fig.~\ref{fig:effmass} shows the effective mass of the
$S$-wave~($\eta$ and $J/\psi$) and $P$-wave~($\chi_{c0}$, $\chi_{c1}$ and $h_{c}$)
charmonium states calculated from the dynamical lattice QCD simulation.
These five charmonium states are classified with quantum numbers $J^{PC}$
and corresponding operators $\Gamma$ as shown in Table~\ref{tab:charmonium_mass}.
A effective mass is defined as 
\begin{equation}
 M_\Gamma(t) = \log\frac{G_\Gamma(t,\tsrc)}{G_\Gamma(t+1,\tsrc)}.
\end{equation}
where $G_\Gamma(t,\tsrc)$ is the two-point function obtained by setting  ${\bm r}$ to be zero 
in the four-point function $G_\Gamma({\bm r}, t,\tsrc)$ defined in Eq.~(\ref{eq_4ptF}).
Each effective mass plot shows a reasonable plateau in the range 
$33 \le t/a \le 47$ for $S$-wave charmonium states 
and $14 \le t/a \le 26$ for $P$-wave charmonium states.
We estimate masses of the five charmonium states by a constant fit to the plateau
over time ranges shown in Table~\ref{tab:charmonium_mass}.
A correlation among effective masses measured at various time slices 
is taken into account by using a covariance matrix in the fit.
An inversion of the covariance matrix is performed once for average
and it is used for each jackknife block.
The statistical uncertainties indicated by shaded bands in Fig.~\ref{fig:effmass}
are estimated by the jackknife method.
In Table~\ref{tab:RHQ_para_charmonium}, we summarize resultant charmonium masses 
together with fit ranges used in the fits and values of $\chi^2$ 
per degrees of freedom ($\text{d.o.f.}$). 
Note that all masses calculated in this study
are obtained from the Coulomb-gauge wall source propagator, while gauge-invariant Gaussian smeared 
source was used for results of charmonium masses compiled in Table~I of
Ref.~\cite{Kawanai:2011xb}. 

Low-lying charmonium masses calculated below $D\bar{D}$ threshold
are all close to the experimental values, though
the hyperfine mass splitting $M_\text{hyp}=0.1124(9)$ GeV
is slightly smaller than the experimental value,
$M_\text{hyp}^{\text{exp}}=0.1166(12)$ GeV~\cite{Beringer:1900zz}.
The similar value of the hyperfine mass splitting is reported even
on the exact physical point in Ref.~\cite{Namekawa:2011wt,Mohler:2011ke}.
Note that here we simply neglect the disconnected diagrams
in all two-point correlation functions.
The several numerical studies reported that the contributions of charm annihilation to
the hyperfine splitting of the $1S$-charmonium state is
sufficiently small, which is of order 1-4~MeV.~\cite{McNeile:2004wu,deForcrand:2004ia,Levkova:2010ft}.
At the charm sector, the effect of the disconnected diagrams on the charmonium, especially on the vector state,
is perturbatively expected to be small due to Okubo-Zweig-Iizuka suppression.

\section{\label{main} Determination of interquark potential}
\subsection{\label{wavefunc} $\QQbar$ BS wave function}
 \begin{figure}
   \centering
   \includegraphics[width=.49\textwidth]{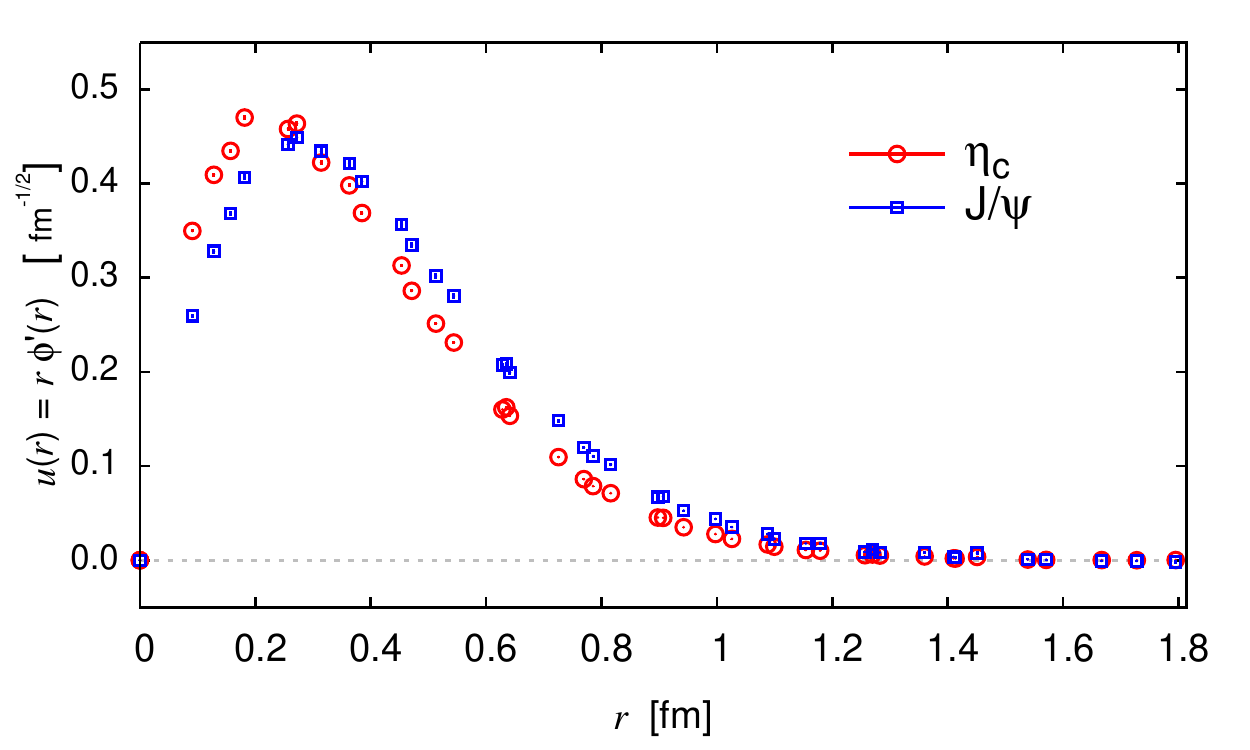}
   \caption{ The reduced $\QQbar$ BS wave functions of the $\eta_c$~(circles)
    and $J/\psi$~(squares) states,  shown as a function of the spatial distance~$r$.
    The data points are taken along $\bm r$ vectors which are multiples of 
   three directions~$(1,0,0)$, $(1,1,0)$ and $(1,1,1)$.
   }
    \label{wavefunction1}
 \end{figure}

We calculate the BS wave functions only for $S$-wave states ($\eta_c$ and $J/\psi$). 
This is simply because the Coulomb-gauge wall source~\footnote{
Clearly, the spatial part of the meson operator constructed from
a local quark bilinear operator with the wall source, 
where the quark operator is summed over all spatial sites at given time slice, 
belongs to the trivial $A_1^+$ irreducible representation of the cubic group.
The plus sign in superscript indicates even parity.
} adopted in this study is not suitable for studying the wave function of $P$-wave states, 
whose spatial part is odd under spatial reflection.

Fig.~\ref{wavefunction1} shows the  $\QQbar$ BS wave functions of
$1S$-charmonium states ($\eta_c$ and $J/\psi$ states).
The BS wave functions are defined by Eq.(\ref{eq_phi}) with a normalization condition of
$\sum \phi_{\Gamma}^2 = 1$.
We use the reduced wave function $u_\Gamma(r)$ for displaying the wave function:
$u_\Gamma(r) =r \phi_\Gamma({\bm r})$.
We practically take a time-average of the BS wave function at fixed $r$ over the 
range $33 \leq t/a \leq 47$,
where effective mass plots for $1S$-charmonium states show excellent plateaus and
excited state contaminations should be negligible. To resolve the strong 
correlations between data of the BS wave function at different time slices, 
we take into account the covariance matrix during the averaging process over the time slice. 

We find that a breaking of rotational symmetry for the $\QQbar$ BS wave functions
is sufficiently small in our calculation.
The resulting wave functions become isotropic
with the help of a projection to the $A_1^+$ sector of the cubic group
that corresponds to the $S$-wave in the continuum theory~(Fig.~\ref{wavefunction1}).
All data points of the $\QQbar$ BS wave functions calculated
in the three different directions fall onto a single curve.

The spatial lattice extent $N_sa \approx 2.9$~fm
is sufficiently large enough to study the $1S$-charmonium system.
Indeed, the BS wave functions shown in Fig.~\ref{wavefunction1}
are localized around the origin and vanished at $r\agt 1.1$~fm.
It suggests that the $Q\bar{Q}$ BS wave functions for the $\eta_c$ and $J/\psi$ states 
fair enough fit into the box $N_s^3$. Needless to say, the localized wave functions 
is interpreted as a sign of bound states. This fact however reminds us that the interquark 
potential can be deduced within the interior of the hadron due to its localized wave function.
This is simply because that the signal-to-noise ratio in the calculation of 
${\bm \nabla}^2\phi/\phi$ of Eq.~(\ref{Eq_potC})-(\ref{eq_quark_mass}) 
is getting worse outside the spatial size of the hadron.

Other important information can be read off from Fig.~\ref{wavefunction1}.
The spatial size of the $J/\psi$ state is slightly larger than that of the $\eta_c$ state.
This indicates that there is a repulsive spin-spin interaction near the origin for the higher spin states.
It is consistent with the pattern of level ordering for the hyperfine multiplet.
The spin-spin charmonium potential will be discussed in more detail later.

\subsection{\label{determination_quarkmass}quark kinetic mass}
 \begin{figure}
   \centering
   \includegraphics[width=.49\textwidth]{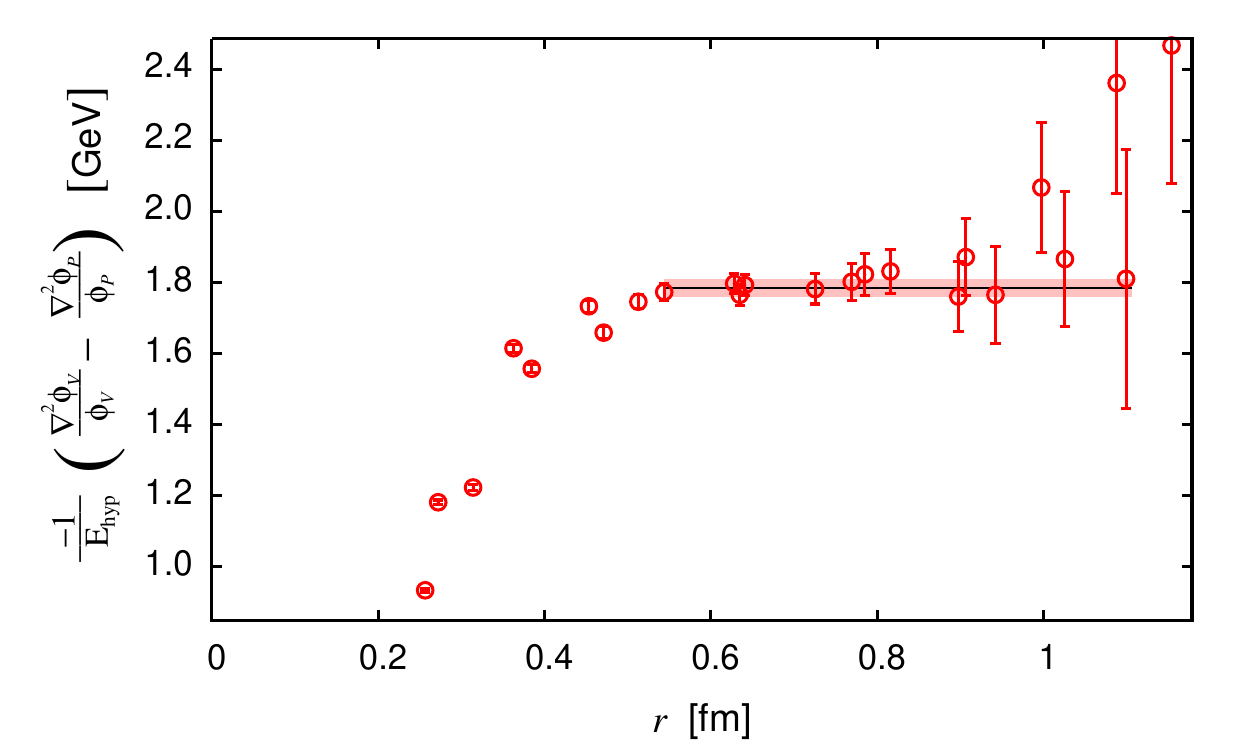}
   \caption{
    The determination of quark kinetic mass within the BS amplitude method.
    The values of $-({\bm \nabla}^2\phi_{\rm V}/\phi_{\rm V}-{\bm \nabla}^2\phi_{\rm PS}/\phi_{\rm PS})/E_\text{hyp}$
    as a function of the spatial distance $r$ are shown in this figure.
    The quark kinetic mass $m_Q$ is obtained from the long-distance asymptotic values of
    $-({\bm \nabla}^2\phi_{\rm V}/\phi_{\rm V}-{\bm \nabla}^2\phi_{\rm PS}/\phi_{\rm PS})/E_\text{hyp}$.
    Horizontal solid line indicates a value of quark kinetic mass
    obtained by fitting a asymptotic constant in the range $0.54~\text{fm}\alt r \alt 1.10~\text{fm}$.
    A shaded band indicates a statistical error estimated by the jackknife method.
  }
  \label{fig:quarkmass_charm}
 \end{figure}
In our formalism, the kinetic mass of the charm quark is
determined self-consistently within the BS amplitude method as well~\cite{Kawanai:2011xb}.
According to Eq.~(\ref{eq_quark_mass}), the quark kinetic mass can be evaluated
from an asymptotic behavior of the quantity 
$-({\bm \nabla}^2\phi_{\rm V}/\phi_{\rm V}-{\bm \nabla}^2\phi_{\rm PS}/\phi_{\rm PS})/E_\text{hyp}$ at long distances.
For the discrete Laplacian operator ${\bm \nabla}^2$, 
we use {\it r-Laplacian}, which is defined in polar coordinates as follows:
\begin{eqnarray}
{\bm \nabla}^2 \phi_\Gamma(r)
&=& \frac{2}{r}\frac{\phi_\Gamma(r+\tilde{a})-\phi_\Gamma(r-\tilde{a})}{2\tilde{a}} \nonumber \\
& & + \frac{\phi_\Gamma(r+\tilde{a})+\phi_\Gamma(r-\tilde{a})-2\phi_\Gamma}{\tilde{a}^2} 
\label{eq_laplacian_polar}
\end{eqnarray}
where $r$ is the absolute value of the relative distance as $r = |{\bm r}|$, and
$\tilde{a}$ is a spacing between grid points along differentiate directions. 
In the on-axis direction~${\bm r} \propto (1,0,0)$ (labeled by ``on-axis''), 
two off-axis directions ${\bm r}\propto  (1,1,0)$ (labeled by ``off-axis I'') and ${\bm r}\propto(1,1,1)$ 
(labeled by ``off-axis II''),
the effective grid spacings correspond to $\tilde{a} = a, \sqrt{2}a$ and $\sqrt{3}a$, respectively.
The difference of ratios ${\bm \nabla}^2 \phi_\Gamma /\phi_\Gamma$ at each ${\bm r}$ are 
obtained by a constant fits to the lattice data  with reasonable $\chi^2/\text{d.o.f.}$ values
over the range of time slices where two-point functions exhibit the plateau behavior~($33 \leq t/a \leq 47$).

Fig.~\ref{fig:quarkmass_charm} illustrates the determination of quark kinetic mass $m_Q$ for the 
charmonium system. The value of $m_Q$ can be determined from an asymptotic value of 
 $-({\bm \nabla}^2\phi_{\rm V}/\phi_{\rm V}-{\bm \nabla}^2\phi_{\rm PS}/\phi_{\rm PS})/E_\text{hyp}$
in the range of  $ 6 \leq r/a \leq 7\sqrt{3}$ ($0.54~\text{fm}\alt r \alt 1.10~\text{fm}$), 
where $V_S(r)$ should vanish.
In this study, three data sets are obtained from three directions: on-axis, off-axis I and off-axis II, 
are separately analyzed so as to expose the size of the possible lattice discretization artifacts.
On each data set, a value of $\mQ$ is obtained by a constant fit to a long-distance asymptotic value
over the range as described above. Finally we average them over three directions, and then obtain 
$\mQ = 1.784(23)(6)(20)$~GeV.
The first error is statistical, given by the jackknife analysis.
In the second error, we quote a systematic uncertainty due to rotational symmetry breaking
by taking the largest difference between average value and individual ones obtained for specific directions.
The third ones are systematic uncertainties stemming from choice of $t_{\rm min}$ in the averaging process 
over the time-slice range $t_{\rm min}/a \le t/a \le 47$. The minimum time-slice $t_{\rm min}/a$ is
varied over range from 33 to 41 and then take a largest difference from the preferred determination of $\mQ$.

The charm quark mass obtained in this study is somewhat heavier than
the usual quark kinetic mass in the NRp models.
For example, the quark kinetic mass adopted in Ref.~\cite{Barnes:2005pb} is about 17\% smaller.
This difference however should not be taken seriously, because
the value of $\mQ$ in the NRp models highly depends on a constant term~$V_0$ of the Cornell potential, 
and $V_0$ is actually forced to be zero in many of the NRp models.
In addition, the spatial profile of the spin-spin potential from lattice QCD is slightly different
from the one used in the NRp models as we will discuss later.

\subsection{\label{sec:potential}Spin-independent interquark potential}
 \begin{figure}
   \centering
   \includegraphics[width=.49\textwidth]{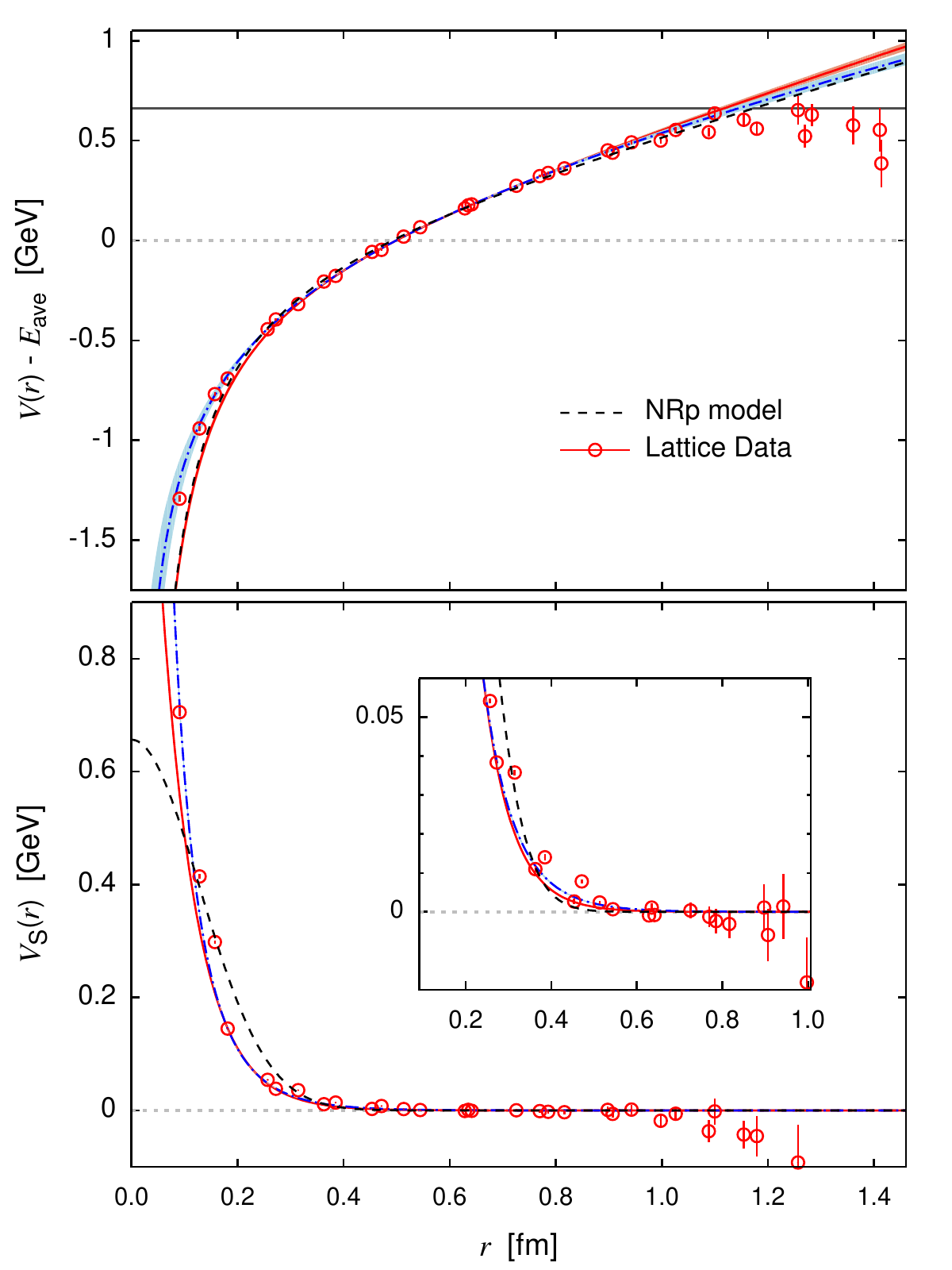}
  \caption{ 
    Central spin-independent and spin-spin
    charmonium potentials calculated from the BS wave functions in the dynamical QCD simulation
    with almost physical quark masses.
    In the upper panel, we show the spin-independent potential $V(r)$.
    A solid (dot-dashed) curve is the fit results
    with the Cornell~(Cornell plus log) form.
    The shaded bands show statistical uncertainties
    in the fitting procedure where the jackknife method is used.
    Note that the spin-averaged mass of $1S$-charmonium states $E_{\rm ave}$
    is not subtracted in this figure.          
    A horizontal line indicates the level of open-charm~($D^0 \bar{D}^0$) threshold $\approx 3729$~MeV.                                
    In the lower panel, we show the spin-spin potential $V_S(r)$.
    A solid~(dot-dashed) curve corresponds to fitting results with exponential~(Yukawa) form.
    The inset shows a magnified view.
    In both panels,  the phenomenological potentials adopted in a NRp model~\cite{Barnes:2005pb}
    are also included as  dashed curves for comparison.
    \label{fig:potential_charm}
  }
 \end{figure}
 \begin{table}
  \centering
   \caption{Summary of the Cornell parameters and the quark mass
     determined by the BS amplitude method. 
     For comparison, 
     ones adopted in a phenomenological NRp model~\cite{Barnes:2005pb}
    and  ones of the static potential obtained from Polyakov line correlations
    are also included.
    In the first column, the quoted errors indicate the sum of the statistical and
    systematic added in quadrature.    
    \label{tab:parameter_charm}
   }
   \begin{ruledtabular}                                                         
     \begin{tabular}{cccccc} 
       & This work& Polyakov lines & NRp model~\cite{Barnes:2005pb}  \\[2pt]        \hline
       $A$                   & 0.713(83)  & 0.476(81)   &0.7281 \\
       $\sqrt{\sigma}$ [GeV] & 0.402(15)  & 0.448(16)   &0.3775 \\
       $\mQ$ [GeV]           & 1.784(31)  & $\infty$   &1.4794 \\ 
     \end{tabular}
   \end{ruledtabular}                                                           
 \end{table}

Once the quark kinetic mass is determined,
we can calculate the central spin-independent and spin-spin
charmonium potentials from the $Q\bar{Q}$ BS wave function
through Eq.~(\ref{Eq_potC}) and Eq.~(\ref{Eq_potS}).
First, we show a result of the spin-independent charmonium potential $V(r)$
in the upper panel of Fig.~\ref{fig:potential_charm}.
The constant energy shift $E_{\rm ave}$ is not subtracted in this figure.
As we reported in Ref~\cite{Kawanai:2011jt},
the charmonium potential calculated by the BS amplitude
method from dynamical lattice QCD simulations
properly exhibits the linearly rising potential
at large distances and the Coulomb-like potential at short distances.
At first glance, the data points of the charmonium potential obtained from
lattice QCD roughly follow the phenomenological potential used in the NRp models, 
which is represented by the dashed curve. Nevertheless, the data points at short 
distances are slightly off the dashed curve. In addition, a {\it string breaking-like behavior} is found 
in the range $r\alt 1.1$ fm, where the charmonium potential reaches
the level of open-charm threshold.  We will discuss this point later,
and for a while we concentrate only on data points in the range $r\alt 1.1$ fm, where
the linearly rising potential is clearly visible. 

For more close comparison, as a first step, we simply adopt the Cornell parametrization 
to fit the data of the spin-independent central potential:
\begin{equation}
 V(r)=-\frac{A}{r}+\sigma r+V_0  \label{eq_Cornell}
\end{equation}
with the Coulombic coefficient $A$, the string tension $\sigma$, and a constant $V_0$.
All fits are performed individually for each three directions over the range 
$[r_{\rm min}/a:  r_{\rm max}/a] = [4: 7\sqrt{3}]$, where $r_{\rm max}\approx 1.1$ fm.
We minimize the $\chi^2/\text{d.o.f.}$ with the covariance matrix and get
the Cornell parameters of the charmonium potential as
$A=0.713(26)(38)(31)(62)$ and $\sqrt{\sigma}=0.402(6)(4)(9)(9)$ MeV
with $\chi^2/\text{d.o.f.} \approx 3.2$. 
The first error is statistical, and the second, third and forth ones are systematic 
uncertainties due to the choice of data points taken from three directions, 
and variations of $t_{\rm min}$ and $r_{\min}$, respectively.

The resulting Cornell parameters are summarized in Table~\ref{tab:parameter_charm}.
We also include both phenomenological ones adopted in a NRp model~\cite{Barnes:2005pb}
and ones of the static potential obtained from Polyakov-line correlators.
The latter is estimated using the same method as in Ref.~\cite{Aoki:2008sm}.
Additionally, we calculate the Sommer parameter defined as $r_0 = \sqrt{(1.65 -A)/\sigma}$, 
and then obtain the value of $r_0 = 0.476(6)(11)(3)(6)$~fm, 
which is fairly consistent with the value quoted in Ref.~\cite{Aoki:2008sm}.

From our previous research in quenched QCD~\cite{Kawanai:2013aca}, 
the finite $\mQ$ corrections could be encoded into the Cornell parameters.
Indeed, as shown in Table~\ref{tab:parameter_charm}, 
in the charmonium potential from the BS wave function, a Coulomb-like behavior is enhanced
and the linearly rising force is slightly reduced due to finite charm quark mass effects
in comparison to the conventional static potential from Wilson loops or Polyakov-line correlators. 
Furthermore, a gap for the Cornell parameters between the static and 
the phenomenological potentials seems 
to close by our new approach, which nonperturbatively accounts for a finite quark mass effect.

Here, we give a few technical remarks on the systematic uncertainties on the Coulombic coefficient $A$,
which highly depends on the choice of minimum value $r_{\min}$ of the fitting window 
compared to the string tension $\sigma$. This is simply because the linear part in the
Cornell potential parametrization is dominated in the region in which we have data points.
Indeed, as shown in the upper panel of Fig.~\ref{fig:potential_charm}, 
a solid curve, which corresponds to fitting results with the Cornell potential form,
does not describe well data points outside the range $[r_{\rm min}:  r_{\rm max}]$ 
used in the fit. 

In order to provide an adequate fit to the data points at shorter distances,
we employed several alternative fitting forms. 
We found that a simple extension of the Cornell potential can describe
the behavior of our charmonium potential reasonably well. 
An alternative fitting form is given such that a $\log$ term is added to the Cornell potential:
\begin{equation}
  V(r)=-\frac{A}{r}+\sigma r+V_0 + B \log(r\Lambda) \label{eq_Cornell_log}
\end{equation}
where the value of $\Lambda$ is simply set to be lattice cutoff $a^{-1}$.
Such a logarithmic $r$-dependence may appear in the leading $1/\mQ$ correction 
to the static potential as reported in Ref.~\cite{Koma:2009ws}.
Moreover, as reported in Ref.~\cite{Laschka:2012cf},
the charmonium potential obtained from the BS amplitude is consistent with the $\QQbar$ potential
obtained in the Wilson-loop approach within errors, when a leading $1/m_Q$ correction calculated 
in Ref.~\cite{Koma:2006si} is added to the static potential from Wilson loops.

A fit with the ``Cornell-plus-log'' form~(\ref{eq_Cornell_log}) leads to the values of 
$A= 0.194(137)(33)(36)(66)$, $\sqrt{\sigma} = 0.300(38)(19)(20)(21)$~GeV 
and $B = 0.390(113)(20)(39)(61)$~GeV with the slightly smaller value of $\chi^2/\text{d.o.f.} \approx 2.3$.
We here chose the fitting range to be $[r_{\rm min}/a : r_{\rm max}/a] = [3 : 7\sqrt{3}]$ and
used a covariance matrix for taking into account the correlation among data points in the fit. 
The quoted errors have the same meaning as described above. 

We also plot the fit result with the Cornell-plus-log form, which is represented by 
a dot-dashed curve, in upper panel of Fig.~\ref{fig:potential_charm}. The shaded band displays 
one standard deviation estimated by the jackknife method.
The short-distance behavior of the charmonium potential is better described 
by the Cornell-plus-log form than the Cornell form (solid curve). 
If compared with the phenomenological potential of the NR models, 
the shape of the fitted curve of the Cornell-plus-log form at long-distances
are much in agreement with the NR models though the string tension $\sigma$ 
becomes a slightly smaller value compared with the phenomenological one.
In this context, the inclusion of the log term into the Cornell form gives only a minor modification
at long-distances as far as the data is accessible in this study.

Finally, we would like to comment on the {\it string breaking-like} behavior appeared
in the range $r\agt 1.1$ fm.
Although in principle, string breaking due to the presence of dynamical quarks is likely to be observed,
the observed feature in this study is suspicious and unreliable.
As mentioned previously, the signal-to-noise ratio on the quantity of 
${\bm \nabla}^2\phi_\Gamma/\phi_\Gamma$ becomes worse rapidly as the spatial distance 
$r$ increases because of the localized nature of the BS wave function $\phi_\Gamma(r)$. 
Moreover, the lattice data of the potential near the spatial boundaries are also sensitive 
to the possible distortion of its spatial profile as finite volume effects. 
Therefore, at least, calculations of the higher charmonium
near the open charm threshold using a larger lattice is necessary
for observing the string breaking in this sense.
Their wave functions are extended until the string breaking sets in.

We also emphasis that there might be another possible reason for no evidence of string breaking
from a view point of studies within the Wilson loop
approach~\cite{Aoki:1998sb, Bali:2000vr,Bolder:2000un,Bernard:2002sb}.
The string breaking in the static heavy quark potential
can be observed only after inserting a operator of light quark-antiquark to create the
heavy-light meson-antimeson state $(Q\bar{q})(q\bar{Q})$,
because the $\QQbar$ creation operator poorly overlaps with
$|(Q\bar{q})(q\bar{Q})\rangle$ state in Fock space~\cite{Pennanen:2000yk,Bernard:2001tz,Bali:2005fu}
(See also Ref.~\cite{Detar:1998qa} in the case of nonzero temperature).
It is worth reminding that the static potential from Wilson loops is regarded 
as the ``energy eigenvalue'' of the considering states. 
There would be nothing to change for the charmonium potential extracted from
the ``stationary'' wave function of the charmonium state, which is well defined in 
the BS amplitude method unless its energy level is above the open-charm threshold. 

\subsection{\label{sec:potential}Spin-Spin potential}
%
 \begin{table}
  \centering
   \caption{
    Results of fitted parameters for the spin-spin potential with the exponential and Yukawa forms.
    The quoted errors are statistical only. 
    In the case of the spin-spin potential, we use only the on-axis data.
     \label{tab:parameter_charm2}
   }
   \begin{ruledtabular}                                                         
     \begin{tabular}{cccccc} 
      Functional form & $\alpha$ & $\beta$ &  $\chi^2/\text{d.o.f.}$ \\[2pt]        \hline
      Exponential     &  2.15(7)~GeV         & 2.93(3)~GeV & 2.0\\
      Yukawa          &  0.815(27)\ \ \ \ \  & 1.97(3)~GeV & 1.7\\
     \end{tabular}
   \end{ruledtabular}                                                           
 \end{table}
 The lower panel in Fig.~\ref{fig:potential_charm} shows the spin-spin
 charmonium potential obtained from the BS amplitude method 
 with almost physical quark masses.
 The spin-spin potential exhibits the short-range {\it repulsive interaction},
 which is required to lead to energy gain for the higher spin state.
 Recall that the Wilson loop approach
 currently dose not achieve  to reproduce the correct behavior of the spin-spin interaction.
 The leading-order spin-spin potential classified in pNRQCD
 becomes attractive at short distances~\cite{Koma:2006fw, Koma:2010zz}.
Their calculation at next-to-leading order is unavailable at present.
 In contrast of the case of the spin-independent potential,
 the spin-spin potential obtained from the BS wave function is absolutely different
 from a repulsive $\delta$-function potential generated
 by perturbative one-gluon exchange~\cite{Eichten:1980mw}.
 Such contact form  $\propto \delta({\bm r})$ of the Fermi-Breit type potential
 is widely adopted in the NRp models~\cite{Buchmuller:1981fr,Godfrey:1985xj}.

 The pointlike spin-spin interaction easily lifts the mass degeneracy between $1^1P_1$ state~($h_c$) and
 spin-weighted average  of $1^3P_J$ states~($\chi_{cJ}$);
 $M(\ovli{1^3P_J}) = (M_{\chi_{c0}} + 3M_{\chi_{c1}} + 5M_{\chi_{c2}})/9$.
 On the other hand, a finite-range interaction gives a non-zero, but small
 finite hyperfine splitting  to the $P$- or  higher-wave charmonia~\cite{Voloshin:2007dx}.
 In the current experiments, however, the splitting $M_{\rm hyp}(1P) = M(\ovli{1^3P_J})- M_{h_c}$
 for $1P$-charmonium states is not appreciably observed within experimental error.
 Here we quote $M_{\rm hyp}(1P) = 0.02 \pm 0.19{\rm (stat)} \pm 0.13{\rm (syst)}$~MeV
 from the CLEO experiment~\cite{Rubin:2005px,Dobbs:2008ec}~(See also Ref.~\cite{Burns:2011fu}).

 The $Q\bar{Q}$ interaction is not entirely due to one-gluon exchange 
 so that the spin-spin potential is not necessary to be a simple contact form
 $\propto \delta({\bm r})$~\cite{Eichten:1975ag, Eichten:1978tg, Barnes:1982eg,Ebert:2005jj}.
This is shown to be true even for the ${\cal O}(1/\mQ^2)$ spin-spin corrections 
in the Wilson-loop approach~\cite{Koma:2006fw, Koma:2010zz}, regardless of the sign issue.  
 In the phenomenological side, the finite-range spin-spin potential described
 by the Gaussian form is adopted by some NRp model in Ref.~\cite{Barnes:2005pb}, where
 many properties of conventional charmonium states at higher masses are predicted.
 This phenomenological spin-spin potential is also plotted
 in the lower plot of Fig.~\ref{fig:potential_charm} for comparison.
 There is a slight difference at very short distances,
 although the range of spin-spin potential calculated from the BS amplitude method 
 is similar  to the phenomenological one.

  To examine an appropriate functional form for the spin-spin potential,
  we try to fit the data with several functional forms,
  and explore which functional form can give a reasonable fit
  over the range of $r/a$ from $2$ to $7\sqrt{3}$.  
  As a results, the long-range screening observed in the spin-spin potential
  is accommodated by the exponential form and the Yukawa form:
  \begin{equation}
    V_{\text{S}}(r)=\left\{
    \begin{array}{lcl}
      \alpha \exp(-\beta r)   &:& \text{Exponential form} \\
      \alpha \exp(-\beta r)/r &:& \text{Yukawa form} 
    \end{array}
    \right.
  \end{equation}
  All results of correlated $\chi^2$ fits are summarized in Table~\ref{tab:parameter_charm2}.
  We also try to fit with the Gaussian form that is often employed in the NRp models, 
  and it however gives an unreasonable $\chi^2/\text{d.o.f.}$ value.
  Note that we here use only the on-axis data which are expected to 
  less suffer from the rotational symmetry breaking and the discretization error, 
  because fit results to the lattice data taken in each direction 
  significantly disagree with each other~\cite{Kawanai:2013aca}.
  We need the finer lattice to have a solid conclusion to the shape of the spin-spin potential 
  and the uncertainties due to the rotational symmetry breaking.

\section{\label{main2} Charmonium mass spectrum from charmonium potential}
Once the quark kinetic mass and the charmonium potentials are determined by first principles of QCD,
we can solve the nonrelativistic Schr\"odinger equation defined with the theoretical inputs
for the bound $c\bar{c}$ systems
as same as calculations in the NRp models~\cite{Eichten:1974af,Eichten:1975ag,Eichten:1978tg,Eichten:1979ms}.

In the non-relativistic description, 
each charmonium state is generally labeled by a symbol $^{2S+1}[L]_J$,
with the spin angular momentum ($S = 0, 1, \cdot\cdot\cdot$), the orbital angular momentum
($[L]=S$, $P$, $D...$ corresponding to $L=0,1,2, \cdot\cdot\cdot$) and
the total angular momentum ($J = S \oplus L$) quantum numbers.
The $J^{PC}$ notation is also used to classify the charmonium state.
The parity ($P$) and the charge-conjugation ($C$) are given by $P=(-1)^{L+1}$ and $C=(-1)^{S+L}$ within the non-relativistic description.

Recall that all of the charmonium states below the open-charm threshold
are experimentally well established~\cite{Beringer:1900zz}.
The last missing $1P$-charmonium state, $h_c$,
and also the first excited state of the pseudoscalar $1S$-charmonium state, $\eta_c(2S)$, have already 
been observed in recent experiments~\cite{Swanson:2006st,Rubin:2005px,Rosner:2005ry,
Dobbs:2008ec,Choi:2002na,Vinokurova:2011dy,delAmoSanchez:2011bt,Aubert:2003pt}.

In this section, we will discuss whether we can get the correct low-lying charmonium spectra within 
{\it the hybrid approach} between lattice QCD simulations and the NRp models
in comparison to experimental data.
In addition, we also perform a consistency check between two different methods in lattice QCD
to verify the validity of our approach.
One is of course the standard lattice spectroscopy, where the mass information is extracted 
from the large-time asymptotic behavior of the two-point correlation functions, while another mainly uses 
the information about the spatial profile of the BS amplitudes.
In this sense, these two methods are essentially different.

\subsection{Nonrelativistic Hamiltonian from lattice QCD}

For solving a nonrelativistic Schr\"odinger equation, the constant energy shift is
irrelevant. We here introduce the energy-shifted potential in the spin-independent part
as $V^\prime(r)=V(r)-E_{\rm ave}$ for the following reason:
The quantity of $V^\prime(r)$ can be directly obtained from the BS amplitudes
of $1S$-charmonium states except the overall factor of $1/\mQ$. 
It ends up with less statistical uncertainties compared to the original potential $V(r)$, 
whose estimation requires the subtraction of $E_{\rm ave}$.
This is simply because the value of $E_{\rm ave} = M_{\rm ave}- 2m_Q$ receives 
somewhat large uncertainties, which arise mainly in the determination of $\mQ$ through Eq.~(\ref{eq_quark_mass}). 
Indeed, when we evaluate a difference between $V_0$ and $E_{\rm ave}$ directly
from the fit with the Cornell functional form to the data of $V^\prime(r)$, this quantity
shows much smaller error such as $V_0 - E_{\rm ave} = -0.146(13)$~GeV, 
in comparison to the values of $E_{\rm ave} = 0.508(69)$~GeV and $\mQ = 1.789(34)$~GeV. 

We therefore adopt the energy-shifted potential of $V^\prime(r)$ to 
reduce uncertainties on the final result for energy eigenvalues,
and then solve the following Schr\"odinger equation for the reduced wave function $u_{SLJ}(r)$~\footnote{
Here, we assume that the reduced wave function vanishes at origin $\lim_{r\to 0} r\phi'(r) = u(r) = 0$.
Indeed, if the potential satisfies $r^2V(r) \xrightarrow{r\to 0} 0$, one can easily show the reduced wave fanction 
asymptotically behaves as $u(r) \xrightarrow{r\to 0} r^{L+1}$.
}:
\begin{equation}
 \left\{  -\frac{{\bm \nabla}^2}{\mQ} + \frac{L(L+1)}{\mQ r^2} + V'_{SLJ}(r)
 \right\} u_{SLJ}(r) 
 =E'_{SLJ} u_{SLJ}(r) 
\label{eq_schre}
\end{equation}
where $V'_{SLJ}(r) = V_{SLJ}(r) - E_{\rm ave}$ and $E'_{SLJ} = E_{SLJ} - E_{\rm ave}$
with angular momentum quantum numbers ($S$, $L$ and $J$).
The interquark potentials $V'_{SLJ}(r)$, which may involve the spin-dependent interactions, clearly depend on 
the charmonium states labeled with specific $S$, $L$ and $J$.
The rest mass energy of the desired charmonium state is obtained simply by adding the energy eigenvalue of $E'_{SLJ}$ to
the spin-averaged $1S$-charmonium mass of $M_{\rm ave}$, which is evaluated by the standard lattice spectroscopy
with high accuracy:~$ M_{SLJ} = M_{\rm ave}+E'_{SLJ} = 2\mQ  + E_{SLJ}$.

The potential calculated from lattice QCD with the BS amplitude method are by definition discretized in space.
In this context, instead of the continuum Schr\"odinger equation, 
we practically solve eigenvalue problems of finite-dimensional vector $u_n = u(n\tilde{a})$ and
finite-dimensional matrix~\cite{Charron:2013paa} as 
\begin{eqnarray}
 \sum_{n>0} H_{m,n} u_n = E^\prime u_m \label{eq_eigen}.
\end{eqnarray}
Note that a summation of $n$ does not include $n=0$ since the reduced wave function is required to vanish at the origin.
In the symmetric matrix $H_{m,n}$ for $n, m>0$,  diagonal and off-diagonal matrix elements are given by
\begin{eqnarray}
  H_{n,n}      &=&  \frac{1}{\tilde{a}^2\mQ}\left[2 + \frac{L(L+1)}{n^2}  \right] + V'(n\tilde{a}),  \\
  H_{n\pm 1,n} &=& - \frac{1}{\tilde{a}^2\mQ},
\end{eqnarray}
and all other elements are zero. Here we omit the labels $SLJ$ for clarity.

In this work, we separately solve Eq.~(\ref{eq_eigen}) in the directions of vectors $\bm r$ 
which are multiples of $(1, 0, 0)$, $(1, 1, 0)$ and $(1, 1, 1)$.
We prefer to use mainly on-axis data, which is expected to receive smallest discretization error
and correction due to rotational symmetry breaking as studied in Ref.~\cite{Kawanai:2013aca},
and take the largest difference between on-axis and off-axis results
as the systematic error due to the choice of the $\bm r$ direction, while statistical errors are estimated by the jackknife method.
Systematic uncertainties stemming from the choice of the fitting window
in the averaging process over the time-slice range are smaller than other errors.

Alternatively, we may solve the continuum Schr\"odinger equation with the parameterized 
charmonium potential by empirical functional forms such as the Cornell form or the Cornell-plus-log form 
as discussed in Sec.~\ref{main}.
This procedure, however, yields large uncertainties in the low-lying energy levels, which
highly depend on the choice of functional forms especially at short distances.
To avoid such model dependence, we adopt the former strategy, which 
does not suffer from the sensitivity to the shape of the potential at short distances.

\subsection{Wave functions solving the Schr\"odinger equation}
\begin{figure}
  \centering
  \includegraphics[width=.49\textwidth]{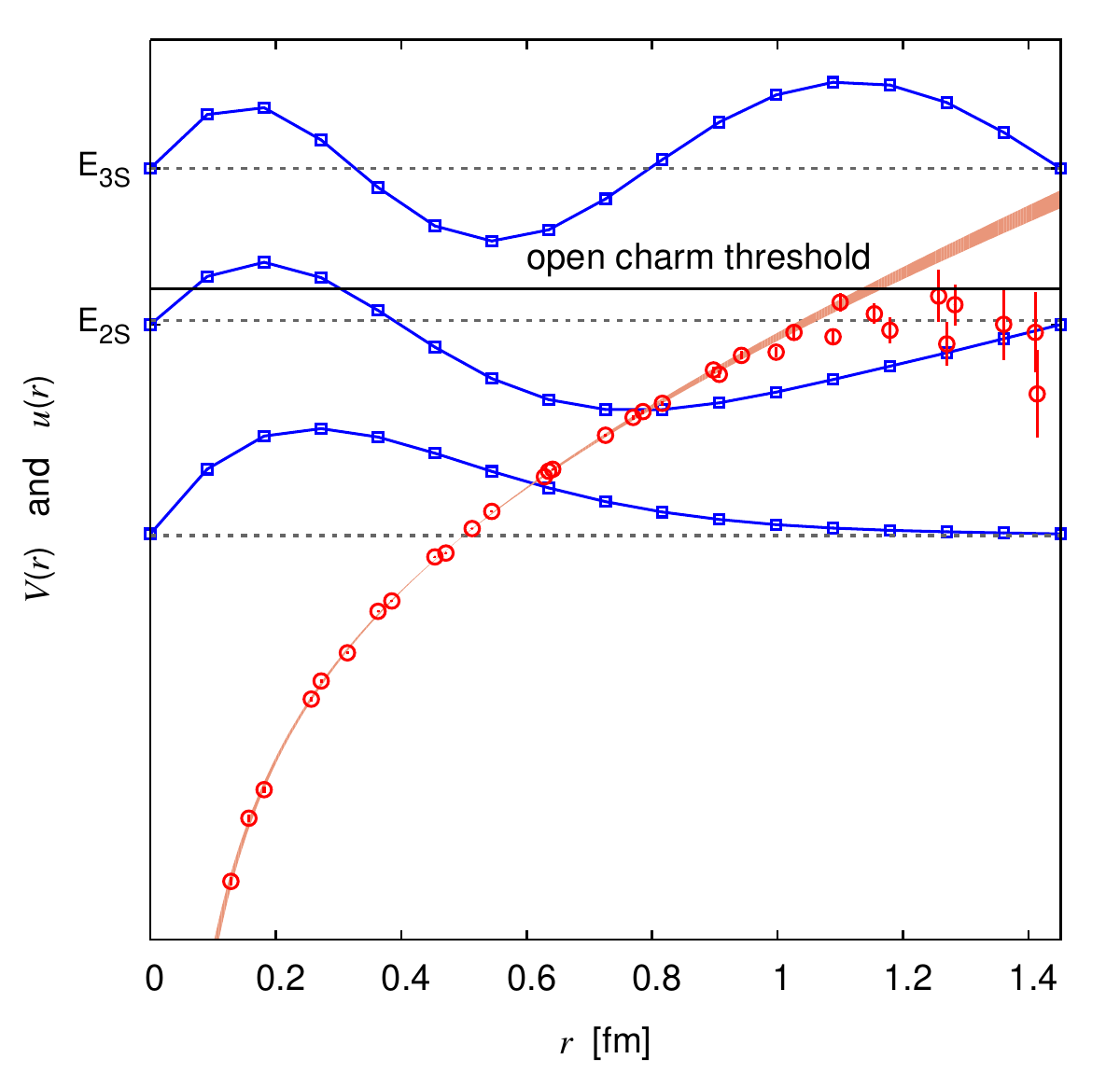}
  \caption{The energy-levels~(dotted  lines)
    and corresponding  reduced wave functions $u(r)$~(squares) of
    spin-averaged $1S$-charmonium states,
    obtained by solving the Schr\"odinger equation with the lattice inputs.
    Only the central (average) values are shown for both quantities.
    A  horizontal line indicates the open-charm threshold.
    The spin-independent charmonium potential obtained from lattice QCD
    and its fitting result with the Cornell-plus-log functional form as a function of $r$
    are also overlaid in the same plot
    as circles and a shaded band, respectively.
}
  \label{fig:wavefunc_charmonium2}
\end{figure}
In Fig~\ref{fig:wavefunc_charmonium2},
we plot energy-levels (dotted lines) and corresponding reduced wave functions (curves with square symbols)
up to the second excited state for the spin-averaged $S$-wave charmonium states,
which are given by the charmonium potential with an expectation value of the spin operator being zero, $\langle \bm{S}_{Q} \cdot \bm{S}_{\bar{Q}} \rangle = 0$. The spin-independent charmonium potential, which is calculated from lattice QCD,  
is also overlaid in the figure as circle symbols together with its fitting result using the Cornell-plus-log form (shaded band). 

We first carefully examine the energy eigenvalue $E'_{\rm ave}$ of the spin-averaged $1S$ charmonium state
whose mass was used as input to calibrate the RHQ parameters for the charm quark.
Recall that the value of $E'_{\rm ave}$ is supposed to be zero because of its definition on the shifted energy 
$E^\prime$ introduced in Eq.~(\ref{eq_schre}). We consequently
obtain $E'_{\rm ave} = 0.2(1.3)(0.5)$~MeV, where the first error is statistical, the second error is systematic error due to rotational symmetry breaking. The obtained value is sufficient for satisfying the condition $E'_{\rm ave}=0$ as a self-consistency check in our approach. We then conclude  
that the spin-averaged $1S$-charmonium state can be well described by our charmonium potential given in the range $r \alt 1.1$~fm.

The boundary condition implemented in the definition of the Hamiltonian matrix defined  
in Eq.~(\ref{eq_eigen}) enforces the wave functions to vanish outside the interval $r \le \tilde{a}N_s/2$.
Although our choice of $N_s/2 \times N_s/2$ for the size of Hamiltonian matrix $H_{n,m}$ is large enough for the $1S$-charmonium states as discussed above,
the higher-lying states that have more extended wave functions seem to suffer from the finite
size effect caused by the boundary condition.
Indeed, the resulting wave functions of the $2S$ and $3S$ charmonium states might be
somewhat squeezed due to the smaller size of $\tilde{a}N_s/2$. Therefore, these energy levels would
be pushed down slightly due to the shrinkage of wave functions being less affected by the confining potential.
As we mentioned above, the lattice data of the spin-independent potential
becomes noisy in the range $r \agt 1.1$~fm, where signal-to-noise ratio of the BS wave function is poor, 
and also suffers from the finite volume effect in lattice QCD simulations.
In order to draw a firm conclusion for properties of higher-lying charmonium states without these effects,
we clearly need to extend the calculation of the charmonium potential derived from the ground state wave 
function to the higher-lying states such as $2S$ and $3S$ states, which have 
more extended wave functions, using a sufficiently large lattice.

\subsection{\label{sec:charm_pot}Chromium mass spectrum}
\begin{figure*}
  \includegraphics[width=.94\textwidth]{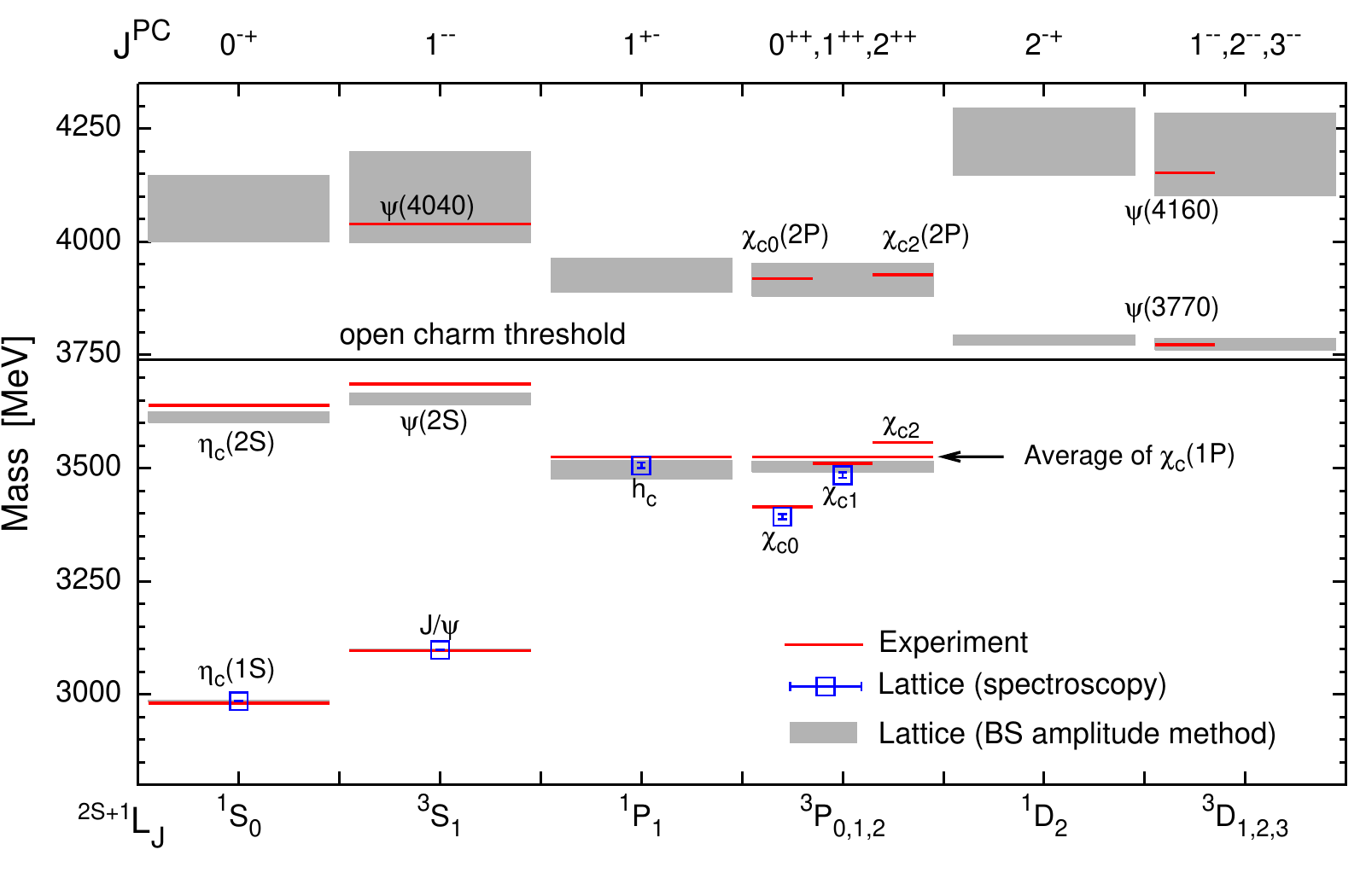}
  \centering
  \caption{Mass spectrum of charmonium states below and near the open-charm threshold.
    The vertical scale is in units of MeV.
    Labels of $^{2S+1}[L]_J$ ($J^{PC}$) are displayed in the lower (upper) horizontal axis.
    Rectangular shaded boxes indicate predictions from the NRp model with 
    purely theoretical inputs based on lattice QCD
    and their errors which are the sum of the statistical and systematic added in quadrature.
    Solid lines indicate experimental values of well established charmonium states, while
    square symbols represent results of the standard lattice spectroscopy.
    A horizontal solid line shows the open-charm threshold.
    A symbol of $\ovli{^3P_J}$ denotes the spin-weighted average of spin-triplet $^3P_J$ states
    whose mass is given by $M_{\ovli{\chi_{cJ}}} = (M_{\chi_{c1}} +3M_{\chi_{c2}} + 5M_{\chi_{c2}})/9$.}
  \label{fig:potential_spec}
\end{figure*}

 \begin{table}
\centering
   \caption{
    Masses of the charmonium states below 4200 MeV are summarized in units of MeV.
    The labels of AVE and HYP in a column of ``state''  for $S$-wave states denotes the spin-averaged mass 
    $(M_{^1S_0}+3M_{^3S_1})/4$ and hyperfine splitting mass $M_{^3S_1}-M_{^1S_0}$.   
    Experimental data~(denoted as Exp.) in the second column are taken from Particle Data Group, 
    rounded to $1$~MeV~\cite{Beringer:1900zz}.
    There are two kinds of lattice QCD results tabulated in the third and fourth columns.
    One is obtained by the standard lattice spectroscopy, while another is evaluated by 
    solving the Schr\"odinger equation with the charmonium potential determined from lattice QCD.
    For the latter, the first error is statistical and the second error systematic as described in text.
    The spin-weighted average mass~(denoted as $\ovli{^3[L]_J}$) are 
    also included for spin triplet states $^3[L]_J$. The last column shows the results from a NRp model
    ~\cite{Barnes:2005pb}.
   }
  \label{tab:higher_charmonium}
   \begin{ruledtabular}
     \begin{tabular}{rcccc} 
       state &Exp. 
      & \multicolumn{2}{c}{Lattice QCD} & NRp model~\cite{Barnes:2005pb} \\
      & & spectroscopy & BS amplitude  & \\[4pt] \hline
$\eta_c$         $(1^1S_0)$     & 2981 & 2985(1)   & 2985(2)(1)  & 2982 \\
$J/\psi$         $(1^3S_1)$     & 3097 & 3099(1)   & 3099(2)(1)  & 3090 \\
AVE                             & 3068 & 3070(9)   & 3070(2)(1)  & 3063 \\
HYP                             & 116  & 114(1)    & 113(1)(0)   & 108  \\[2pt]
		     	            		       	    
$\eta_c$         $(2^1S_0)$     & 3639 &             & 3612(9)(7)  & 3630 \\
$\psi$           $(2^3S_1)$     & 3686 &             & 3653(12)(5) & 3672 \\
AVE                             & 3674 &             & 3643(11)(5) & 3662 \\
HYP                             & 47   &             & 41(6)(3)    & 42   \\[2pt]
		     	            		       	    
$\eta_c$         $(3^1S_0)$     &      &             & 4074(20)(70)& 4043 \\
$\psi$           $(3^3S_1)$     & 4039 &             & 4099(24)(98)& 4072 \\
AVE                             &      &             & 4092(22)(91)& 4065 \\
HYP                             &      &             & 25(15)(28)  & 29   \\[2pt] \hline

$h_c$            $(1^1P_1)$     & 3525 & 3506(6) & 3496(7)(19) & 3516\\
$\ovli{\chi_{cJ}}$ $(\ovli{1^3P_J})$& 3525 &         & 3503(7)(10) & 3524\\
$\chi_{c0}$      $(1^3P_0)$     & 3415 & 3393(6) &             & 3424\\
$\chi_{c1}$      $(1^3P_1)$     & 3511 & 3485(6) &             & 3505\\
$\chi_{c2}$      $(1^3P_2)$     & 3556 &             &             & 3556\\[2pt]
		     	            
$h_c$            $(2^1P_1)$     &      &             & 3927(16)(34) & 3934\\
$\ovli{\chi_{cJ}}$ $(\ovli{2^3P_J})$&      &         & 3916(19)(31) & 3943\\
$\chi_{c0}$      $(2^3P_0)$     & 3918 &             &              & 3852\\
$\chi_{c1}$      $(2^3P_1)$     &      &             &              & 3925\\
$\chi_{c2}$      $(2^3P_2)$     & 3927 &             &              & 3972\\[2pt] \hline
$\eta_{c2}$      $(1^1D_2)$     &      &             & 3783(12)(4)  & 3799   \\
$\ovli{\psi}$      $(\ovli{1^3D_J})$&      &         & 3774(13)(2)  & 3800   \\
$\psi$           $(1^3D_1)$     & 3773 &             &              & 3785   \\
$\psi$           $(1^3D_2)$     &      &             &              & 3800   \\
$\psi$           $(1^3D_3)$     &      &             &              & 3806   \\[2pt]
		     	            
$\eta_{c2}$      $(2^1D_2)$     &      &             & 4221(21)(72) & 4158   \\
$\ovli{\psi}$      $(\ovli{2^3D_J})$&      &         & 4193(25)(88) & 4159   \\
$\psi$           $(2^3D_1)$     & 4153 &             &              & 4142   \\
$\psi$           $(2^3D_2)$     &      &             &              & 4158   \\
$\psi$           $(2^3D_3)$     &      &             &              & 4167  \\ 
     \end{tabular}
   \end{ruledtabular}
 \end{table}

We show the charmonium spectrum below 4200~MeV in Fig.~\ref{fig:potential_spec}.
Theoretical spectra plotted as rectangular shaded boxes are given
by solving the discrete nonrelativistic Schr\"odinger
equations with the theoretical inputs. 
Vertical box length represents the level of uncertainty, which is given by adding 
statistical and systematic errors in quadrature.
For the purpose of comparison, we plot both experimental values (horizontal lines) and results of the standard lattice 
spectroscopy (square symbols) together. The experimental values are taken from 
Particle Data Group~\cite{Beringer:1900zz}.
At first glance, one can find that below the open charm threshold, 
our theoretical calculations from the NRp model with the lattice inputs 
excellently agree with not only the lattice spectroscopy, but also experiments.
Especially a agreement between two lattice results provides a strong check
for the validity of our new method.
All results including the lattice spectroscopy results are also summarized 
together with the experimental values in Table~\ref{tab:higher_charmonium}.

In this study, we have succeeded in extracting {\it only the spin-spin potential}
among the spin-dependent parts of the interquark potential.
Thus at this stage we cannot predict the spin-orbit splitting
which is led by the tensor and spin-orbit terms of the spin-dependent potential.
In other words, we can compute only the {\it spin-averaged mass}
for higher-wave charmonium states like $P$-wave charmonium $\chi_{cJ}$ state.

\begin{figure}
  \centering
  \includegraphics[width=.23\textwidth]{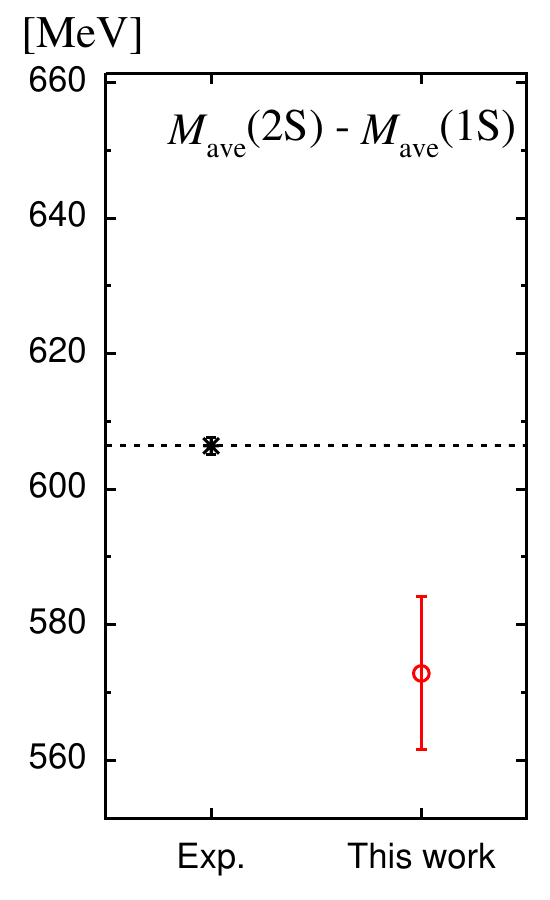}
  \includegraphics[width=.23\textwidth]{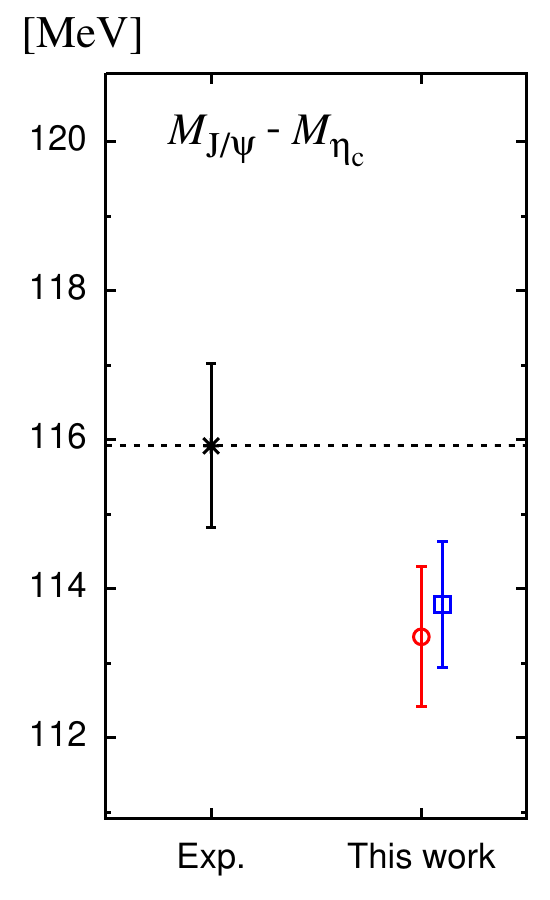}
  \includegraphics[width=.23\textwidth]{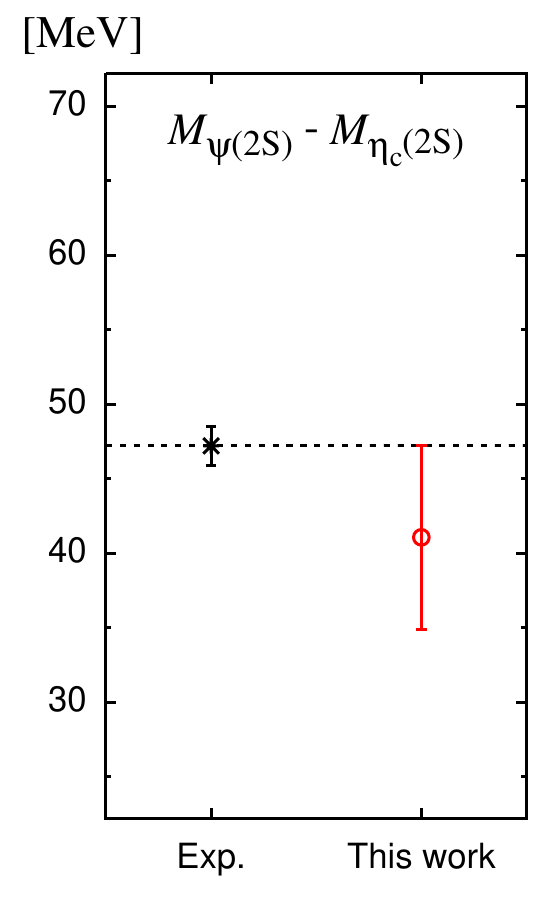}
  \includegraphics[width=.23\textwidth]{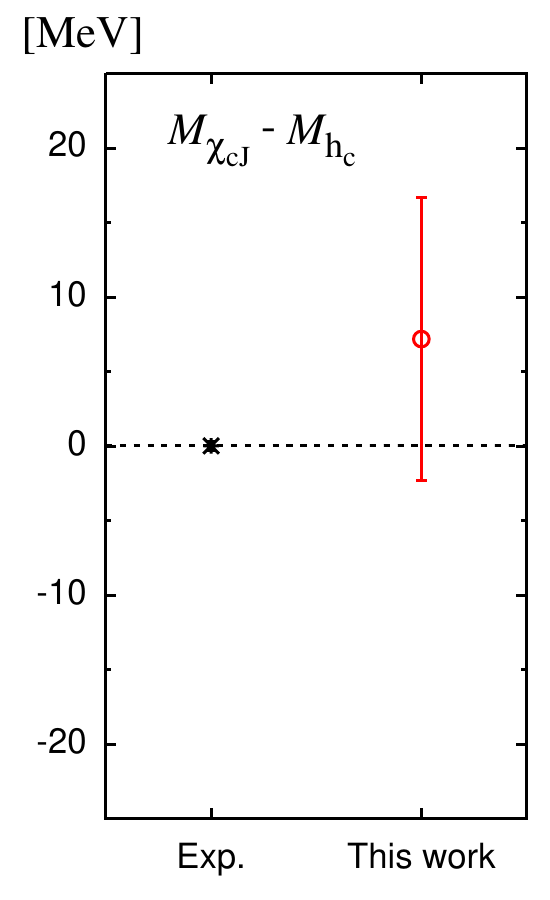}
  \caption{Mass splittings of states lying below the open charm threshold in units of MeV,
    compared to the physical mass splitting.
    Upper panels show
    the mass splitting $M_{\rm ave}(2S)- M_{\rm ave}(1S)$  between the spin-averaged $1S$- and $2S$-states~(left),
    and the hyperfine mass splittings $M_{J/\psi}- M_{\eta_c}$ between $1S$-states~(right),
    while lower panels show mass splitting of 
    $M_{\psi(2S)}- M_{\eta_c(2S)}$ between $2S$-states~(left) 
    and mass splitting of $M_{\ovli{\chi_{cJ}}}- M_{h_c}$ between $1P$-states~(right).
    In each plot,  cross and  circle  symbols indicate
    the experimental data and theoretical results obtained from the NRp model with the lattice inputs,
    respectively.
    The quoted errors indicate the sum of the statistical and systematic errors added in quadrature.
    Dashed lines represent the central value of the experiment.
    Only for the hyperfine mass splitting between $1S$-states,
    the results of the lattice spectroscopy is shown as a square.
  }
  \label{fig:split_charm}
\end{figure}

The mass splitting between the radial excitations 
and the ground state also provides an important validity check on our new approach.
Fig.~\ref{fig:split_charm} shows several mass splittings theoretically predicted
by the hybrid approach in comparison to physical values of the corresponding splittings.
In the top left panel of Fig.~\ref{fig:split_charm}, the radial excitation mass splitting of 
the spin-averaged $1S$ and $2S$ states
is $M_{\rm ave}(2S)-M_{\rm ave}(1S) = 573(10)(5)$~MeV,
of which value is slightly smaller than the experimental value of $606(1)$~MeV~\cite{Beringer:1900zz}.
This deviation~($\sim 30$~MeV) from the experiment can be attributed to the finite volume effect,
which is caused by the fact that more extended $2S$ state than $1S$ state is forced to fit in the 
spatial volume~$\sim (3\;{\rm fm})^3$.
Note that the $S {\text -} D$ mixing due to the tensor force is simply ignored in our calculations.
Thus, the spin-spin potential solely gives the mass splitting in hyperfine multiplets.
The top right panel of Fig.~\ref{fig:split_charm} shows
the hyperfine-splitting energy for the $1S$ charmonium states: 
$M_{1S,\rm hyp} = M_{J/\psi}-M_{\eta_c}$.
It is found that there is a good agreement between two lattice results; 
$113.4(9)(1)$~GeV from the NRp model with lattice inputs and 
$113.8(8)$~GeV from the standard lattice spectroscopy.
This simply suggests that there is no additional uncertainties induced by
both determining the charmonium potential and the charm quark mass 
by the BS amplitude method and solving the nonrelativistic Schr\"odinger equation with them.

We remark that the values of the hyperfine-splitting are slightly smaller 
than the experimental one $M_{1S,\rm hyp}^{\rm exp} = 116.6(1.2)$ GeV.
This would be simply due to insufficient calibration of the RHQ parameters and also other
possible systematic uncertainties including the remnant discretization artifact. 
On the other hand, the hyperfine splitting energy for the $2S$ charmonium states, which 
is plotted in the bottom left panel of Fig.~\ref{fig:split_charm}, shows that
the value of $M_{\psi^\prime(2S)}-M_{\eta_c(2S)} = 41(6)(3)$~MeV obtained from the hybrid approach is 
roughly consistent with the experimental value $47(1)$~MeV, within its error range.


The bottom right panel of Fig.~\ref{fig:split_charm} shows 
the $1P$ hyperfine mass splitting which is given by an energy difference between 
between the $h_c$ and spin-averaged $\chi_{cJ}$ states: $M_{1P, \rm hyp} = M_{\ovli{\chi_{cJ}}}-M_{h_c}$.
Experimentally, the value of $M_{1P, \rm hyp}^{\text{exp}}$ is known to be zero with high accuracy as
$M_{1P, \rm hyp}^{\text{exp}}= 0.02(23)$~MeV~\cite{Rubin:2005px,Dobbs:2008ec}.
The hybrid approach yields a small splitting energy as $M_{1P, \rm hyp} = 7.2(1.6)(9.3)$~MeV,
which is consistent with the zero value within a large error.
Of course, however, the spin-spin charmonium potential determined in the BS amplitude method 
is not still enough to describe the tiny $1P$ hyperfine splitting measured in experiment.
As we mentioned in the previous subsection, a finite-range spin-spin potential gives a
nonzero value of hyperfine mass splitting even in the case of higher-wave states such
as $P$-wave state, while zero hyperfine splitting measured in experiments is easily 
reproduced by the {\it point-like} spin-spin potential widely adopted in phenomenological quark potential models.
Here, we stress that the spin-spin potential from the BS amplitude method is {\it finite-range} and
therefore the value of $M_{1P, \rm hyp}^{\text{exp}}$ is highly sensitive to both shapes of the spin-spin potential
and wave functions of $P$-wave states. According to our systematic study of the BS amplitude method performed
in quenched lattice QCD~\cite{Kawanai:2013aca}, the spin-spin potential receives large uncertainties due to
the discretization artifacts more than the spin-independent central potential. To make a firm conclusion,
it is necessary to perform the present calculation on the finer lattice.

Our theoretical calculations for the charmonium mass spectrum 
below the open-charm threshold are basically in good agreement with the experimental measurements.
The point we wish to emphasize here is that our novel approach has
no free parameters in solving the Schr\"odinger equation in contrast to the phenomenological NRp models.
All of the parameters are fixed by lattice QCD simulations, where
three light hadron masses ({\it e.g.} pion, kaon and $\Omega$ baryon)
are used for setting the lattice spacing $a$ and hopping parameters of the light and strange quarks 
({\it i.e.} the light and strange quark masses). 
In this study, the charm quark was treated in the quenched approximation. 
Then the experimental values of $\eta_c$ and $J/\psi$ charmonium masses
are used to determine the charm quark parameters appeared in the RHQ action.
In this sense, the hybrid approach proposed here is distinctly different from
existing calculations in the phenomenological quark potential models.

Let us now attempt to straightforwardly extend the hybrid approach to above the open-charm threshold.
Only the spin-averaged mass is considered for the $P$ and $D$ spin-triplet states:
$M(\ovli{n {}^3P_J}) = (M_{n {}^3P_0} +3M_{n {}^3P_1} + 5M_{n {}^3P_2})/9$ and
$M(\ovli{n {}^3D_J}) = (3M_{n {}^3D_1} +5M_{n {}^3D_2} + 7M_{n {}^3D_3})/15$.
In order to provide mass splittings among these spin-triplet states,
the tensor and spin-orbit potentials are inevitably required. Since, in this paper, we
succeeded in extracting the spin-spin potential solely for the spin-dependent potentials, 
we should focus on the spin-averaged masses. 

First of all, one can observe that the values obtained from the hybrid approach above the open-charm threshold 
fairly agree with the existing experimental data, although errors are relatively large
as shown in Table~\ref{tab:higher_charmonium}.
We, however, are not in a position to give a realistic description
to the higher-lying charmonium states, which are located above the open-charm threshold.
This is because there are the following remarks in our calculations
including the higher-lying charmonium states.


\begin{enumerate}
\item The higher-lying charmonium states significantly suffer from systematic uncertainties,
which are mainly due to the less knowledge of the long-range part of the spin-independent potential.
We have no reasonable data for the charmonium potential at longer distances than about $1.1$ fm
since the wave function of  the $1S$ ground-state possesses the highly localized nature. 
Therefore we need to calculate the potential form the higher-lying charmonium states.
Alternatively, we simply extrapolate the long-range behavior of the potential 
outside the region, where the charmonium potential is really determined from the localized wave function.
In the latter case, the higher-lying spectrum of the charmonium is more sensitive to the choice of 
the adopted functional form in the fitting procedure.   

\item The possible mass shift due to mixing the $Q\bar{Q}$ states with $D\bar{D}$ continuum
  is completely neglected in this study.
  One may expect that the NRp models without such mixing works well 
  to describe the low-lying charmonium systems far below its threshold.
  On the other hand, such coupled channel effects might not be negligible
  near and above the threshold and then the potential description may lose the accuracy 
  of theoretical prediction, though the naive treatment of the NRp models even for higher-lying 
  charmonium states was phenomenologically successful despite the absence of coupled channel effects~\cite{Godfrey:1985xj,Barnes:2005pb}

\item  For the higher-lying excitations of the spin-1 charmonium state, 
the $S \text{-} D$ mixing becomes severe since the level spacings between $(n+1)^3S_1$ and $n^3D_1$ get 
narrower~\cite{Badalian:2008dv}. However, $S \text{-} D$ mixing effects on $J/\psi$, $\psi(2S)$ 
and $\psi(3S)$ states are not taken into account in the present calculation since the tensor term
in the spin-dependent potentials is not determined in this study. 
Similarly, the mass estimations of $\ovli{\chi_{cJ}}(nP)$ and $\ovli{\psi}(nD)$ tabulated 
in Table~\ref{tab:higher_charmonium} are calculated without consideration of 
possible partial-wave mixings such as $S \text{-} D$, $F \text{-} P$ and  $D \text{-} G$ mixings. 

\end{enumerate}

To calculate the BS wave function of $1P$-states, better source operators with respect to odd-parity wave 
function~\cite{Murano:2011aa} are required. Meanwhile some extension of the variational 
method~\cite{Michael:1985ne,Luscher:1990ck}  
to the four-point correlation functions is necessary for extracting the BS wave function of the radial excitation of the $S$-wave states. These new calculations can give more realistic prediction especially to the higher-lying charmonia. The former provides information of the spin-orbit and also tensor potentials~\cite{Murano:2011aa}.
The latter can provide not only the tensor potential, but also the mixing angle between $2^3S_1$ and $1^3D_1$ states in the same way as the nuclear force~\cite{Aoki:2009ji}.
Furthermore more data points of the charmonium potential at large distances can be
accessible from such excited states of the charmonium, which have more extended wave function 
than that of $1S$ ground states. Such kind of studies is now under way~\cite{SSandTK}.

\section{\label{Ds} Application to heavy-light system}
In the charmonium (heavy-heavy) system, 
the spectrum below the open charm threshold are well described 
by {\it potential description} with our charmonium potential including the spin-spin interaction, 
which was determined by the BS amplitude method in dynamical lattice QCD simulations.
In this section, we apply the new method to the $D_s$ heavy-light meson system, which 
represents the case of mesons with non-degenerate quark masses.
Apart from the phenomenological interest, we also would like to examine the validity range of the new method
in terms of the size of quark kinetic mass. 

\subsection{Lattice setup for charmed-strange mesons}
\begin{table}
  \caption{
    Masses of low-lying $D_s$ meson states,
   the spin-averaged mass and hyperfine splitting energy of $1S$ charmonium states.
 The colums have the same meaning as in Table~\ref{tab:higher_charmonium}.
    Results are given in units of GeV.  
    \label{tab:Ds_mass}
      }
      \begin{ruledtabular}                                                                              
      \begin{tabular}{cccccc} 
       state & ($J^{P}$) &  $\Gamma$ & fit range & mass~[GeV] & $\chi^2/{\rm d.o.f.}$\\ \hline
      $D_s$      & ($0^{-}$)  & $\gamma_5$ & [30:47] & 1.9780(12) & 1.08\\
      $D_s^*$  &  ($1^{-}$) & $\gamma_i$ & [30:47] & 2.1230(42) & 0.61\\ 
      $M_{\text{ave}}(1S)$ & ---&      & --- & 2.0865(33) & --- \\ 
      $E_{\text{hyp}}(1S)$  & ---&     & --- & 0.1461(37) & --- \\[2pt] \hline
      $D_{s0}^*$ & ($0^{+}$) & $1$  & [14:26] & 2.3536(77) & 1.45\\  
      $D_{s1}$   & ($1^{+}$)  & $\gamma_5\gamma_i$ & [14:26] & 2.4689(83) & 1.14\\ 
      $D_{s1}$   & ($1^{+}$)  & $\gamma_i\gamma_j$ &[14:22] & 2.4893(87) & 1.20\\[2pt] 
    \end{tabular}
      \end{ruledtabular}                                                                                
\end{table}
\begin{figure}
  \includegraphics[width=.49\textwidth]{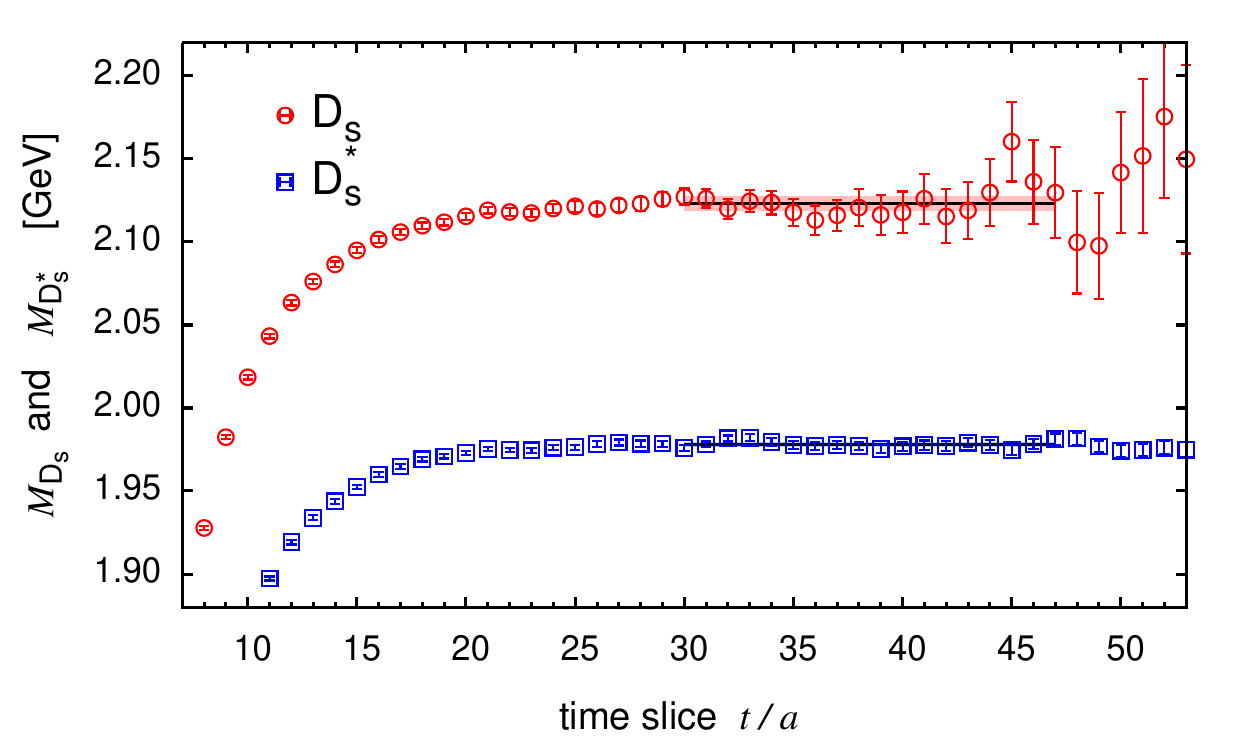}
  \includegraphics[width=.49\textwidth]{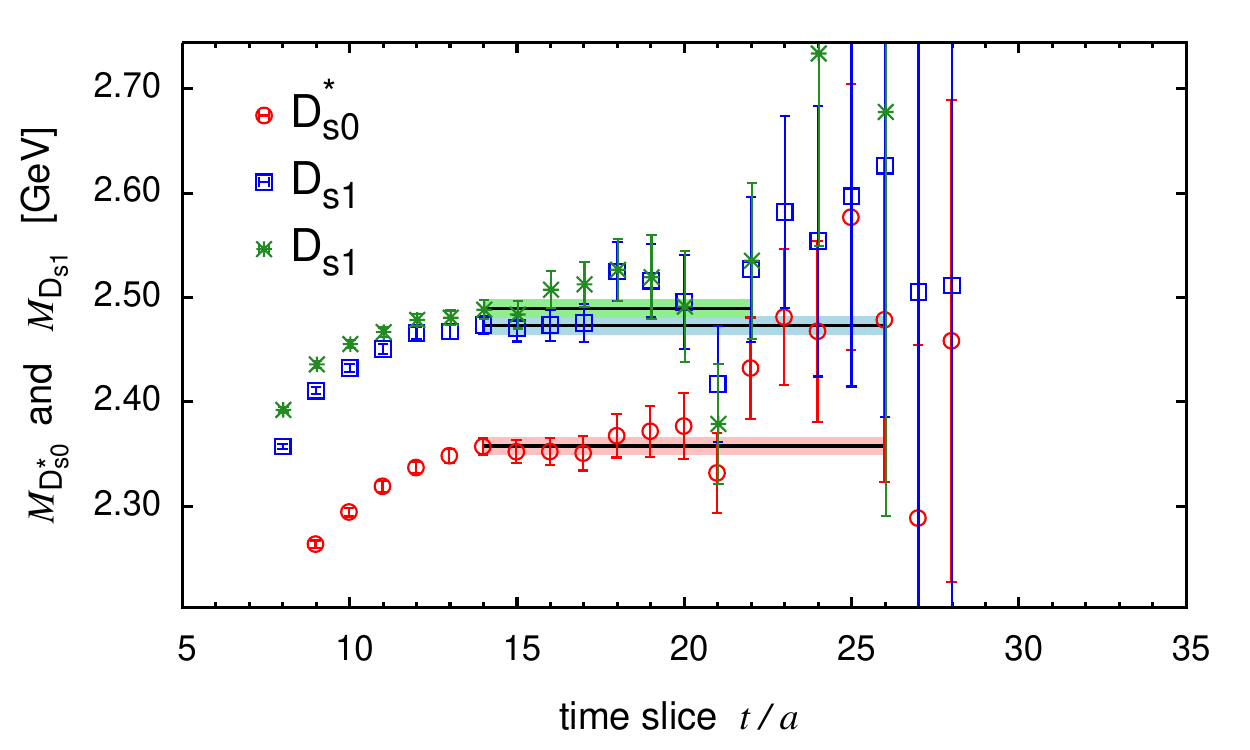}
  \caption{Effective mass plots for low-lying $S$ and $P$-wave $D_s$ meson states. 
    $S$-wave states ($D_s$ and $D_s^{*}$) and 
    $P$-wave states ($D_{s0}^{*}(2317)$, $D_{s1}(2460)$ and $D_{s1}(2536)$) 
    are shown in the upper and lower panels, respectively. 
    Each $D_s$ meson state is specified in legend.    
    Horizontal lines and shaded bands denote fit results with 
    statistical errors estimated by the jackknife method and their fit range.
    }
  \label{fig:effmass_Ds}
\end{figure}
The numerical setup for the charmed-strange ($D_s$) mesons system is basically same as for the 
calculation of the charmonium system.
For the charm quark, we use the RHQ action with the parameters calibrated by $1S$-charmonium states.
In addition to the computation of the charm quark propagator, 
another fermion matrix inversion for a strange quark is required to compute the $D_s$-meson 
correlation functions.
The non-perturbatively $\mathcal{O}(a)$-improved Wilson quark action~($c_{SW} = 1.715$)
is used for the strange quark.
A hopping parameter of the strange quark is chosen to be $\kappa_{s}=0.13640$,
which is the same as the sea strange quark used in gauge field generation.
Simultaneously, $s\bar{s}$-mesons are supplementarily calculated, and
we use $s\bar{s}$-meson data for a consistency check on the kinetic masses of the strange quark 
determined through  $s\bar{s}$ and $c\bar{s}$ systems as we will discuss later.

Fig.~\ref{fig:effmass_Ds} shows the effective mass plots for $S$ and $P$-wave $D_s$-meson states. 
The $D_s$-meson masses are determined by constant fits to the plateaus observed in the effective mass plots
with covariance matrices accounting for the data correlation among different time slices. 
Results of the $D_s$-meson masses together with fit ranges used in the fits and the values of 
$\chi^2/{\rm d.o.f}$ are summarized in Table~\ref{tab:Ds_mass}.
The quoted errors represent only the statistical errors given by the jackknife analysis.
The RHQ action for the charm quark works well even for the low-lying $D_s$-mesons.
The spin-averaged and hyperfine splitting $D_s$-meson masses, $M_{\rm ave}^{1S} = 2.0865(33)$~GeV
and  $M_{\rm hyp}^{1S} = 0.1461(37)$~GeV, are obtained from the standard lattice spectroscopy.
Although the simulated strange quarks are slightly off the physical point, these results are quite close to the experimental data of 
$M_\text{ave}^{\text{exp}}(1S) = 2.07635(27)$~GeV and
$M_\text{hyp}^{\text{exp}}(1S) = 0.1438(4)$~GeV. 
The deviations from the experimental results are within about $0.5\%$.
Furthermore, results of $P$-wave $D_s$-meson states from the lattice spectroscopy
marginally reproduce the experimental data. Similar results are reported by the PACS-CS collaboration 
using $2+1$ flavor dynamical gauge configurations generated with the physical strange quark~\cite{Namekawa:2011wt,Aoki:2009ix}.

The two-point correlation functions of both pseudoscalar and vector $s\bar{s}$-mesons, 
{\it i.e.}  $\eta_{s\bar{s}} (0^{-})$ and $\phi (1^{-})$ mesons, 
are also calculated in this study.
We obtain results of $M_{\eta_{s\bar{s}}} = 0.7699(9)$~GeV and $M_\phi = 1.0827(68)$~GeV.
The similar values are reported in Ref.~\cite{Aoki:2008sm}.
The fit range was chosen to be $24 \leq t \leq 39$ for both states.
The the $\phi$ meson mass is somewhat heavier than the experimental values of 
$M_\phi^{\rm exp} =  1.019455(20)$~GeV. 
It should be attributed to the fact that the simulated strange quarks are slightly off the physical point.
Although the systematic uncertainty due to slightly heavier strange quark mass 
is expected to be extremely small in the charmonium spectrum, 
we should take into account some corrections for the $D_s$-meson spectrum~\cite{Mohler:2011ke}.

\subsection{BS wave function}
\begin{figure}
  \centering
  \includegraphics[width=.49\textwidth]{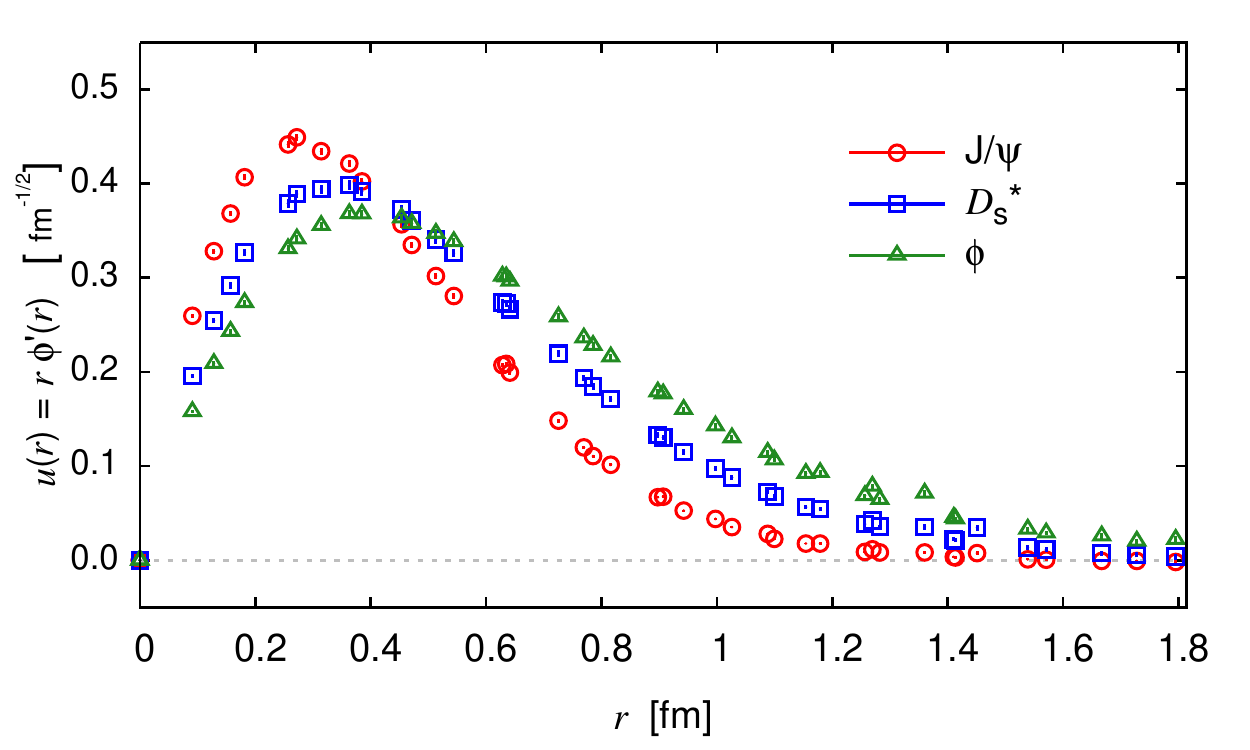}
  \caption{The reduced BS wave function $u(r)=r\phi(r)$ for the $1S$ vector $c\bar{c}$~(circles), 
    $c\bar{s}$~(squares) and $s\bar{s}$~(triangles)  states,
     as a function of spatial distance $r$. 
    They are normalized as $\sum \phi(\vec{x}) = 1$.
   }
  \label{fig:wavefunc_Ds}
\end{figure}
In Fig.~\ref{fig:wavefunc_Ds}, we show
the reduced BS wave functions for the $1S$ vector $c\bar{c}$, $c\bar{s}$ and $s\bar{s}$  states
corresponding to  $J/\psi$, $D_s^*$ and $\phi$ mesons, respectively.
It is found that the $D_s^*$ wave function is spatially extended to at least 
the half of the spatial extent of lattice volume ($N_sa/2\sim 1.5$~fm).
Although the amplitude of the wave function of the $D_s^*$ meson is considerably small at $r\sim 1.5$ fm, 
it still seems to remain non-zero values in the range of $r > 1.5$ fm, where only off-axis data points are available.
The wrap round effect would cause the rotational symmetry breaking at longer distances.
Therefore, in the $D_s$ system, the interquark potential
could be more affected by the finite volume effect than the charmonium system.
In the case of the $s\bar{s}$ system, which is more spatially extended than the $c\bar{s}$ system
as shown in Fig.~\ref{fig:wavefunc_Ds}, this problem could become more severe.


\subsection{quark kinetic mass}
\begin{figure}
  \centering
  \includegraphics[width=.49\textwidth]{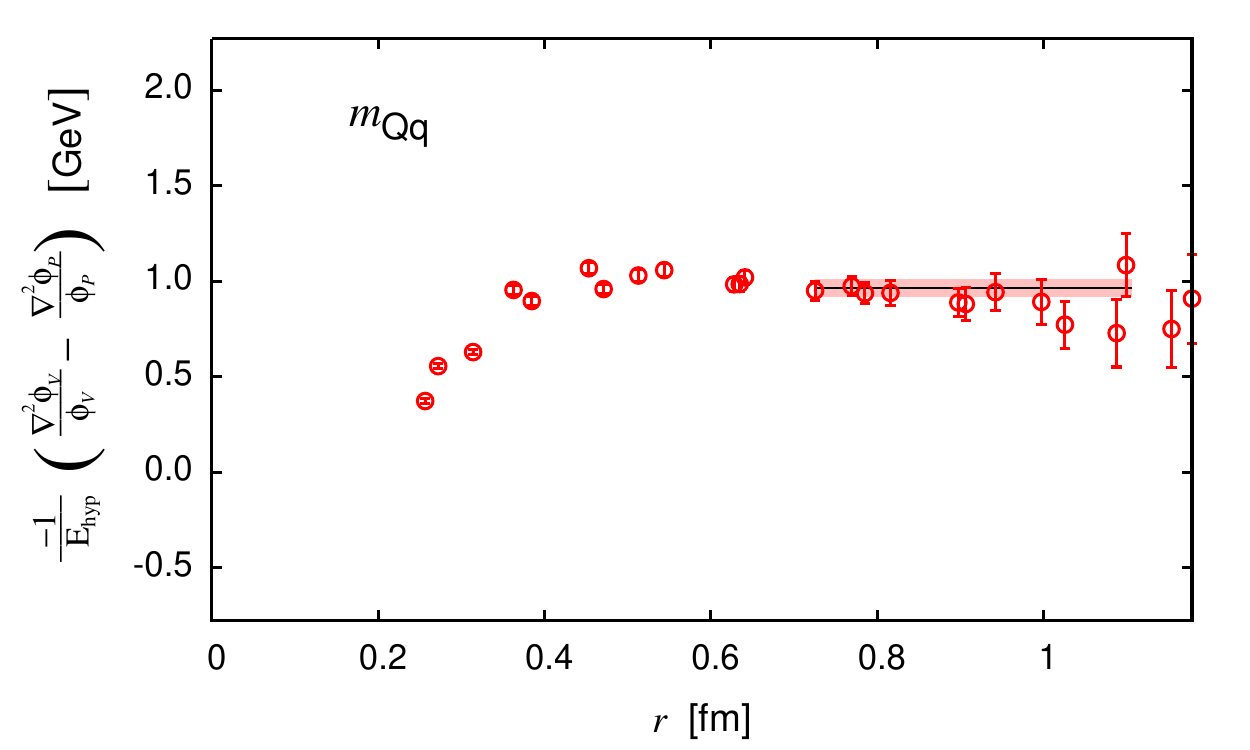}
  \includegraphics[width=.49\textwidth]{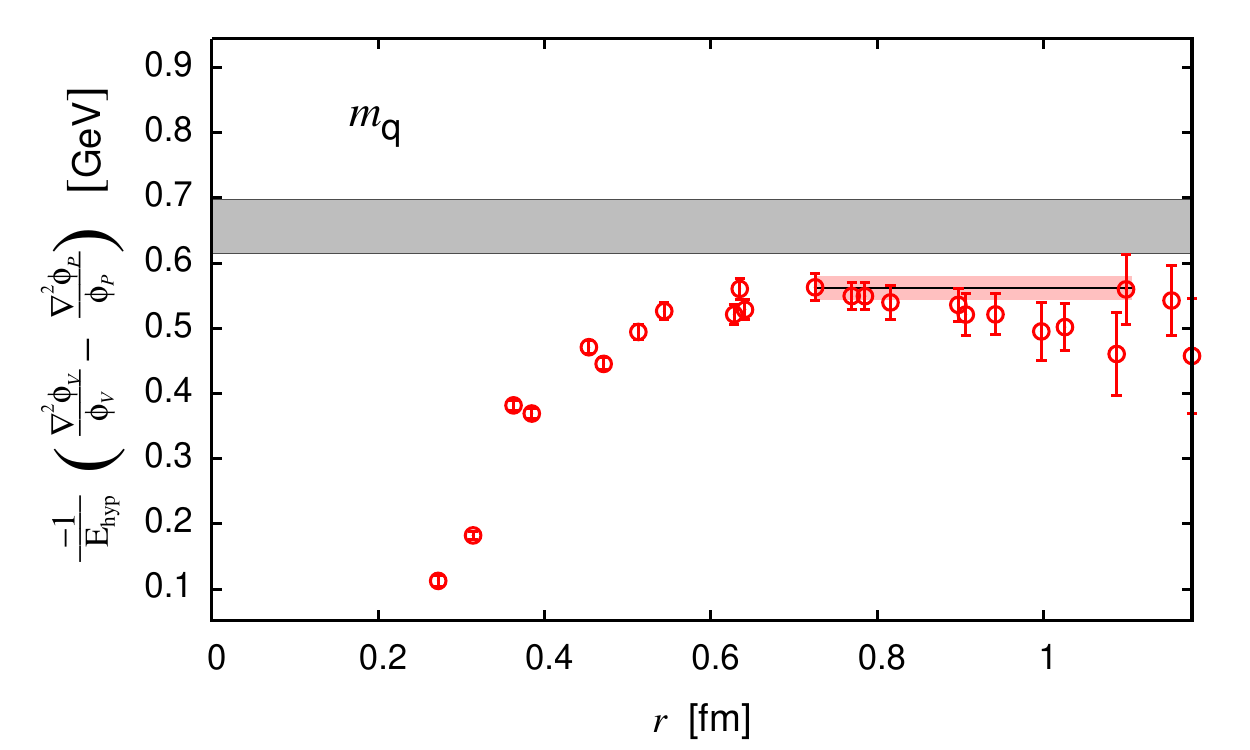}
  \caption{
    The determination of the reduced mass $m_{Qq}$ from the $D_s$($c\bar{s}$) system~(upper) and 
    strange quark mass $m_q$ from the $s\bar{s}$ system~(lower) in the BS amplitude method.
    We obtain the quark kinetic masses of $m_{Qq}$ and $m_q$ from 
    the asymptotic behavior of the right-hand side of Eq.~(\ref{eq_quark_mass}) in long-distance region.
    Solid lines with shaded bands represent the fitting results and fit ranges with the statistical error estimated by the jackknife method.
    In the lower plot, a horizontal shaded band indicates a kinetic mass of the strange quark, which is 
    independently evaluated 
    by the relation $m_q = m_{Qq}m_{Q}/(2m_{Q}-m_{Qq})$ with the values of $m_{Qq}$ and $m_Q$
    from the $c\bar{s}$ and $c\bar{c}$ systems. 
  }
  \label{fig:quarkmass_Ds}
\end{figure}
 Fig.~\ref{fig:quarkmass_Ds} illustrates the determination of quark kinetic  
 mass from the $c\bar{s}$ and $s\bar{s}$ meson systems in the BS amplitude method.
 A quantity $m_{Qq}$ is defined as twice the reduced mass of the $D_s$($c\bar{s}$) system:
 $m_{Qq}=2m_Qm_q/(m_Q+m_q)$, while $m_q$ corresponds to the strange quark mass.

 As shown in the upper panel of Fig.~\ref{fig:quarkmass_Ds}, we fit the data points 
 of the $c\bar{s}$ meson system at relatively large distances, where the reasonable plateau is 
 found in the region of $r \agt 0.7$~fm. We then obtain the value of 
 $m_{Qq}=0.959(45)(34)(36)$~GeV.  
 The first error is statistical, and the second and third ones are systematic uncertainties 
 due to the choice of data points taken from three directions and a variation of $t_{\rm min}$, respectively.
The strange quark mass $m_q$ can be evaluated by two data sets of $m_{Qq}$ and $m_Q$ 
through the relation $m_q = m_{Qq}m_{Q}/(2m_{Q}-m_{Qq})$. The value of $m_Q$
corresponds to the charm mass, which was already determined in the previous section.
When combined with results obtained from the $D_s(c\bar{s})$ and charmonium ($c\bar{c}$) system,
we obtain the value of $m_q = 656(41)$~MeV for the strange quark mass.  
Quoted error is statistical one, which was determined by the jackknife method.

Independently, the strange quark mass $m_q$ can be determined through 
the $s\bar{s}$ system as depicted by circle symbols in the lower panel of Fig.~\ref{fig:quarkmass_Ds}. 
Similar to the upper figure, the reasonable plateau is found in the region of $r \agt 0.7$~fm.
We compute a weighted average of the data points in the range of $8\leq r/a \leq 7\sqrt{3}$ with 
a covariance matrix accounting for the correlation, and then obtain the value of
$m_q=554(19)(6)(8)$~MeV, which is close to a typical value of constitute 
strange quark mass ($\sim M_\phi/2 \approx 500$~MeV) adopted 
in $SU(6)$ quark models~\cite{Close:1979bt}.
The meaning of the three quoted errors is explained above.

For comparison, the previously estimated value of $m_q = 656(41)$~MeV 
from the data sets of $m_{Qq}$ and $m_Q$ is also displayed by a 
horizontal shaded band in the lower figure.
There is $2\sigma$ discrepancy between this band and the plateau behavior 
of $-({\bm \nabla}^2\phi_{\rm V}/\phi_{\rm V}-{\bm \nabla}^2\phi_{\rm PS}/\phi_{\rm PS})/E_\text{hyp}$ 
for the $s\bar{s}$ system at large distances.
Although this discrepancy may imply that 
nonrelativistic treatment is no longer valid for the heavy-light system, we would like to remind of the
fact that the BS wave function of the $s\bar{s}$-meson system at large distances is likely affected 
by the finite volume effect as discussed previously.

\begin{figure}
   \centering
  \includegraphics[width=.49\textwidth]{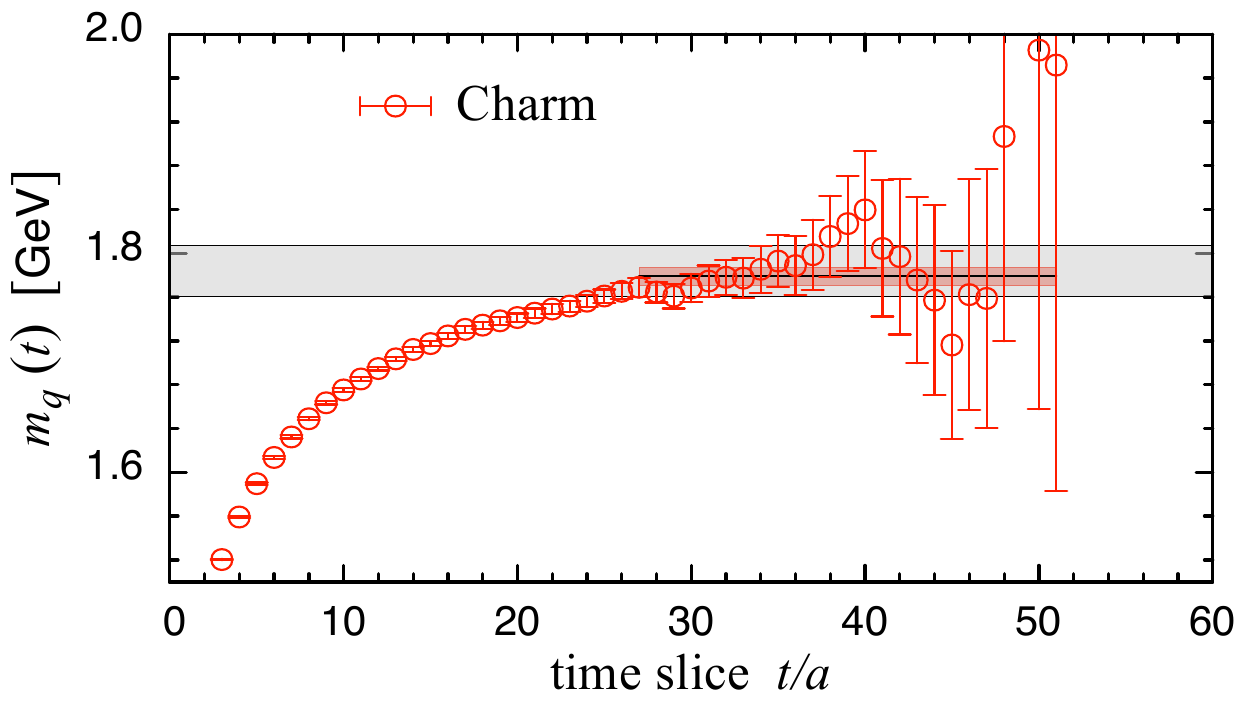}
  \includegraphics[width=.49\textwidth]{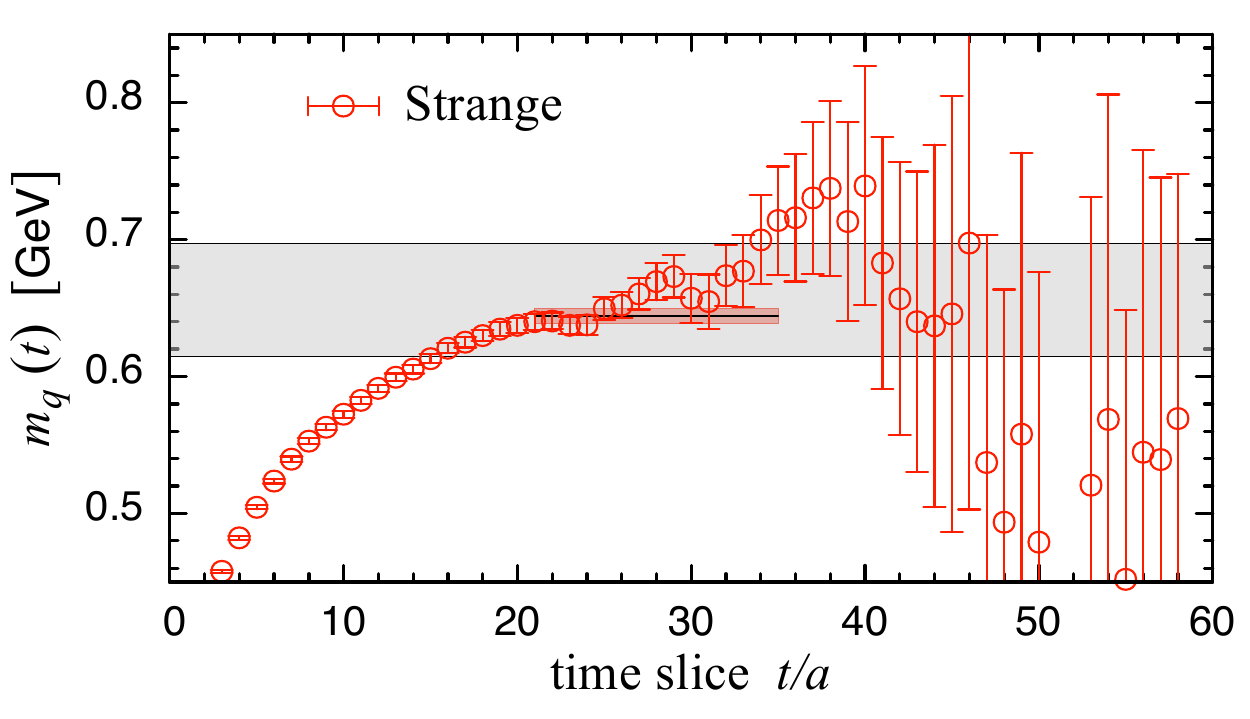}
  \caption{
  Effective masses of gauge-variant quark two-point correlator in the Landau gauge for charm (upper panel) and strange (lower panel).
  Solid lines indicate fit results for ``pole'' masses of charm and strange quarks and shaded bands display
  the fitting ranges and one standard deviations estimated by the jackknife method.
  In each panel, a wider and horizontal shaded band indicates a kinetic mass evaluated from the $c\bar{s}$ and/or $c\bar{c}$ systems
  within the BS amplitude method.
  }
  \label{fig:quarkmass_pole}
 \end{figure}

The following discussion shows that the above speculation is likely to be true.
The strange quark mass determined from the BS wave function of the $s\bar{s}$-meson states
is indeed underestimated compared to a ``pole mass'' determined from the effective mass of 
gauge-variant quark two-point correlator {\it in the Landau gauge}, while two different
calculations show remarkable consistency for the charm quark as summarized in Table~\ref{tab:quark_mass}.
Fig.~\ref{fig:quarkmass_pole} shows effective mass plots and comparisons with ones obtained within the BS amplitude method.
The Landau-gauge pole mass is an alternative way of measuring the quark mass in lattice QCD.
For details of how to calculate it, see Ref.~\cite{Sasaki:2006jn}.

If one may choose the $c\bar{s}$ result rather than the $s\bar{s}$ result
for the BS amplitude method, the strange quark masses from two estimation methods become consistent again.
Although the physics behind the consistency discussed here is beyond the scope of this paper, we may 
simply conclude that the discrepancy between results from the $c\bar{s}$ and $s\bar{s}$ mesons 
is mainly attributed to the finite volume effect on the $s\bar{s}$ wave function.

 \begin{table}
  \centering
   \caption{
   Charm and strange quark masses, which are determined from the Coulomb-gauge quark-antiquark BS amplitude and
   the Landau-gauge quark propagator, are summarized in units of GeV. 
     \label{tab:quark_mass}
   }
   \begin{ruledtabular}                                                         
     \begin{tabular}{cccccc} 
     & \multicolumn{2}{c}{BS amplitude} & quark propagator \\
     flavor  & $Q\bar{Q}$ or $q\bar{q}$ & $Q\bar{q}$ & Landau gauge \\ 
     \hline
     charm    &  1.784(23) &  ---  & 1.776(8) \\
     strange  &  0.554(19) &  0.656(41) & 0.643(5)\\
     \end{tabular}
   \end{ruledtabular}                                                           
 \end{table}

\subsection{charmed-strange potential}
\begin{figure}
  \centering
  \includegraphics[width=.49\textwidth]{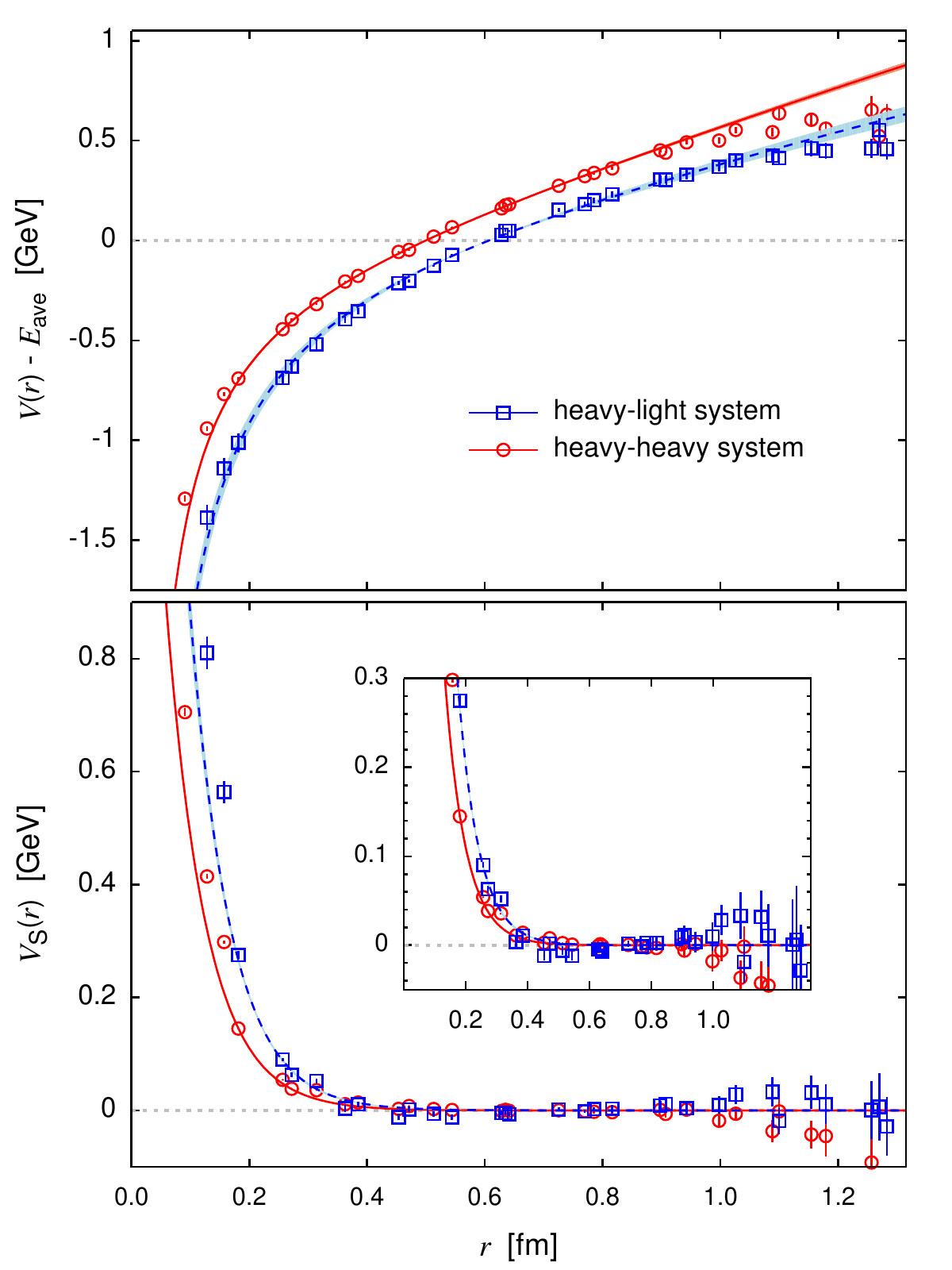}
  \caption{The spin-independent and spin-spin 
    interquark potentials for the $c\bar{c}$~(circles) and $c\bar{s}$~(squares) systems,
    calculated from the BS wave functions in dynamical lattice QCD simulations
    with almost physical quark masses.
    In the upper panel, the spin-independent parts of both the charmonium 
    and $c\bar{s}$ potentials are plotted. For clarity of the figure, the constant
    energy shift $E_{\rm ave}$, which is given by the spin-averaged mass of $1S$ states,
    is not subtracted. Solid and dashed curves represent the fit results with the Cornell parametrization. 
    The shaded bands show statistical uncertainties in the fitting procedure where
    the jackknife method is employed.
    In the lower panel, we show the spin-spin potential $V_S(r)$. 
 The exponential form is used for fitting the resultant spin-spin potentials for the $c\bar{c}$  and $c\bar{s}$ systems.
    The inset shows a magnified view.
  }
  \label{fig:potential_Ds}
\end{figure}
Fig~\ref{fig:potential_Ds} shows results of the spin-independent and 
spin-spin interquark potentials obtained from the $D_s$ and $D_s^*$ meson states (hereafter  
called {\it $c\bar{s}$ potential}) in the dynamical QCD simulation.
For purpose of comparison, the charmonium potentials are also displayed. 

At first glance, a shape of the $c\bar{s}$ potential is basically similar 
to that  of the charmonium potential,
so that we similarly adopt the Cornell potential form
for the spin-independent $c\bar{s}$ potential and also 
exponential~(Yukawa) form for the spin-spin $c\bar{s}$ potential
as is the case in the charmonium potential.
We obtain the Cornell parameters of the $c\bar{s}$ potential as 
$A=1.30(8)(22)(21)(21)$ and $\sqrt{\sigma}=324(16)(34)(26)(4)$ MeV
with a reasonably small value of $\chi^2/\text{d.o.f.} \approx 1.9$.
The first error is statistical and the second, third and forth ones are systematic 
uncertainties due to the choice of data points taken from three directions, 
and variations of $t_{\rm min}$ and $r_{\rm min}$, respectively.

The appropriate fitting range was determined to minimize a $\chi^2/\text{d.o.f.}$ value 
taking into account the data correlation among different spatial distances $r$.
We choose the fit range of $[r_{\rm min}/a: r_{\rm max}/a] = [4: 7\sqrt{3}]$, which corresponds
to the same range in the case of the charmonium potential.
Although the string tension has weak dependence on quark kinetic mass~\cite{Kawanai:2013aca},
the Coulomb coefficient of the $c\bar{s}$ potential significantly grows 
in comparison with that of the charmonium potential.
For the spin-spin potential, we obtain 
$\alpha = 3.79(36)$~GeV and $\beta = 2.89(9)$~GeV~($\alpha = 1.48(14)$ and $\beta = 1.97(9)$~GeV) 
from the exponential (Yukawa) form fit with $\chi^2/\text{d.o.f.} \approx 1.49$~($\chi^2/\text{d.o.f.} \approx 1.86$).
We quote only the statistical errors, which are determined by
the jackknife method. We find that the size of the finite-range of the spin-spin potential for the $c\bar{s}$ system
is almost consistent with the one obtained from the charmonium spin-spin potential within statistical uncertainties.

\begin{figure*}
  \centering
  \includegraphics[width=.90\textwidth]{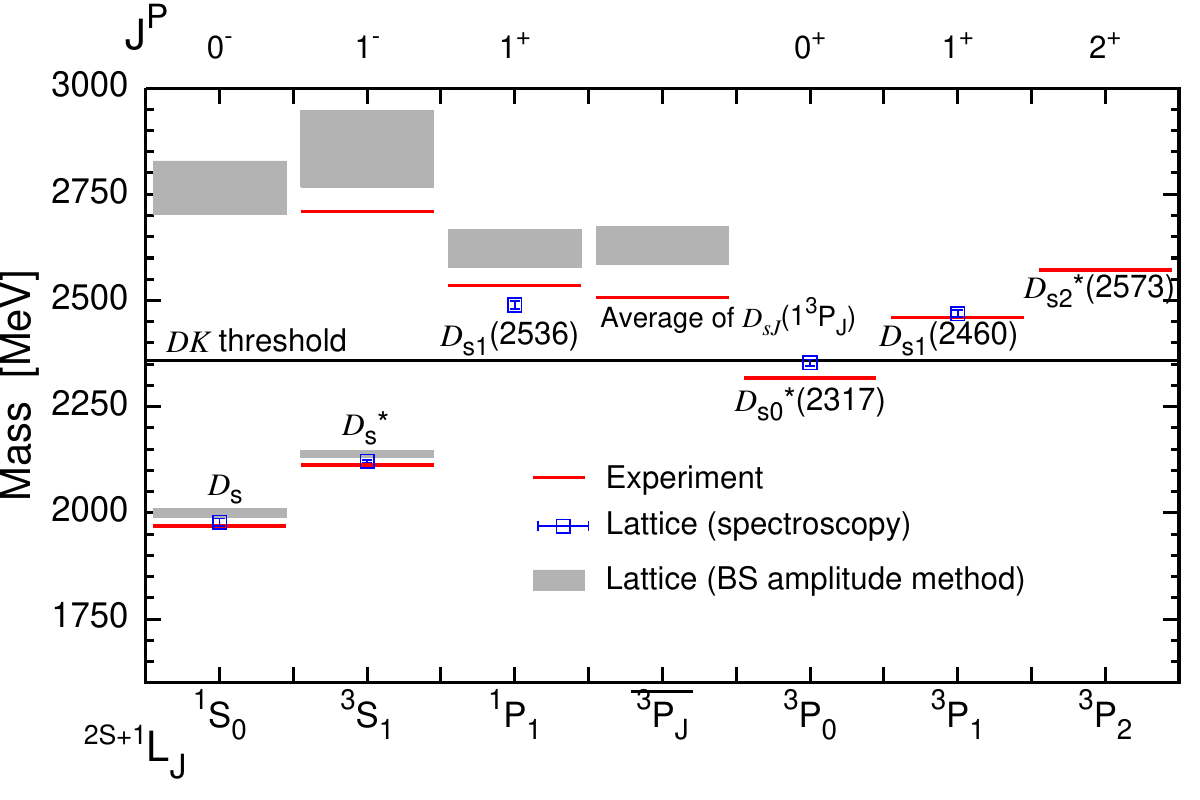}
  \caption{
    Mass spectrum of the charmed-strange mesons around the $DK$ threshold.
    The vertical scale is in units of MeV.
    Labels of $^{2S+1}[L]_J$ ($J^{P}$) are displayed in lower (upper) horizontal axis.    
    Solid lines indicate experimental values of well established $D_s$ meson states, 
    while square symbols represent results of the standard lattice spectroscopy. 
    Rectangular shaded boxes indicate predictions
    from the NRp models with purely theoretical inputs based on lattice QCD
    and their errors which are the sum of the statistical and systematic errors added in quadrature. 
    A horizontal solid line shows the $DK$ threshold.  
    A symbol of $\ovli{^3P_J}$ denotes the spin-weighted average of 
    spin-triplet $^3P_J$ states.
}
    \label{fig:spec_Ds}
\end{figure*}
\begingroup
 \begin{table}
\centering
   \caption{
    Masses of low-lying $D_s$ mesons are summarized in units of MeV.
    The labels of AVE and HYP in a column of ``state''  for $S$-wave states denotes the spin-averaged mass 
    $(M_{^1S_0}+3M_{^3S_1})/4$ and hyperfine splitting mass $M_{^3S_1}-M_{^1S_0}$.   
    Experimental values in the second column are taken from Particle Data Group, 
    rounded to $1$~MeV~\cite{Beringer:1900zz}.
    There are two kinds of lattice QCD results tabulated in the third and fourth columns.
    One is obtained by the standard lattice spectroscopy, while another is evaluated by 
    solving the Schr\"odinger equation with the $c\bar{s}$ potential determined from lattice QCD.
    For the latter, the first error is statistical and the second error systematic as described in text.
    The spin-weighted average mass~(denoted as $\ovli{^3[L]_J}$) are 
    also included for spin triplet states $^3[L]_J$. The last column gives
    the results from a NRp model~\cite{Lucha:2003gs}.  }
    \label{tab:Ds_spec}
     \begin{tabular}{ccccc} \hline  \hline 
     state &Exp. & \multicolumn{2}{c}{Lattice QCD~(This work)} & NRp model~\cite{Lucha:2003gs} \\
      & & spectroscopy & BS amplitude    & \\ \hline
      $D_s$        $(1^1S_0)$      & 1968  & 1978(1) & 2000(10)(3)  & 1963  \\
      $D_s^*$      $(1^3S_1)$      & 2112  & 2123(4) & 2138(8)(3)   & 2099  \\
      AVE                          & 2076  & 2087(3) & 2103(8)(3)   & 2065  \\
      HYP                          & 144   &  146(4) & 138(5)(1)    & 136\smallskip  \\ 
      
      $D_s$        $(2^1S_0)$      &       &         & 2766(38)(50) & \\
      $D_s^*$      $(2^3S_1)$      & 2709  &         & 2857(42)(80) & \\
      AVE                          &       &         & 2834(40)(73) & \\
      HYP                          &       &         & 92(9)(30)    & \smallskip  \\ \hline
      
      $D_{s1}$     $(1^1P_1)$           & 2535  & 2489(9) & 2623(30)(32) & 2527 \\
      $\ovli{D_{sJ}}$ $(\ovli{1^3P_J})$ & 2506  &         & 2629(29)(32) & 2532\\
      $D_{s0}^*$   $(1^3P_0)$           & 2318  & 2354(8) &              & 2446 \\
      $D_{s1}$     $(1^3P_1)$           & 2460  & 2469(8) &              & 2515 \\
      $D_{s2}^*$   $(1^3P_2)$           & 2572  &         &              & \ 2561 \smallskip \\ \hline \hline
     \end{tabular}
 \end{table}
\endgroup

\subsection{charmed-strange meson mass spectrum}
Using the $c\bar{s}$ potential calculated from lattice QCD, 
we obtain the spectrum of the charmed-strange mesons
in the same footing as the charmonium spectrum discussed in Sec.~\ref{sec:charm_pot}.
The resulting $D_s$ meson spectrum together with the experimental values and 
the results from the standard lattice spectroscopy are summarized in Table~\ref{tab:Ds_spec} and also Fig.~\ref{fig:spec_Ds}.

The results of $1S$ states in the $D_s$-meson family by solving 
the discrete nonrelativistic Schr\"odinger equation with lattice inputs 
show a good agreement with both experiments and standard lattice calculations below the $DK$ threshold.
Above the $DK$ threshold, the $c\bar{s}$ potential obtained from the BS amplitude method
can properly reproduce the mass ordering of the $D_s$ mesons, while absolute values of the masses are 
systematically larger than the experimental values; for example, the  
corresponding $D_{s1}(2536)$ state is overestimated by about $90$~MeV,
and then observe a large systematic discrepancy between 
two kinds of lattice QCD results. 

Recall that the results obtained from the standard lattice spectroscopy
near the $DK$ threshold are also slightly deviated from the experimental values. 
Although this implies that the observed discrepancies among them are not solely attributed 
to the BS amplitude approach, there are three possible sources of above mentioned discrepancies in our method. 
The first is, as we have noted repeatedly, 
associated with uncertainties in the long-range part of the spin-independent potential.
The fact that the $D_s$ state has the wider wave function leads to the large finite volume
effect on the interquark potential determined within the BS amplitude method at long distances.
Furthermore, information of the $c\bar{s}$ potential is accessible only within 
the ``size'' of the $S$-wave $D_s$ mesons. 
We therefore force the wave function to be zero at the spatial boundary
in the process of solving the discrete nonrelativistic Schr\"odinger equation
with present limited data of the $c\bar{s}$ potential.

The second possibility is that there are the $DK$ and $D^*K$ threshold effects. 
Since $D_{s0}(2310)$ and $D_{s1}(2460)$ are located near the $DK$ and  $D^*K$
thresholds respectively. Therefore, the coupling of these states to 
the $DK$ and $D^*K$ two-hadron states could not be negligible~\cite{Lang:2014yfa,Torres:2014vna}.
Such channel couplings may cause level repulsion and thus 
mass shift by the threshold effect might be happened~\cite{Barnes:2007xu,Godfrey:1986wj,Close:2005se}.

Finally, it is worth pointing out that we assumed that there is no $S {\text -} D$ mixing due to the tensor force
because of large energy gap between $1S$ and $1D$ states.
Strictly speaking, however, the $D_s^*$ meson which is specified by the quantum numbers
$J^P=1^-$ is not purely composed of the $^3S_1$ wave function. 
This approximation could introduce a small correction to the intermediate and short-range parts of the $c\bar{s}$ potential calculated in the BS amplitude method.

Therefore, we need further development of our approach to take into account 
both the coupled channel effect and $S \text{-} D$ mixing.  
A simulation with sufficiently large volume is also required for precise prediction of 
masses of the $D_{s}$-meson states near and above the $DK$ ($D^*K$) threshold within our approach.

 %
 %
\section{\label{summary}Summary}
We have calculated the interquark potentials for both the charmonium ($c\bar{c}$) and charmed-strange
 ($c\bar{s}$) mesons at almost physical point.
The interquark potential with finite quark masses are defined through the 
equal-time Bethe-Salpeter wave functions of the pseudoscalar and vector mesons. 
Our simulations have been done in the vicinity of the physical light quark masses, 
which corresponds to $M_\pi \approx 156$~MeV,
using the PACS-CS Iwasaki gauge configurations with $2+1$ flavors of dynamical clover light quarks. 
We use the relativistic charm quark tuned to reproduce the experimental values of the 
$\eta_c$ and $J/\psi$ masses. 

We first investigated the charmonium potential. 
The resulting spin-independent potential has the Coulomb-plus-linear form, 
and their parameters are close to the values used in the phenomenological NRp models.
The string breaking due to the presence of dynamical sea quarks is not apparently observed.
The spin-spin potential obtained from the dynamical simulations exhibits 
the finite-range repulsive interaction. 
Its shape is quite different from a repulsive $\delta$-function
potential induced by the one-gluon exchange, which is often adopted in the NRp models.

Our ultimate goal is to reveal the mystery behind 
rich structures recently observed in the heavy-heavy and heavy-light systems 
including the newly discovered charmonium-like mesons.
As a first step, we calculated the charmonium mass spectrum 
by solving nonrelativistic Schr\"odinger equation with purely theoretical inputs of 
the spin-independent and spin-spin potentials, and also the quark kinetic mass. 
To avoid any model dependence from fitting,
we practically solve the discrete Schr\"odinger equation in finite volume 
with Dirichlet boundary condition, and thus can handle direct lattice data 
of the charmonium potential without any parameterization.

We found an excellent agreement of low-lying charmonium masses between our results and the experimental data.
We here emphasize that our novel approach has no free parameters 
in solving the Schr\"odinger equation in contrast to
the phenomenological NRp models. In our calculations, three light hadron masses ({\it e.g.} the pion,
kaon and $\Omega$ baryon are chosen in the PACS-CS collaboration) and two charmonium masses
(the $\eta_c$ and $J/\psi$) are used for fixing the lattice spacing and quark mass parameters in the lattice QCD action including
the RHQ parameters for the charm quarks.

In order to precisely predict the mass spectrum above the open charm threshold, 
we should take into account both coupled-channel effect with the $D\bar{D}$ continuum
and $S {\text -} D$ mixing due to the presence of the tensor force. 
In this study, we simply ignore these effects and then apply the hybrid approach
to higher-lying charmonium states above the open-charm threshold. 
We found that the theoretical predictions of the NRp model calculation with lattice inputs 
are remarkably consistent with well-established experimental data for the conventional 
charmonium states.

For an application, we straightforwardly extend our method to calculate 
the charmed-strange meson system, which represents the case of
mesons with non-degenerate quark masses and also heavy-light system. 
A shape of the interquark $c\bar{s}$ potential in the $D_s$-meson system is 
basically similar to that of the charmonium potential.
Using the resulting $c\bar{s}$ potential as theoretical inputs, 
we obtain the spectrum of the charmed-strange mesons. 
Below the $DK$ threshold, our new method works well in spite of 
the fact that the $D_s$ mesons contain a strange quark or strange anti-quark. 
Although above the $DK$ threshold our $c\bar{s}$ potential can 
reproduce the mass ordering of the $D_s$ mesons, 
absolute values of the masses are consistent
with neither the experimental values nor the results from the standard 
lattice spectroscopy. 

Although it is difficult to draw any firm conclusion at this stage,
the discrepancies observed above the $DK$ threshold suggest that
all of coupled-channel effect, $S {\text -} D$ mixing and higher-order 
relativistic corrections, that are omitted in this study, are possibly 
important to understand the properties of the heavy-light mesons 
near the threshold. To disentangle these effects, simulations in
a larger lattice is necessary.
Conversely, one might say that by taking into account these effects properly,
our approach can give a systematic way to examine 
the validity of the potential description even for the charmed-strange system.
Such full analysis finally sheds light 
on the detailed properties of the $D_{s}$ mesons.

At least, in this study, the charmonium and charmed-strange potentials obtained from the BS amplitude method 
have been succeed in reproducing the low-lying masses below the open charm threshold and $DK$ threshold
respectively. Furthermore we showed that our new analysis can potentially shed light on the detailed 
properties of the heavy quarkonium system.  While only energy eigenvalues are evaluated from 
temporal information of meson correlation functions in the standard lattice spectroscopy, the new method takes 
an advantage of full spatial information together with temporal information.
The BS wave functions can be identified with the eigenstates of the Hamiltonian. 
Hence, without knowing the details of an explicit form of the Hamiltonian, 
lots of physical quantities could be calculated directly by the BS wave functions 
as studied in the NRp models. For examples, $E1$ and $M1$ radiative partial widths are supposed to be evaluated 
with the BS wave functions of the charmonium states. Such information is important to reveal the structure of hadrons.  

To derive a complete nonrelativistic Hamiltonian of the heavy-heavy and heavy-light systems
from lattice QCD, we must calculate all spin-dependent terms (spin-spin, tensor and spin-orbit forces),
which are required for more realistic predictions for the higher-lying states.
We now develop the BS amplitude method to calculate the BS wave functions 
of $P$-wave mesons, which provides information of the spin-orbit and also tensor potentials.

Once all spin-dependent terms of the interquark potential are determined and also
all systematic uncertainties are well understood, we will gain new and valuable insight on 
the  mesons newly discovered in the heavy-heavy and heavy-light systems.
It is also important to examine the validity of the potential description with the BS amplitude method
from the viewpoint of the $v$-expansion for a nonlocal and energy-independent interquark potential
$U$ originally appeared in Eq.~(\ref{Eq_schr}).
For this purpose, we would like to examine whether the same interquark potential is obtained
from the BS wave function of the radial excitation of the $S$-wave states. 
All of the above-mentioned extensions of the new method are now in progress~\cite{SSandTK}.

 %
\begin{acknowledgments}
 We would like to thank T. Hatsuda for helpful suggestions, H. Iida, Y. Ikeda and B. Charron for fruitful discussions. 
 This work was partially supported by JSPS/MEXT Grants-in- Aid 
 (No. 22-7653, No. 19540265, No. 21105504 and No. 23540284).
 T. Kawanai was partially supported by JSPS Strategic Young Researcher Overseas Visits Program
  for Accelerating Brain Circulation (No.R2411).
\end{acknowledgments}

\bibliography{paper} 

\begin{thebibliography}{89}
\expandafter\ifx\csname natexlab\endcsname\relax\def\natexlab#1{#1}\fi
\expandafter\ifx\csname bibnamefont\endcsname\relax
  \def\bibnamefont#1{#1}\fi
\expandafter\ifx\csname bibfnamefont\endcsname\relax
  \def\bibfnamefont#1{#1}\fi
\expandafter\ifx\csname citenamefont\endcsname\relax
  \def\citenamefont#1{#1}\fi
\expandafter\ifx\csname url\endcsname\relax
  \def\url#1{\texttt{#1}}\fi
\expandafter\ifx\csname urlprefix\endcsname\relax\def\urlprefix{URL }\fi
\providecommand{\bibinfo}[2]{#2}
\providecommand{\eprint}[2][]{\url{#2}}

\bibitem[{\citenamefont{Close}()}]{Close:1979bt}
\bibinfo{author}{\bibfnamefont{F.}~\bibnamefont{Close}}, \bibinfo{note}{{\it An
  introduction to quarks and partons}, Academic Press/london 1979, 481p}.

\bibitem[{\citenamefont{Eichten et~al.}(1975)\citenamefont{Eichten, Gottfried,
  Kinoshita, Kogut, Lane et~al.}}]{Eichten:1974af}
\bibinfo{author}{\bibfnamefont{E.}~\bibnamefont{Eichten}},
  \bibinfo{author}{\bibfnamefont{K.}~\bibnamefont{Gottfried}},
  \bibinfo{author}{\bibfnamefont{T.}~\bibnamefont{Kinoshita}},
  \bibinfo{author}{\bibfnamefont{J.~B.} \bibnamefont{Kogut}},
  \bibinfo{author}{\bibfnamefont{K.}~\bibnamefont{Lane}}, \bibnamefont{et~al.},
  \bibinfo{journal}{Phys.Rev.Lett.} \textbf{\bibinfo{volume}{34}},
  \bibinfo{pages}{369} (\bibinfo{year}{1975}).

\bibitem[{\citenamefont{Godfrey and Isgur}(1985)}]{Godfrey:1985xj}
\bibinfo{author}{\bibfnamefont{S.}~\bibnamefont{Godfrey}} \bibnamefont{and}
  \bibinfo{author}{\bibfnamefont{N.}~\bibnamefont{Isgur}},
  \bibinfo{journal}{Phys.Rev.} \textbf{\bibinfo{volume}{D32}},
  \bibinfo{pages}{189} (\bibinfo{year}{1985}).

\bibitem[{\citenamefont{Barnes et~al.}(2005)\citenamefont{Barnes, Godfrey, and
  Swanson}}]{Barnes:2005pb}
\bibinfo{author}{\bibfnamefont{T.}~\bibnamefont{Barnes}},
  \bibinfo{author}{\bibfnamefont{S.}~\bibnamefont{Godfrey}}, \bibnamefont{and}
  \bibinfo{author}{\bibfnamefont{E.}~\bibnamefont{Swanson}},
  \bibinfo{journal}{Phys.Rev.} \textbf{\bibinfo{volume}{D72}},
  \bibinfo{pages}{054026} (\bibinfo{year}{2005}), \eprint{hep-ph/0505002}.

\bibitem[{\citenamefont{Eichten and Feinberg}(1981)}]{Eichten:1980mw}
\bibinfo{author}{\bibfnamefont{E.}~\bibnamefont{Eichten}} \bibnamefont{and}
  \bibinfo{author}{\bibfnamefont{F.}~\bibnamefont{Feinberg}},
  \bibinfo{journal}{Phys.Rev.} \textbf{\bibinfo{volume}{D23}},
  \bibinfo{pages}{2724} (\bibinfo{year}{1981}).

\bibitem[{\citenamefont{Brambilla et~al.}(2005)\citenamefont{Brambilla, Pineda,
  Soto, and Vairo}}]{Brambilla:2004jw}
\bibinfo{author}{\bibfnamefont{N.}~\bibnamefont{Brambilla}},
  \bibinfo{author}{\bibfnamefont{A.}~\bibnamefont{Pineda}},
  \bibinfo{author}{\bibfnamefont{J.}~\bibnamefont{Soto}}, \bibnamefont{and}
  \bibinfo{author}{\bibfnamefont{A.}~\bibnamefont{Vairo}},
  \bibinfo{journal}{Rev.Mod.Phys.} \textbf{\bibinfo{volume}{77}},
  \bibinfo{pages}{1423} (\bibinfo{year}{2005}), \eprint{hep-ph/0410047}.

\bibitem[{\citenamefont{Voloshin}(2008)}]{Voloshin:2007dx}
\bibinfo{author}{\bibfnamefont{M.}~\bibnamefont{Voloshin}},
  \bibinfo{journal}{Prog.Part.Nucl.Phys.} \textbf{\bibinfo{volume}{61}},
  \bibinfo{pages}{455} (\bibinfo{year}{2008}), \eprint{0711.4556}.

\bibitem[{\citenamefont{Bondar et~al.}(2012)}]{Belle:2011aa}
\bibinfo{author}{\bibfnamefont{A.}~\bibnamefont{Bondar}} \bibnamefont{et~al.}
  (\bibinfo{collaboration}{Belle}), \bibinfo{journal}{Phys. Rev. Lett.}
  \textbf{\bibinfo{volume}{108}}, \bibinfo{pages}{122001}
  (\bibinfo{year}{2012}), \eprint{1110.2251}.

\bibitem[{\citenamefont{Godfrey and Olsen}(2008)}]{Godfrey:2008nc}
\bibinfo{author}{\bibfnamefont{S.}~\bibnamefont{Godfrey}} \bibnamefont{and}
  \bibinfo{author}{\bibfnamefont{S.~L.} \bibnamefont{Olsen}},
  \bibinfo{journal}{Ann.Rev.Nucl.Part.Sci.} \textbf{\bibinfo{volume}{58}},
  \bibinfo{pages}{51} (\bibinfo{year}{2008}), \eprint{0801.3867}.

\bibitem[{\citenamefont{Wegner}(1971)}]{Wegner:1984qt}
\bibinfo{author}{\bibfnamefont{F.~J.} \bibnamefont{Wegner}},
  \bibinfo{journal}{J. Math. Phys.} \textbf{\bibinfo{volume}{12}},
  \bibinfo{pages}{2259} (\bibinfo{year}{1971}).

\bibitem[{\citenamefont{Wilson}(1974)}]{Wilson:1974sk}
\bibinfo{author}{\bibfnamefont{K.~G.} \bibnamefont{Wilson}},
  \bibinfo{journal}{Phys.Rev.} \textbf{\bibinfo{volume}{D10}},
  \bibinfo{pages}{2445} (\bibinfo{year}{1974}).

\bibitem[{\citenamefont{Bali}(2001)}]{Bali:2000gf}
\bibinfo{author}{\bibfnamefont{G.~S.} \bibnamefont{Bali}},
  \bibinfo{journal}{Phys.Rept.} \textbf{\bibinfo{volume}{343}},
  \bibinfo{pages}{1} (\bibinfo{year}{2001}), \eprint{hep-ph/0001312}.

\bibitem[{\citenamefont{Bali et~al.}(1997{\natexlab{a}})\citenamefont{Bali,
  Schilling, and Wachter}}]{Bali:1996cj}
\bibinfo{author}{\bibfnamefont{G.~S.} \bibnamefont{Bali}},
  \bibinfo{author}{\bibfnamefont{K.}~\bibnamefont{Schilling}},
  \bibnamefont{and} \bibinfo{author}{\bibfnamefont{A.}~\bibnamefont{Wachter}},
  \bibinfo{journal}{Phys.Rev.} \textbf{\bibinfo{volume}{D55}},
  \bibinfo{pages}{5309} (\bibinfo{year}{1997}{\natexlab{a}}),
  \eprint{hep-lat/9611025}.

\bibitem[{\citenamefont{Bali et~al.}(1997{\natexlab{b}})\citenamefont{Bali,
  Schilling, and Wachter}}]{Bali:1997am}
\bibinfo{author}{\bibfnamefont{G.~S.} \bibnamefont{Bali}},
  \bibinfo{author}{\bibfnamefont{K.}~\bibnamefont{Schilling}},
  \bibnamefont{and} \bibinfo{author}{\bibfnamefont{A.}~\bibnamefont{Wachter}},
  \bibinfo{journal}{Phys.Rev.} \textbf{\bibinfo{volume}{D56}},
  \bibinfo{pages}{2566} (\bibinfo{year}{1997}{\natexlab{b}}),
  \eprint{hep-lat/9703019}.

\bibitem[{\citenamefont{Koike}(1989)}]{Koike:1989jf}
\bibinfo{author}{\bibfnamefont{Y.}~\bibnamefont{Koike}},
  \bibinfo{journal}{Phys.Lett.} \textbf{\bibinfo{volume}{B216}},
  \bibinfo{pages}{184} (\bibinfo{year}{1989}).

\bibitem[{\citenamefont{Born et~al.}(1994)\citenamefont{Born, Laermann, Walsh,
  and Zerwas}}]{Born:1993cp}
\bibinfo{author}{\bibfnamefont{K.}~\bibnamefont{Born}},
  \bibinfo{author}{\bibfnamefont{E.}~\bibnamefont{Laermann}},
  \bibinfo{author}{\bibfnamefont{T.}~\bibnamefont{Walsh}}, \bibnamefont{and}
  \bibinfo{author}{\bibfnamefont{P.}~\bibnamefont{Zerwas}},
  \bibinfo{journal}{Phys.Lett.} \textbf{\bibinfo{volume}{B329}},
  \bibinfo{pages}{332} (\bibinfo{year}{1994}).

\bibitem[{\citenamefont{Koma and Koma}(2007)}]{Koma:2006fw}
\bibinfo{author}{\bibfnamefont{Y.}~\bibnamefont{Koma}} \bibnamefont{and}
  \bibinfo{author}{\bibfnamefont{M.}~\bibnamefont{Koma}},
  \bibinfo{journal}{Nucl.Phys.} \textbf{\bibinfo{volume}{B769}},
  \bibinfo{pages}{79} (\bibinfo{year}{2007}), \eprint{hep-lat/0609078}.

\bibitem[{\citenamefont{Koma and Koma}(2010)}]{Koma:2010zz}
\bibinfo{author}{\bibfnamefont{Y.}~\bibnamefont{Koma}} \bibnamefont{and}
  \bibinfo{author}{\bibfnamefont{M.}~\bibnamefont{Koma}},
  \bibinfo{journal}{Prog.Theor.Phys.Suppl.} \textbf{\bibinfo{volume}{186}},
  \bibinfo{pages}{205} (\bibinfo{year}{2010}).

\bibitem[{\citenamefont{Kawanai and Sasaki}(2011)}]{Kawanai:2011xb}
\bibinfo{author}{\bibfnamefont{T.}~\bibnamefont{Kawanai}} \bibnamefont{and}
  \bibinfo{author}{\bibfnamefont{S.}~\bibnamefont{Sasaki}},
  \bibinfo{journal}{Phys.Rev.Lett.} \textbf{\bibinfo{volume}{107}},
  \bibinfo{pages}{091601} (\bibinfo{year}{2011}), \eprint{hep-lat/1102.3246}.

\bibitem[{\citenamefont{Kawanai and Sasaki}(2012)}]{Kawanai:2011jt}
\bibinfo{author}{\bibfnamefont{T.}~\bibnamefont{Kawanai}} \bibnamefont{and}
  \bibinfo{author}{\bibfnamefont{S.}~\bibnamefont{Sasaki}},
  \bibinfo{journal}{Phys.Rev.} \textbf{\bibinfo{volume}{D85}},
  \bibinfo{pages}{091503} (\bibinfo{year}{2012}), \eprint{1110.0888}.

\bibitem[{\citenamefont{Kawanai and Sasaki}(2014)}]{Kawanai:2013aca}
\bibinfo{author}{\bibfnamefont{T.}~\bibnamefont{Kawanai}} \bibnamefont{and}
  \bibinfo{author}{\bibfnamefont{S.}~\bibnamefont{Sasaki}},
  \bibinfo{journal}{Phys. Rev.} \textbf{\bibinfo{volume}{D89}},
  \bibinfo{pages}{054507} (\bibinfo{year}{2014}), \eprint{1311.1253}.

\bibitem[{\citenamefont{Aoki et~al.}(2009)}]{Aoki:2008sm}
\bibinfo{author}{\bibfnamefont{S.}~\bibnamefont{Aoki}} \bibnamefont{et~al.}
  (\bibinfo{collaboration}{PACS-CS Collaboration}),
  \bibinfo{journal}{Phys.Rev.} \textbf{\bibinfo{volume}{D79}},
  \bibinfo{pages}{034503} (\bibinfo{year}{2009}), \eprint{0807.1661}.

\bibitem[{\citenamefont{Aoki et~al.}(2003)\citenamefont{Aoki, Kuramashi, and
  Tominaga}}]{Aoki:2001ra}
\bibinfo{author}{\bibfnamefont{S.}~\bibnamefont{Aoki}},
  \bibinfo{author}{\bibfnamefont{Y.}~\bibnamefont{Kuramashi}},
  \bibnamefont{and} \bibinfo{author}{\bibfnamefont{S.-i.}
  \bibnamefont{Tominaga}}, \bibinfo{journal}{Prog.Theor.Phys.}
  \textbf{\bibinfo{volume}{109}}, \bibinfo{pages}{383} (\bibinfo{year}{2003}),
  \eprint{hep-lat/0107009}.

\bibitem[{\citenamefont{Ishii et~al.}(2007)\citenamefont{Ishii, Aoki, and
  Hatsuda}}]{Ishii:2006ec}
\bibinfo{author}{\bibfnamefont{N.}~\bibnamefont{Ishii}},
  \bibinfo{author}{\bibfnamefont{S.}~\bibnamefont{Aoki}}, \bibnamefont{and}
  \bibinfo{author}{\bibfnamefont{T.}~\bibnamefont{Hatsuda}},
  \bibinfo{journal}{Phys.Rev.Lett.} \textbf{\bibinfo{volume}{99}},
  \bibinfo{pages}{022001} (\bibinfo{year}{2007}), \eprint{nucl-th/0611096}.

\bibitem[{\citenamefont{Aoki et~al.}(2010{\natexlab{a}})\citenamefont{Aoki,
  Hatsuda, and Ishii}}]{Aoki:2009ji}
\bibinfo{author}{\bibfnamefont{S.}~\bibnamefont{Aoki}},
  \bibinfo{author}{\bibfnamefont{T.}~\bibnamefont{Hatsuda}}, \bibnamefont{and}
  \bibinfo{author}{\bibfnamefont{N.}~\bibnamefont{Ishii}},
  \bibinfo{journal}{Prog.Theor.Phys.} \textbf{\bibinfo{volume}{123}},
  \bibinfo{pages}{89} (\bibinfo{year}{2010}{\natexlab{a}}), \eprint{0909.5585}.

\bibitem[{\citenamefont{Nemura et~al.}(2009)\citenamefont{Nemura, Ishii, Aoki,
  and Hatsuda}}]{Nemura:2008sp}
\bibinfo{author}{\bibfnamefont{H.}~\bibnamefont{Nemura}},
  \bibinfo{author}{\bibfnamefont{N.}~\bibnamefont{Ishii}},
  \bibinfo{author}{\bibfnamefont{S.}~\bibnamefont{Aoki}}, \bibnamefont{and}
  \bibinfo{author}{\bibfnamefont{T.}~\bibnamefont{Hatsuda}},
  \bibinfo{journal}{Phys.Lett.} \textbf{\bibinfo{volume}{B673}},
  \bibinfo{pages}{136} (\bibinfo{year}{2009}), \eprint{0806.1094}.

\bibitem[{\citenamefont{Ikeda}(2010)}]{Ikeda:2010zz}
\bibinfo{author}{\bibfnamefont{Y.}~\bibnamefont{Ikeda}}
  (\bibinfo{collaboration}{HAL QCD Collaboration}),
  \bibinfo{journal}{Prog.Theor.Phys.Suppl.} \textbf{\bibinfo{volume}{186}},
  \bibinfo{pages}{228} (\bibinfo{year}{2010}).

\bibitem[{\citenamefont{Kawanai and Sasaki}(2010)}]{Kawanai:2010ev}
\bibinfo{author}{\bibfnamefont{T.}~\bibnamefont{Kawanai}} \bibnamefont{and}
  \bibinfo{author}{\bibfnamefont{S.}~\bibnamefont{Sasaki}},
  \bibinfo{journal}{Phys.Rev.} \textbf{\bibinfo{volume}{D82}},
  \bibinfo{pages}{091501} (\bibinfo{year}{2010}), \eprint{1009.3332}.

\bibitem[{\citenamefont{Doi et~al.}(2012)}]{Doi:2011gq}
\bibinfo{author}{\bibfnamefont{T.}~\bibnamefont{Doi}} \bibnamefont{et~al.}
  (\bibinfo{collaboration}{HAL QCD Collaboration}),
  \bibinfo{journal}{Prog.Theor.Phys.} \textbf{\bibinfo{volume}{127}},
  \bibinfo{pages}{723} (\bibinfo{year}{2012}), \eprint{1106.2276}.

\bibitem[{\citenamefont{Aoki et~al.}(2012)}]{Aoki:2012tk}
\bibinfo{author}{\bibfnamefont{S.}~\bibnamefont{Aoki}} \bibnamefont{et~al.}
  (\bibinfo{collaboration}{HAL QCD Collaboration}) (\bibinfo{year}{2012}),
  \eprint{1206.5088}.

\bibitem[{\citenamefont{Murano et~al.}(2011)\citenamefont{Murano, Ishii, Aoki,
  and Hatsuda}}]{Murano:2011nz}
\bibinfo{author}{\bibfnamefont{K.}~\bibnamefont{Murano}},
  \bibinfo{author}{\bibfnamefont{N.}~\bibnamefont{Ishii}},
  \bibinfo{author}{\bibfnamefont{S.}~\bibnamefont{Aoki}}, \bibnamefont{and}
  \bibinfo{author}{\bibfnamefont{T.}~\bibnamefont{Hatsuda}},
  \bibinfo{journal}{Prog.Theor.Phys.} \textbf{\bibinfo{volume}{125}},
  \bibinfo{pages}{1225} (\bibinfo{year}{2011}), \eprint{1103.0619}.

\bibitem[{\citenamefont{Aoki et~al.}(2011)}]{Aoki:2011gt}
\bibinfo{author}{\bibfnamefont{S.}~\bibnamefont{Aoki}} \bibnamefont{et~al.}
  (\bibinfo{collaboration}{HAL QCD Collaboration}),
  \bibinfo{journal}{Proc.Japan Acad.} \textbf{\bibinfo{volume}{B87}},
  \bibinfo{pages}{509} (\bibinfo{year}{2011}), \eprint{1106.2281}.

\bibitem[{\citenamefont{Ishii et~al.}(2012)}]{HALQCD:2012aa}
\bibinfo{author}{\bibfnamefont{N.}~\bibnamefont{Ishii}} \bibnamefont{et~al.}
  (\bibinfo{collaboration}{HAL QCD Collaboration}),
  \bibinfo{journal}{Phys.Lett.} \textbf{\bibinfo{volume}{B712}},
  \bibinfo{pages}{437} (\bibinfo{year}{2012}), \eprint{1203.3642}.

\bibitem[{\citenamefont{Velikson and Weingarten}(1985)}]{Velikson:1984qw}
\bibinfo{author}{\bibfnamefont{B.}~\bibnamefont{Velikson}} \bibnamefont{and}
  \bibinfo{author}{\bibfnamefont{D.}~\bibnamefont{Weingarten}},
  \bibinfo{journal}{Nucl.Phys.} \textbf{\bibinfo{volume}{B249}},
  \bibinfo{pages}{433} (\bibinfo{year}{1985}).

\bibitem[{\citenamefont{Gupta et~al.}(1993)\citenamefont{Gupta, Daniel, and
  Grandy}}]{Gupta:1993vp}
\bibinfo{author}{\bibfnamefont{R.}~\bibnamefont{Gupta}},
  \bibinfo{author}{\bibfnamefont{D.}~\bibnamefont{Daniel}}, \bibnamefont{and}
  \bibinfo{author}{\bibfnamefont{J.}~\bibnamefont{Grandy}},
  \bibinfo{journal}{Phys.Rev.} \textbf{\bibinfo{volume}{D48}},
  \bibinfo{pages}{3330} (\bibinfo{year}{1993}), \eprint{hep-lat/9304009}.

\bibitem[{\citenamefont{Luscher}(1991)}]{Luscher:1990ux}
\bibinfo{author}{\bibfnamefont{M.}~\bibnamefont{Luscher}},
  \bibinfo{journal}{Nucl.Phys.} \textbf{\bibinfo{volume}{B354}},
  \bibinfo{pages}{531} (\bibinfo{year}{1991}).

\bibitem[{\citenamefont{Caswell and Lepage}(1978)}]{Caswell:1978mt}
\bibinfo{author}{\bibfnamefont{W.~E.} \bibnamefont{Caswell}} \bibnamefont{and}
  \bibinfo{author}{\bibfnamefont{G.~P.} \bibnamefont{Lepage}},
  \bibinfo{journal}{Phys.Rev.} \textbf{\bibinfo{volume}{A18}},
  \bibinfo{pages}{810} (\bibinfo{year}{1978}).

\bibitem[{\citenamefont{Ikeda and Iida}(2011)}]{Ikeda:2011bs}
\bibinfo{author}{\bibfnamefont{Y.}~\bibnamefont{Ikeda}} \bibnamefont{and}
  \bibinfo{author}{\bibfnamefont{H.}~\bibnamefont{Iida}}
  (\bibinfo{year}{2011}), \eprint{1102.2097}.

\bibitem[{\citenamefont{Aoki et~al.}(2006)}]{Aoki:2005et}
\bibinfo{author}{\bibfnamefont{S.}~\bibnamefont{Aoki}} \bibnamefont{et~al.}
  (\bibinfo{collaboration}{CP-PACS Collaboration, JLQCD Collaboration}),
  \bibinfo{journal}{Phys.Rev.} \textbf{\bibinfo{volume}{D73}},
  \bibinfo{pages}{034501} (\bibinfo{year}{2006}), \eprint{hep-lat/0508031}.

\bibitem[{\citenamefont{Iwasaki}(1983)}]{Iwasaki:2011np}
\bibinfo{author}{\bibfnamefont{Y.}~\bibnamefont{Iwasaki}}
  (\bibinfo{year}{1983}), \eprint{1111.7054}.

\bibitem[{\citenamefont{Kayaba et~al.}(2007)}]{Kayaba:2006cg}
\bibinfo{author}{\bibfnamefont{Y.}~\bibnamefont{Kayaba}} \bibnamefont{et~al.}
  (\bibinfo{collaboration}{CP-PACS Collaboration}), \bibinfo{journal}{JHEP}
  \textbf{\bibinfo{volume}{0702}}, \bibinfo{pages}{019} (\bibinfo{year}{2007}),
  \eprint{hep-lat/0611033}.

\bibitem[{\citenamefont{El-Khadra et~al.}(1997)\citenamefont{El-Khadra,
  Kronfeld, and Mackenzie}}]{ElKhadra:1996mp}
\bibinfo{author}{\bibfnamefont{A.~X.} \bibnamefont{El-Khadra}},
  \bibinfo{author}{\bibfnamefont{A.~S.} \bibnamefont{Kronfeld}},
  \bibnamefont{and} \bibinfo{author}{\bibfnamefont{P.~B.}
  \bibnamefont{Mackenzie}}, \bibinfo{journal}{Phys.Rev.}
  \textbf{\bibinfo{volume}{D55}}, \bibinfo{pages}{3933} (\bibinfo{year}{1997}),
  \eprint{hep-lat/9604004}.

\bibitem[{\citenamefont{Christ et~al.}(2007)\citenamefont{Christ, Li, and
  Lin}}]{Christ:2006us}
\bibinfo{author}{\bibfnamefont{N.~H.} \bibnamefont{Christ}},
  \bibinfo{author}{\bibfnamefont{M.}~\bibnamefont{Li}}, \bibnamefont{and}
  \bibinfo{author}{\bibfnamefont{H.-W.} \bibnamefont{Lin}},
  \bibinfo{journal}{Phys.Rev.} \textbf{\bibinfo{volume}{D76}},
  \bibinfo{pages}{074505} (\bibinfo{year}{2007}), \eprint{hep-lat/0608006}.

\bibitem[{\citenamefont{Namekawa et~al.}(2011)}]{Namekawa:2011wt}
\bibinfo{author}{\bibfnamefont{Y.}~\bibnamefont{Namekawa}} \bibnamefont{et~al.}
  (\bibinfo{collaboration}{PACS-CS Collaboration}),
  \bibinfo{journal}{Phys.Rev.} \textbf{\bibinfo{volume}{D84}},
  \bibinfo{pages}{074505} (\bibinfo{year}{2011}), \eprint{1104.4600}.

\bibitem[{\citenamefont{Beringer et~al.}(2012)}]{Beringer:1900zz}
\bibinfo{author}{\bibfnamefont{J.}~\bibnamefont{Beringer}} \bibnamefont{et~al.}
  (\bibinfo{collaboration}{Particle Data Group}), \bibinfo{journal}{Phys.Rev.}
  \textbf{\bibinfo{volume}{D86}}, \bibinfo{pages}{010001}
  (\bibinfo{year}{2012}).

\bibitem[{\citenamefont{Mohler and Woloshyn}(2011)}]{Mohler:2011ke}
\bibinfo{author}{\bibfnamefont{D.}~\bibnamefont{Mohler}} \bibnamefont{and}
  \bibinfo{author}{\bibfnamefont{R.}~\bibnamefont{Woloshyn}},
  \bibinfo{journal}{Phys.Rev.} \textbf{\bibinfo{volume}{D84}},
  \bibinfo{pages}{054505} (\bibinfo{year}{2011}), \eprint{1103.5506}.

\bibitem[{\citenamefont{McNeile and Michael}(2004)}]{McNeile:2004wu}
\bibinfo{author}{\bibfnamefont{C.}~\bibnamefont{McNeile}} \bibnamefont{and}
  \bibinfo{author}{\bibfnamefont{C.}~\bibnamefont{Michael}}
  (\bibinfo{collaboration}{UKQCD Collaboration}), \bibinfo{journal}{Phys.Rev.}
  \textbf{\bibinfo{volume}{D70}}, \bibinfo{pages}{034506}
  (\bibinfo{year}{2004}), \eprint{hep-lat/0402012}.

\bibitem[{\citenamefont{de~Forcrand et~al.}(2004)}]{deForcrand:2004ia}
\bibinfo{author}{\bibfnamefont{P.}~\bibnamefont{de~Forcrand}}
  \bibnamefont{et~al.} (\bibinfo{collaboration}{QCD-TARO Collaboration}),
  \bibinfo{journal}{JHEP} \textbf{\bibinfo{volume}{0408}}, \bibinfo{pages}{004}
  (\bibinfo{year}{2004}), \eprint{hep-lat/0404016}.

\bibitem[{\citenamefont{Levkova and DeTar}(2011)}]{Levkova:2010ft}
\bibinfo{author}{\bibfnamefont{L.}~\bibnamefont{Levkova}} \bibnamefont{and}
  \bibinfo{author}{\bibfnamefont{C.}~\bibnamefont{DeTar}},
  \bibinfo{journal}{Phys.Rev.} \textbf{\bibinfo{volume}{D83}},
  \bibinfo{pages}{074504} (\bibinfo{year}{2011}), \eprint{1012.1837}.

\bibitem[{\citenamefont{Koma and Koma}(2009)}]{Koma:2009ws}
\bibinfo{author}{\bibfnamefont{Y.}~\bibnamefont{Koma}} \bibnamefont{and}
  \bibinfo{author}{\bibfnamefont{M.}~\bibnamefont{Koma}},
  \bibinfo{journal}{PoS} \textbf{\bibinfo{volume}{LAT2009}},
  \bibinfo{pages}{122} (\bibinfo{year}{2009}), \eprint{0911.3204}.

\bibitem[{\citenamefont{Laschka et~al.}(2012)\citenamefont{Laschka, Kaiser, and
  Weise}}]{Laschka:2012cf}
\bibinfo{author}{\bibfnamefont{A.}~\bibnamefont{Laschka}},
  \bibinfo{author}{\bibfnamefont{N.}~\bibnamefont{Kaiser}}, \bibnamefont{and}
  \bibinfo{author}{\bibfnamefont{W.}~\bibnamefont{Weise}},
  \bibinfo{journal}{Phys.Lett.} \textbf{\bibinfo{volume}{B715}},
  \bibinfo{pages}{190} (\bibinfo{year}{2012}), \eprint{1205.3390}.

\bibitem[{\citenamefont{Koma et~al.}(2006)\citenamefont{Koma, Koma, and
  Wittig}}]{Koma:2006si}
\bibinfo{author}{\bibfnamefont{Y.}~\bibnamefont{Koma}},
  \bibinfo{author}{\bibfnamefont{M.}~\bibnamefont{Koma}}, \bibnamefont{and}
  \bibinfo{author}{\bibfnamefont{H.}~\bibnamefont{Wittig}},
  \bibinfo{journal}{Phys.Rev.Lett.} \textbf{\bibinfo{volume}{97}},
  \bibinfo{pages}{122003} (\bibinfo{year}{2006}), \eprint{hep-lat/0607009}.

\bibitem[{\citenamefont{Aoki et~al.}(1999)}]{Aoki:1998sb}
\bibinfo{author}{\bibfnamefont{S.}~\bibnamefont{Aoki}} \bibnamefont{et~al.}
  (\bibinfo{collaboration}{CP-PACS Collaboration}),
  \bibinfo{journal}{Nucl.Phys.Proc.Suppl.} \textbf{\bibinfo{volume}{73}},
  \bibinfo{pages}{216} (\bibinfo{year}{1999}), \eprint{hep-lat/9809185}.

\bibitem[{\citenamefont{Bali et~al.}(2000)}]{Bali:2000vr}
\bibinfo{author}{\bibfnamefont{G.~S.} \bibnamefont{Bali}} \bibnamefont{et~al.}
  (\bibinfo{collaboration}{TXL Collaboration, T(X)L Collaboration}),
  \bibinfo{journal}{Phys.Rev.} \textbf{\bibinfo{volume}{D62}},
  \bibinfo{pages}{054503} (\bibinfo{year}{2000}), \eprint{hep-lat/0003012}.

\bibitem[{\citenamefont{Bolder et~al.}(2001)\citenamefont{Bolder, Struckmann,
  Bali, Eicker, Lippert et~al.}}]{Bolder:2000un}
\bibinfo{author}{\bibfnamefont{B.}~\bibnamefont{Bolder}},
  \bibinfo{author}{\bibfnamefont{T.}~\bibnamefont{Struckmann}},
  \bibinfo{author}{\bibfnamefont{G.~S.} \bibnamefont{Bali}},
  \bibinfo{author}{\bibfnamefont{N.}~\bibnamefont{Eicker}},
  \bibinfo{author}{\bibfnamefont{T.}~\bibnamefont{Lippert}},
  \bibnamefont{et~al.}, \bibinfo{journal}{Phys.Rev.}
  \textbf{\bibinfo{volume}{D63}}, \bibinfo{pages}{074504}
  (\bibinfo{year}{2001}), \eprint{hep-lat/0005018}.

\bibitem[{\citenamefont{Bernard et~al.}(2003)}]{Bernard:2002sb}
\bibinfo{author}{\bibfnamefont{C.}~\bibnamefont{Bernard}} \bibnamefont{et~al.}
  (\bibinfo{collaboration}{MILC Collaboration}),
  \bibinfo{journal}{Nucl.Phys.Proc.Suppl.} \textbf{\bibinfo{volume}{119}},
  \bibinfo{pages}{598} (\bibinfo{year}{2003}), \eprint{hep-lat/0209051}.

\bibitem[{\citenamefont{Pennanen and Michael}(2000)}]{Pennanen:2000yk}
\bibinfo{author}{\bibfnamefont{P.}~\bibnamefont{Pennanen}} \bibnamefont{and}
  \bibinfo{author}{\bibfnamefont{C.}~\bibnamefont{Michael}}
  (\bibinfo{collaboration}{UKQCD Collaboration}) (\bibinfo{year}{2000}),
  \eprint{hep-lat/0001015}.

\bibitem[{\citenamefont{Bernard et~al.}(2001)\citenamefont{Bernard, DeGrand,
  Detar, Lacock, Gottlieb et~al.}}]{Bernard:2001tz}
\bibinfo{author}{\bibfnamefont{C.~W.} \bibnamefont{Bernard}},
  \bibinfo{author}{\bibfnamefont{T.~A.} \bibnamefont{DeGrand}},
  \bibinfo{author}{\bibfnamefont{C.~E.} \bibnamefont{Detar}},
  \bibinfo{author}{\bibfnamefont{P.}~\bibnamefont{Lacock}},
  \bibinfo{author}{\bibfnamefont{S.~A.} \bibnamefont{Gottlieb}},
  \bibnamefont{et~al.}, \bibinfo{journal}{Phys.Rev.}
  \textbf{\bibinfo{volume}{D64}}, \bibinfo{pages}{074509}
  (\bibinfo{year}{2001}), \eprint{hep-lat/0103012}.

\bibitem[{\citenamefont{Bali et~al.}(2005)\citenamefont{Bali, Neff, Duessel,
  Lippert, and Schilling}}]{Bali:2005fu}
\bibinfo{author}{\bibfnamefont{G.~S.} \bibnamefont{Bali}},
  \bibinfo{author}{\bibfnamefont{H.}~\bibnamefont{Neff}},
  \bibinfo{author}{\bibfnamefont{T.}~\bibnamefont{Duessel}},
  \bibinfo{author}{\bibfnamefont{T.}~\bibnamefont{Lippert}}, \bibnamefont{and}
  \bibinfo{author}{\bibfnamefont{K.}~\bibnamefont{Schilling}}
  (\bibinfo{collaboration}{SESAM Collaboration}), \bibinfo{journal}{Phys.Rev.}
  \textbf{\bibinfo{volume}{D71}}, \bibinfo{pages}{114513}
  (\bibinfo{year}{2005}), \eprint{hep-lat/0505012}.

\bibitem[{\citenamefont{Detar et~al.}(1999)\citenamefont{Detar, Kaczmarek,
  Karsch, and Laermann}}]{Detar:1998qa}
\bibinfo{author}{\bibfnamefont{C.~E.} \bibnamefont{Detar}},
  \bibinfo{author}{\bibfnamefont{O.}~\bibnamefont{Kaczmarek}},
  \bibinfo{author}{\bibfnamefont{F.}~\bibnamefont{Karsch}}, \bibnamefont{and}
  \bibinfo{author}{\bibfnamefont{E.}~\bibnamefont{Laermann}},
  \bibinfo{journal}{Phys.Rev.} \textbf{\bibinfo{volume}{D59}},
  \bibinfo{pages}{031501} (\bibinfo{year}{1999}), \eprint{hep-lat/9808028}.

\bibitem[{\citenamefont{Buchmuller}(1982)}]{Buchmuller:1981fr}
\bibinfo{author}{\bibfnamefont{W.}~\bibnamefont{Buchmuller}},
  \bibinfo{journal}{Phys.Lett.} \textbf{\bibinfo{volume}{B112}},
  \bibinfo{pages}{479} (\bibinfo{year}{1982}).

\bibitem[{\citenamefont{Rubin et~al.}(2005)}]{Rubin:2005px}
\bibinfo{author}{\bibfnamefont{P.}~\bibnamefont{Rubin}} \bibnamefont{et~al.}
  (\bibinfo{collaboration}{CLEO Collaboration}), \bibinfo{journal}{Phys.Rev.}
  \textbf{\bibinfo{volume}{D72}}, \bibinfo{pages}{092004}
  (\bibinfo{year}{2005}), \eprint{hep-ex/0508037}.

\bibitem[{\citenamefont{Dobbs et~al.}(2008)}]{Dobbs:2008ec}
\bibinfo{author}{\bibfnamefont{S.}~\bibnamefont{Dobbs}} \bibnamefont{et~al.}
  (\bibinfo{collaboration}{CLEO Collaboration}),
  \bibinfo{journal}{Phys.Rev.Lett.} \textbf{\bibinfo{volume}{101}},
  \bibinfo{pages}{182003} (\bibinfo{year}{2008}), \eprint{0805.4599}.

\bibitem[{\citenamefont{Burns}(2011)}]{Burns:2011fu}
\bibinfo{author}{\bibfnamefont{T.}~\bibnamefont{Burns}},
  \bibinfo{journal}{Phys.Rev.} \textbf{\bibinfo{volume}{D84}},
  \bibinfo{pages}{034021} (\bibinfo{year}{2011}), \eprint{1105.2533}.

\bibitem[{\citenamefont{Eichten et~al.}(1976)\citenamefont{Eichten, Gottfried,
  Kinoshita, Lane, and Yan}}]{Eichten:1975ag}
\bibinfo{author}{\bibfnamefont{E.}~\bibnamefont{Eichten}},
  \bibinfo{author}{\bibfnamefont{K.}~\bibnamefont{Gottfried}},
  \bibinfo{author}{\bibfnamefont{T.}~\bibnamefont{Kinoshita}},
  \bibinfo{author}{\bibfnamefont{K.}~\bibnamefont{Lane}}, \bibnamefont{and}
  \bibinfo{author}{\bibfnamefont{T.-M.} \bibnamefont{Yan}},
  \bibinfo{journal}{Phys.Rev.Lett.} \textbf{\bibinfo{volume}{36}},
  \bibinfo{pages}{500} (\bibinfo{year}{1976}).

\bibitem[{\citenamefont{Eichten et~al.}(1978)\citenamefont{Eichten, Gottfried,
  Kinoshita, Lane, and Yan}}]{Eichten:1978tg}
\bibinfo{author}{\bibfnamefont{E.}~\bibnamefont{Eichten}},
  \bibinfo{author}{\bibfnamefont{K.}~\bibnamefont{Gottfried}},
  \bibinfo{author}{\bibfnamefont{T.}~\bibnamefont{Kinoshita}},
  \bibinfo{author}{\bibfnamefont{K.}~\bibnamefont{Lane}}, \bibnamefont{and}
  \bibinfo{author}{\bibfnamefont{T.-M.} \bibnamefont{Yan}},
  \bibinfo{journal}{Phys.Rev.} \textbf{\bibinfo{volume}{D17}},
  \bibinfo{pages}{3090} (\bibinfo{year}{1978}).

\bibitem[{\citenamefont{Barnes and Ghandour}(1982)}]{Barnes:1982eg}
\bibinfo{author}{\bibfnamefont{T.}~\bibnamefont{Barnes}} \bibnamefont{and}
  \bibinfo{author}{\bibfnamefont{G.}~\bibnamefont{Ghandour}},
  \bibinfo{journal}{Phys.Lett.} \textbf{\bibinfo{volume}{B118}},
  \bibinfo{pages}{411} (\bibinfo{year}{1982}).

\bibitem[{\citenamefont{Ebert et~al.}(2005)\citenamefont{Ebert, Faustov, and
  Galkin}}]{Ebert:2005jj}
\bibinfo{author}{\bibfnamefont{D.}~\bibnamefont{Ebert}},
  \bibinfo{author}{\bibfnamefont{R.}~\bibnamefont{Faustov}}, \bibnamefont{and}
  \bibinfo{author}{\bibfnamefont{V.}~\bibnamefont{Galkin}},
  \bibinfo{journal}{Mod.Phys.Lett.} \textbf{\bibinfo{volume}{A20}},
  \bibinfo{pages}{875} (\bibinfo{year}{2005}), \eprint{hep-ph/0503012}.

\bibitem[{\citenamefont{Eichten et~al.}(1980)\citenamefont{Eichten, Gottfried,
  Kinoshita, Lane, and Yan}}]{Eichten:1979ms}
\bibinfo{author}{\bibfnamefont{E.}~\bibnamefont{Eichten}},
  \bibinfo{author}{\bibfnamefont{K.}~\bibnamefont{Gottfried}},
  \bibinfo{author}{\bibfnamefont{T.}~\bibnamefont{Kinoshita}},
  \bibinfo{author}{\bibfnamefont{K.}~\bibnamefont{Lane}}, \bibnamefont{and}
  \bibinfo{author}{\bibfnamefont{T.-M.} \bibnamefont{Yan}},
  \bibinfo{journal}{Phys.Rev.} \textbf{\bibinfo{volume}{D21}},
  \bibinfo{pages}{203} (\bibinfo{year}{1980}).

\bibitem[{\citenamefont{Swanson}(2006)}]{Swanson:2006st}
\bibinfo{author}{\bibfnamefont{E.~S.} \bibnamefont{Swanson}},
  \bibinfo{journal}{Phys.Rept.} \textbf{\bibinfo{volume}{429}},
  \bibinfo{pages}{243} (\bibinfo{year}{2006}), \eprint{hep-ph/0601110}.

\bibitem[{\citenamefont{Rosner et~al.}(2005)}]{Rosner:2005ry}
\bibinfo{author}{\bibfnamefont{J.}~\bibnamefont{Rosner}} \bibnamefont{et~al.}
  (\bibinfo{collaboration}{CLEO Collaboration}),
  \bibinfo{journal}{Phys.Rev.Lett.} \textbf{\bibinfo{volume}{95}},
  \bibinfo{pages}{102003} (\bibinfo{year}{2005}), \eprint{hep-ex/0505073}.

\bibitem[{\citenamefont{Choi et~al.}(2002)}]{Choi:2002na}
\bibinfo{author}{\bibfnamefont{S.}~\bibnamefont{Choi}} \bibnamefont{et~al.}
  (\bibinfo{collaboration}{BELLE collaboration}),
  \bibinfo{journal}{Phys.Rev.Lett.} \textbf{\bibinfo{volume}{89}},
  \bibinfo{pages}{102001} (\bibinfo{year}{2002}), \eprint{hep-ex/0206002}.

\bibitem[{\citenamefont{Vinokurova et~al.}(2011)}]{Vinokurova:2011dy}
\bibinfo{author}{\bibfnamefont{A.}~\bibnamefont{Vinokurova}}
  \bibnamefont{et~al.} (\bibinfo{collaboration}{Belle collaboration}),
  \bibinfo{journal}{Phys.Lett.} \textbf{\bibinfo{volume}{B706}},
  \bibinfo{pages}{139} (\bibinfo{year}{2011}), \eprint{1105.0978}.

\bibitem[{\citenamefont{del Amo~Sanchez et~al.}(2011)}]{delAmoSanchez:2011bt}
\bibinfo{author}{\bibfnamefont{P.}~\bibnamefont{del Amo~Sanchez}}
  \bibnamefont{et~al.} (\bibinfo{collaboration}{BABAR Collaboration}),
  \bibinfo{journal}{Phys.Rev.} \textbf{\bibinfo{volume}{D84}},
  \bibinfo{pages}{012004} (\bibinfo{year}{2011}), \eprint{1103.3971}.

\bibitem[{\citenamefont{Aubert et~al.}(2004)}]{Aubert:2003pt}
\bibinfo{author}{\bibfnamefont{B.}~\bibnamefont{Aubert}} \bibnamefont{et~al.}
  (\bibinfo{collaboration}{BABAR Collaboration}),
  \bibinfo{journal}{Phys.Rev.Lett.} \textbf{\bibinfo{volume}{92}},
  \bibinfo{pages}{142002} (\bibinfo{year}{2004}), \eprint{hep-ex/0311038}.

\bibitem[{\citenamefont{Charron}(2014)}]{Charron:2013paa}
\bibinfo{author}{\bibfnamefont{B.}~\bibnamefont{Charron}}
  (\bibinfo{collaboration}{HAL QCD}), \bibinfo{journal}{PoS}
  \textbf{\bibinfo{volume}{LATTICE2013}}, \bibinfo{pages}{223}
  (\bibinfo{year}{2014}), \eprint{1312.1032}.

\bibitem[{\citenamefont{Badalian et~al.}(2009)\citenamefont{Badalian, Bakker,
  and Danilkin}}]{Badalian:2008dv}
\bibinfo{author}{\bibfnamefont{A.}~\bibnamefont{Badalian}},
  \bibinfo{author}{\bibfnamefont{B.}~\bibnamefont{Bakker}}, \bibnamefont{and}
  \bibinfo{author}{\bibfnamefont{I.}~\bibnamefont{Danilkin}},
  \bibinfo{journal}{Phys.Atom.Nucl.} \textbf{\bibinfo{volume}{72}},
  \bibinfo{pages}{638} (\bibinfo{year}{2009}), \eprint{0805.2291}.

\bibitem[{\citenamefont{Murano}(2011)}]{Murano:2011aa}
\bibinfo{author}{\bibfnamefont{K.}~\bibnamefont{Murano}}
  (\bibinfo{collaboration}{HALQCD Collaboration}), \bibinfo{journal}{PoS}
  \textbf{\bibinfo{volume}{LATTICE2011}}, \bibinfo{pages}{319}
  (\bibinfo{year}{2011}), \eprint{1112.2051}.

\bibitem[{\citenamefont{Michael}(1985)}]{Michael:1985ne}
\bibinfo{author}{\bibfnamefont{C.}~\bibnamefont{Michael}},
  \bibinfo{journal}{Nucl.Phys.} \textbf{\bibinfo{volume}{B259}},
  \bibinfo{pages}{58} (\bibinfo{year}{1985}).

\bibitem[{\citenamefont{Luscher and Wolff}(1990)}]{Luscher:1990ck}
\bibinfo{author}{\bibfnamefont{M.}~\bibnamefont{Luscher}} \bibnamefont{and}
  \bibinfo{author}{\bibfnamefont{U.}~\bibnamefont{Wolff}},
  \bibinfo{journal}{Nucl.Phys.} \textbf{\bibinfo{volume}{B339}},
  \bibinfo{pages}{222} (\bibinfo{year}{1990}).

\bibitem[{\citenamefont{Kawanai and Sasaki}()}]{SSandTK}
\bibinfo{author}{\bibfnamefont{T.}~\bibnamefont{Kawanai}} \bibnamefont{and}
  \bibinfo{author}{\bibfnamefont{S.}~\bibnamefont{Sasaki}}, \bibinfo{note}{{in
  progress}}.

\bibitem[{\citenamefont{Aoki et~al.}(2010{\natexlab{b}})}]{Aoki:2009ix}
\bibinfo{author}{\bibfnamefont{S.}~\bibnamefont{Aoki}} \bibnamefont{et~al.}
  (\bibinfo{collaboration}{PACS-CS Collaboration}),
  \bibinfo{journal}{Phys.Rev.} \textbf{\bibinfo{volume}{D81}},
  \bibinfo{pages}{074503} (\bibinfo{year}{2010}{\natexlab{b}}),
  \eprint{0911.2561}.

\bibitem[{\citenamefont{Sasaki and Yamazaki}(2006)}]{Sasaki:2006jn}
\bibinfo{author}{\bibfnamefont{S.}~\bibnamefont{Sasaki}} \bibnamefont{and}
  \bibinfo{author}{\bibfnamefont{T.}~\bibnamefont{Yamazaki}},
  \bibinfo{journal}{Phys.Rev.} \textbf{\bibinfo{volume}{D74}},
  \bibinfo{pages}{114507} (\bibinfo{year}{2006}), \eprint{hep-lat/0610081}.

\bibitem[{\citenamefont{Lucha and Schoberl}(2003)}]{Lucha:2003gs}
\bibinfo{author}{\bibfnamefont{W.}~\bibnamefont{Lucha}} \bibnamefont{and}
  \bibinfo{author}{\bibfnamefont{F.~F.} \bibnamefont{Schoberl}},
  \bibinfo{journal}{Mod.Phys.Lett.} \textbf{\bibinfo{volume}{A18}},
  \bibinfo{pages}{2837} (\bibinfo{year}{2003}), \eprint{hep-ph/0309341}.

\bibitem[{\citenamefont{Lang et~al.}(2014)\citenamefont{Lang, Leskovec, Mohler,
  Prelovsek, and Woloshyn}}]{Lang:2014yfa}
\bibinfo{author}{\bibfnamefont{C.}~\bibnamefont{Lang}},
  \bibinfo{author}{\bibfnamefont{L.}~\bibnamefont{Leskovec}},
  \bibinfo{author}{\bibfnamefont{D.}~\bibnamefont{Mohler}},
  \bibinfo{author}{\bibfnamefont{S.}~\bibnamefont{Prelovsek}},
  \bibnamefont{and} \bibinfo{author}{\bibfnamefont{R.}~\bibnamefont{Woloshyn}},
  \bibinfo{journal}{Phys.Rev.} \textbf{\bibinfo{volume}{D90}},
  \bibinfo{pages}{034510} (\bibinfo{year}{2014}), \eprint{1403.8103}.

\bibitem[{\citenamefont{Mart?nez~Torres
  et~al.}(2015)\citenamefont{Mart?nez~Torres, Oset, Prelovsek, and
  Ramos}}]{Torres:2014vna}
\bibinfo{author}{\bibfnamefont{A.}~\bibnamefont{Mart?nez~Torres}},
  \bibinfo{author}{\bibfnamefont{E.}~\bibnamefont{Oset}},
  \bibinfo{author}{\bibfnamefont{S.}~\bibnamefont{Prelovsek}},
  \bibnamefont{and} \bibinfo{author}{\bibfnamefont{A.}~\bibnamefont{Ramos}},
  \bibinfo{journal}{JHEP} \textbf{\bibinfo{volume}{05}}, \bibinfo{pages}{153}
  (\bibinfo{year}{2015}), \eprint{1412.1706}.

\bibitem[{\citenamefont{Barnes and Swanson}(2008)}]{Barnes:2007xu}
\bibinfo{author}{\bibfnamefont{T.}~\bibnamefont{Barnes}} \bibnamefont{and}
  \bibinfo{author}{\bibfnamefont{E.}~\bibnamefont{Swanson}},
  \bibinfo{journal}{Phys.Rev.} \textbf{\bibinfo{volume}{C77}},
  \bibinfo{pages}{055206} (\bibinfo{year}{2008}), \eprint{0711.2080}.

\bibitem[{\citenamefont{Godfrey and Kokoski}(1991)}]{Godfrey:1986wj}
\bibinfo{author}{\bibfnamefont{S.}~\bibnamefont{Godfrey}} \bibnamefont{and}
  \bibinfo{author}{\bibfnamefont{R.}~\bibnamefont{Kokoski}},
  \bibinfo{journal}{Phys.Rev.} \textbf{\bibinfo{volume}{D43}},
  \bibinfo{pages}{1679} (\bibinfo{year}{1991}).

\bibitem[{\citenamefont{Close and Swanson}(2005)}]{Close:2005se}
\bibinfo{author}{\bibfnamefont{F.}~\bibnamefont{Close}} \bibnamefont{and}
  \bibinfo{author}{\bibfnamefont{E.}~\bibnamefont{Swanson}},
  \bibinfo{journal}{Phys.Rev.} \textbf{\bibinfo{volume}{D72}},
  \bibinfo{pages}{094004} (\bibinfo{year}{2005}), \eprint{hep-ph/0505206}.

\end{thebibliography}
\end{document}